\documentclass{ws-procs975x65}

\usepackage{graphicx,here,rotate,epsf,epsfig}
\usepackage{axodraw}
 
\begin{document}

\title{Particle Physics and Cosmology}
 
\author{John Ellis}
 
\address{Theoretical Physics Division, CERN, CH- 1211 Geneva 23, 
Switzerland\\
E-mail: John.Ellis@cern.ch}

%
%
%
 
\newcommand{\mycomm}[1]{\hfill\break{ \tt===$>$ \bf #1}\hfill\break}

\def\ga{\mathrel{\raise.3ex\hbox{$>$\kern-.75em\lower1ex\hbox{$\sim$}}}}
\def\la{\mathrel{\raise.3ex\hbox{$<$\kern-.75em\lower1ex\hbox{$\sim$}}}}
\def\gev{{\rm \, Ge\kern-0.125em V}}
\def\tev{{\rm \, Te\kern-0.125em V}}
\def\beq{\begin{equation}}
\def\eeq{\end{equation}}
\def\st{\scriptstyle}
\def\ss{\scriptscriptstyle}
\def\mb{m_{\widetilde B}}
\def\msf{m_{\tilde f}}
\def\mst{m_{\tilde t}}
\def\mf{m_{\ss{f}}}
\def\mpar{m_{\ss\|}^2}
\def\mpl{M_{\rm Pl}}
\def\mchi{m_{\chi}}
\def\ohsq{\Omega_{\chi} h^2}
\def\msn{m_{\tilde\nu}}
\def\m12{m_{1\!/2}}
\def\mstpl{m_{\tilde t_{\ss 1}}^2}
\def\mstpr{m_{\tilde t_{\ss 2}}^2}

\def\sm{Standard Model}

\def\ga{\mathrel{\raise.3ex\hbox{$>$\kern-.75em\lower1ex\hbox{$\sim$}}}}
\def\la{\mathrel{\raise.3ex\hbox{$<$\kern-.75em\lower1ex\hbox{$\sim$}}}}
\def\gyr{{\rm \, G\kern-0.125em yr}}
\def\gev{{\rm \, Ge\kern-0.125em V}}
\def\tev{{\rm \, Te\kern-0.125em V}}
\def\beq{\begin{equation}}
\def\eeq{\end{equation}}
\def\ss{\scriptscriptstyle}
\def\scs{\scriptstyle}
\def\mb{m_{\widetilde B}}
\def\mst{m_{\tilde\tau_R}}
\def\mstop{m_{\tilde t_1}}
\def\msl{m_{\tilde{\ell}_1}}
\def\stau{\tilde \tau}
\def\stop{\tilde t}
\def\sbot{\tilde b}
\def\mchi{m_{\tilde \chi}}
\def\mxi{m_{\tilde{\chi}_i^0}}
\def\mxj{m_{\tilde{\chi}_j^0}}
\def\mchari{m_{\tilde{\chi}_i^+}}
\def\mcharj{m_{\tilde{\chi}_j^+}}
\def\mgluino{m_{\tilde g}}
\def\msf{m_{\tilde f}}
\def\m12{m_{1\!/2}}
\def\mtb{\overline{m}_{\ss t}}
\def\mbb{\overline{m}_{\ss b}}
\def\mfb{\overline{m}_{\ss f}}
\def\mf{m_{\ss{f}}}
\def\gt{\gamma_t}
\def\gb{\gamma_b}
\def\gf{\gamma_f}
\def\thm{\theta_\mu}
\def\tha{\theta_A}
\def\thb{\theta_B}
\def\mgl{m_{\ss \tilde g}}
\def\cp{C\!P}
\def\ch{{\widetilde \chi}} 
\def\st{{\widetilde \tau}_{\scriptscriptstyle\rm 1}}
\def\sel{{\widetilde e}_{\scriptscriptstyle\rm R}}
\def\sl{{\widetilde \ell}_{\scriptscriptstyle\rm R}}
\def\msn{m_{\ch}}
\def\tsq{|{\cal T}|^2}
\def\tcm{\theta_{\rm\scriptscriptstyle CM}}
\def\half{{\textstyle{1\over2}}}
\def\neq{n_{\rm eq}}
\def\qeq{q_{\rm eq}}
\def\slash#1{\rlap{\hbox{$\mskip 1 mu /$}}#1}%
\def\mw{m_W}
\def\mz{m_Z}
\def\mhb{m_{H}}
\def\mhl{m_{h}}
\newcommand\f[1]{f_#1}
\def\nl{\hfill\nonumber\\&&}

\def\gappeq{\mathrel{\rlap {\raise.5ex\hbox{$>$}}
{\lower.5ex\hbox{$\sim$}}}}

\def\lappeq{\mathrel{\rlap{\raise.5ex\hbox{$<$}}
{\lower.5ex\hbox{$\sim$}}}}

\def\Toprel#1\over#2{\mathrel{\mathop{#2}\limits^{#1}}}
\def\FF{\Toprel{\hbox{$\scriptscriptstyle(-)$}}\over{$\nu$}}

\newcommand{\Zee}{$Z^0$}


\def\Yi{\eta^{\ast}_{11} \left( \frac{y_{i}}{2} g' Z_{\chi 1} + 
        g T_{3i} Z_{\chi 2} \right) + \eta^{\ast}_{12} 
        \frac{g m_{q_{i}} Z_{\chi 5-i}}{2 m_{W} B_{i}}}

\def\Xii{\eta^{\ast}_{11} 
        \frac{g m_{q_{i}}Z_{\chi 5-i}^{\ast}}{2 m_{W} B_{i}} - 
        \eta_{12}^{\ast} e_{i} g' Z_{\chi 1}^{\ast}}

\def\Wi{\eta_{21}^{\ast}
        \frac{g m_{q_{i}}Z_{\chi 5-i}^{\ast}}{2 m_{W} B_{i}} -
        \eta_{22}^{\ast} e_{i} g' Z_{\chi 1}^{\ast}}

\def\Vi{\eta_{22}^{\ast} \frac{g m_{q_{i}} Z_{\chi 5-i}}{2 m_{W} B_{i}}
        + \eta_{21}^{\ast}\left( \frac{y_{i}}{2} g' Z_{\chi 1}
        + g T_{3i} Z_{\chi 2} \right)}

\def\zthree{\delta_{1i} [g Z_{\chi 2} - g' Z_{\chi 1}]}

\def\zfour{\delta_{2i} [g Z_{\chi 2} - g' Z_{\chi 1}]}


\maketitle
 
\abstracts{
In the first Lecture, the Big Bang and the Standard Model of particle
physics are introduced, as well as the structure of the latter and open
issues beyond it. Neutrino physics is discussed in the second Lecture,
with emphasis on models for neutrino masses and oscillations. The third
Lecture is devoted to supersymmetry, including the prospects for
discovering it at accelerators or as cold dark matter. Inflation is
reviewed from the viewpoint of particle physics in the fourth Lecture,
including simple models with a single scalar inflaton field: the
possibility that this might be a sneutrino is proposed. Finally, the fifth
Lecture is devoted to topics further beyond the Standard Model, such as
grand unification, baryo- and leptogenesis - that might be due to
sneutrino inflaton decays - and ultra-high-energy cosmic rays - that might
be due to the decays of metastable superheavy dark matter particles.}

\begin{center}
{\it Lectures presented at the Australian National University Summer 
School on the New Cosmology, Canberra, January 2003}\\
\end{center}
\begin{center}
CERN-TH/2003-098 $\; \; \; \;$ {\tt astro-ph/0305038}
\end{center}

\section{Introduction to the Standard Models}

\subsection{The Big Bang and Particle Physics}

The Universe is currently expanding almost homogeneously and 
isotropically, as discovered by Hubble, and the radiation it contains is 
cooling as it expands adiabatically:
\beq
a \times T \; \simeq \; {\rm Constant},
\label{adiabatic}
\eeq
where $a$ is the scale factor of the Universe and $T$ is the temperature. 
There are two important pieces of evidence that the scale factor 
of the Universe was once much smaller than it is today, and 
correspondingly that its temperature was much higher. One is the {\it 
Cosmic Microwave Background}~\cite{CMB}, which bathes us in photons with a 
density
\beq
n_\gamma \; \simeq \; 400~{\rm cm}^{-3},
\label{photons}
\eeq
with an effective temperature $T \simeq 2.7$~K. These photons
were released when electrons and nuclei combined to form atoms, when the 
Universe was some 3000 times hotter and the
scale factor correspondingly 3000 times smaller than it is today. The
second is the agreement of the {\it Abundances of Light 
Elements}~\cite{BBN}, in
particular those of $^4$He, Deuterium and $^6$Li, with calculations of
cosmological nucelosynthesis. For these elements to have been produced by
nuclear fusion, the Universe must once have been some $10^9$ times hotter
and smaller than it is today.

During this epoch of the history of the Universe, its energy density would 
have been dominated by relativistic particles such as photons and 
neutrinos, in which case the age $t$ of the Universe is given 
approximately by
\beq
t \; \propto \; a^2 \; \propto \; {1 \over T^2}.
\label{expnrate}
\eeq
The constant of proportionality between time and temperature is such that 
$t \simeq 1$~second when the temperature $T \simeq 1$~MeV, near the start 
of cosmological nucleosynthesis. Since typical particle energies in a 
thermal plasma are ${\cal O}(T)$, and the Boltzmann distribution 
guarantees 
large densities of particles weighing ${\cal O}(T)$, the history of the 
earlier Universe when $ T > {\cal O}(1)$~MeV was dominated by elementary 
particles weighing an MeV or more~\cite{KT}.

The landmarks in the history of the Universe during its first second
presumably included the epoch when protons and neutrons were created out
of quarks, when $T \sim 200$~MeV and $t \sim 10^{-5}$~s. Prior to that,
there was an epoch when the symmetry between weak and electromagnetic
interactions was broken, when $T \sim 100$~GeV and $t \sim 10^{-10}$~s.  
Laboratory experiments with accelerators have already explored physics at
energies $E \lappeq 100$~GeV, and the energy range $E \lappeq 1000$~GeV,
corresponding to the history of the Universe when $t \gappeq 10^{-12}$~s,
will be explored at CERN's LHC accelerator that is scheduled to start
operation in 2007~\cite{LHC}. Our ideas about physics at earlier epochs
are necessarily more speculative, but one possibility is that there was an
inflationary epoch when the age of the Universe was somewhere between
$10^{-40}$ and $10^{-30}$~s.

We return later to possible experimental probes of the physics of these
early epochs, but first we review the Standard Model of particle physics,
which underlies our description of the Universe since it was $10^{-10}$~s
old.

\subsection{Summary of the Standard Model of Particle Physics}

The \sm~ of particle physics has been established by a series of 
experiments and theoretical developments over the past century~\cite{SM}, 
including:
\begin{itemize}
\item
1897 - The discovery of the electron;
\item
1910 - The discovery of the nucleus;
\item
1930 - The nucleus found to be made of protons and neutrons; neutrino 
postulated;
\item
1936 - The muon discovered;
\item
1947 - Pion and strange particles discovered;
\item
1950's - Many strongly-interacting particles discovered;
\item
1964 - Quarks proposed;
\item
1967 - The Standard Model proposed;
\item
1973 - Neutral weak interactions discovered;
\item
1974 - The charm quark discovered;
\item
1975 - The $\tau$ lepton discovered;
\item
1977 - The bottom quark discovered;
\item
1979 - The gluon discovered;
\item
1983 - The intermediate $W^\pm, Z^0$ bosons discovered;
\item
1989 - Three neutrino species counted;
\item
1994 - The top quark discovered;
\item
1998 - Neutrino oscillations discovered.
\end{itemize}

All the above historical steps, apart from the last (which was made with 
neutrinos from astrophysical sources), fit within the \sm, 
and the \sm~ continues to survive all experimental tests at accelerators. 

The \sm~ contains  the following set of spin-1/2 matter 
particles:
\bea
& {\rm Leptons}:&
\left(
\begin{array}{c}
\nu_e \\
e
\end{array}
\right),
\left(
\begin{array}{c}
\nu_\mu \\
\mu
\end{array}
\right),
\left(
\begin{array}{c}
\nu_\tau \\
\tau
\end{array}
\right) \\
& {\rm Quarks}:& \;
\left(
\begin{array}{c}
u \\
d
\end{array}
\right), \; \;
\left(
\begin{array}{c}
c \\
s
\end{array}
\right), \; \;
\left(
\begin{array}{c}
b \\
t
\end{array}
\right)
\label{matter}
\eea
We know from experiments at CERN's LEP accelerator in 1989 that there can 
there can only be three neutrinos~\cite{LEPEWWG}:
\beq
N_\nu = 2.9841 \pm 0.0083,
\label{oneforteenone}
\eeq
which is a couple of standard deviations below 3, but that cannot be 
considered a significant discrepancy.
I had always hoped that $N_\nu$ might turn out to be non-integer: $N_\nu =
\pi$ would have been good, and $N_\nu = e$ would have been even better,
but this was not to be! The constraint (\ref{oneforteenone}) is also
important for possible physics beyond the \sm, such as supersymmetry as we
discuss later. The measurement (\ref{oneforteenone}) implies, by
extension, that there can only be three charged leptons and hence
no more quarks, by analogy and in order to preserve the 
calculability of the \sm~\cite{BIM}.

The forces between these matter particles are carried by spin-1 bosons:
electromagnetism by the familiar massless photon $\gamma$, the weak
interactions by the massive intermediate $W^\pm$ and $Z^0$ bosons that
weigh $\simeq 80, 91$~GeV, respectively, and the strong interactions by
the massless gluon. {\it Among the key objectives of particle physics are
attempts to unify these different interactions, and to explain the very
different masses of the various matter particles and spin-1 bosons.}

Since the \sm~ is the rock on which our quest for new physics must be
built, we now review its basic features~\cite{SM} and examine whether its
successes offer any hint of the direction in which to search for new
physics. Let us first recall the structure of the
charged-current weak interactions, which have the current-current form:
\beq
{1 \over 4}{\cal L}_{cc} = {G_F\over
\sqrt{2}}~~J^+_\mu~J^{-\mu},
\label{onetwelve}
\eeq
where the charged currents violate parity maximally:
\beq
J^+_\mu \; = \; \Sigma_{\ell = e, \mu, \tau} {\bar \ell} \gamma_\mu (1 - 
\gamma_5 ) \nu_\ell \; + \; {\rm~similarly~for~quarks}.
\label{chargedcurrent}
\eeq
The charged current (\ref{chargedcurrent}) can be interpreted as a 
generator of a weak SU(2) isospin symmetry acting on the matter-particle 
doublets in (\ref{matter}). The matter fermions with left-handed 
helicities are doublets of this weak SU(2), whereas the right-handed 
matter fermions are singlets. It was suggested already in the 1930's, 
and with more conviction in the 1960's, that the structure 
(\ref{chargedcurrent}) could most naturally be obtained by exchanging 
massive $W^\pm$ vector bosons with coupling $g$ and mass $m_W$:
\beq
{G_F\over\sqrt{2}} \equiv {g^2\over 8 m^2_W}.
\label{GFW}
\eeq
In 1973, neutral weak interactions with an analogous current-current 
structure were discovered at CERN:
\beq
{1 \over 4} {\cal L}_{NC} = {G^{NC}_F\over \sqrt{2}}~~J^0_\mu~J^{\mu
0},
\label{onethirteen}
\eeq
and it was natural to suggest that these might also be carried by massive 
neutral vector bosons $Z^0$. 

The $W^\pm$ and $Z^0$ bosons were discovered at CERN in 1983, so 
let us now review the theory of them, as well as the Higgs mechanism of
spontaneous symmetry breaking by which we believe they acquire
masses~\cite{SSB}. The vector bosons are described by the Lagrangian
\beq
{\cal L} = -\frac{1}{4}~G^i_{\mu\nu} G^{i\mu\nu} - {1\over 4}
F_{\mu\nu}F^{\mu\nu}
\label{oneone}
\eeq
where $G^i_{\mu\nu} \equiv \partial_\mu W^i_\nu - \partial_\nu W^i_\mu +
i g \epsilon_{ijk}W^j_\mu W^k_\nu$ is the field strength for the SU(2)
vector boson $W^i_\mu$, and $F_{\mu\nu} \equiv \partial_\mu W^i_\nu - 
\partial_\nu W^i_\mu$ is the field strength for a U(1) vector boson
$B_\mu$ that is needed when we incorporate electromagnetism. The 
Lagrangian (\ref{oneone}) contains bilinear terms that yield the
boson propagators, and also trilinear and quartic vector-boson 
interactions. 

The vector bosons couple to quarks and leptons via
\beq
{\cal L}_F = -\sum_f i~~\left[ \bar f_L \gamma^\mu D_\mu f_L +
\bar f_R \gamma^\mu D_\mu f_R \right]
\label{onetwo}
\eeq
where the $D_\mu$ are covariant derivatives:
\beq
D_\mu \equiv \partial_\mu - i~ g ~\sigma_i~ W^i_\mu - i~g^\prime~ Y~ B_\mu
\label{onethree}
\eeq
The SU(2) piece appears only for the left-handed fermions $f_L$, whereas
the U(1) vector boson $B_\mu$ couples to both left- and right-handed 
compnents, via their respective hypercharges $Y$.

The origin of all the masses in the \sm~ is postulated to be a weak 
doublet of scalar Higgs fields, whose kinetic term in the Lagrangian is
\beq
{\cal L}_\phi = -\vert D_\mu \phi\vert^2
\label{onefour}
\eeq
and which has the magic potential:
\beq
{\cal L}_V = -V(\phi ) : V(\phi ) = -\mu^2\phi^{\dagger}\phi +
{\lambda \over 2} (\phi^{\dagger}\phi)^2
\label{onefive}
\eeq
Because of the negative sign for the quadratic term in (\ref{onefive}),
the symmetric solution $<0\vert\phi\vert 0> = 0$ is unstable, and if
$\lambda > 0$ the favoured solution has a non-zero vacuum expectation
value which we may write in the form:
\beq
<0\vert\phi\vert 0> = <0\vert\phi^{\dagger}\vert 0> = v\left(\matrix{0\cr
{1\over
\sqrt{2}}}\right) : v^2 = {\mu^2\over 2\lambda}
\label{onesix}
\eeq
corresponding to spontaneous breakdown of the electroweak symmetry.

Expanding around the vacuum: $\phi = <0\vert\phi\vert 0> + \,\hat\phi$,
the
kinetic term (\ref{onefour}) for the Higgs field yields mass terms for
the vector bosons:
\beq
{\cal L}_\phi \ni -{g^2v^2\over 2}~~W^+_\mu ~W^{\mu -} - g^{\prime 2}
~{v^2\over 2}~B_\mu~B^\mu + g~g^\prime  v^2~B_\mu~W^{\mu 3} -
g^2~{v^2 \over 2}~W^3_\mu~W^{\mu 3}
\label{oneseven}
\eeq
corresponding to masses
\beq
m_{W^\pm} = {gv\over 2}
\label{oneeight}
\eeq
for the charged vector bosons.
The neutral vector bosons $(W^3_\mu , B_\mu)$ have a 2$\times$2
mass-squared matrix:
\beq
\left(\matrix{
{g^2\over 2} & {-gg^\prime\over 2} \cr \cr
{-gg^\prime\over 2} & {g^{\prime 2}\over 2}}\right) v^2
\label{onenine}
\eeq
This is easily diagonalized to yield the mass eigenstates:
\beq
Z_\mu = {gW^3_\mu - g^\prime B_\mu\over \sqrt{g^2+g^{\prime 2}}}~ :~~ m_Z
= {1\over 2} \sqrt{g^2+g^{\prime 2}} v~;~~
A_\mu = {g^\prime W^3_\mu + g B_\mu\over \sqrt{g^2+g^{\prime 2}}} ~:~~
m_A = 0
\label{oneten}
\eeq
that we identify with the massive $Z^0$ and massless $\gamma$, 
respectively. It is useful to
introduce the electroweak mixing angle $\theta_W$ defined by
\beq
\sin\theta_W = {g^\prime\over \sqrt{g^2+g^{\prime 2}}} 
\label{oneeleven}
\eeq
in terms of the  weak SU(2) coupling $g$ and the weak U(1) 
coupling $g^\prime$. Many other
quantities can be expressed in terms of $\sin\theta_W$ (\ref{oneeleven}):
for example, $m^2_W/ m^2_Z = \cos^2\theta_W$. 

With these boson masses, one indeed obtains charged-current
interactions of the current-current form (\ref{chargedcurrent}) shown 
above, and the neutral currents take the form:
\beq
J^0_\mu \equiv J^3_\mu - \sin^2\theta_W~J^{em}_\mu~,~~G^{NC}_F \equiv
{g^2+g^{\prime 2}\over 8 m^2_Z}
\label{neutralcurrent}
\eeq
The ratio of neutral- and charged-current interaction strengths is often
expressed as
\beq
\rho = {G^{NC}_F\over G_F} = {m^2_W\over m^2_Z \cos^2\theta_W}
\label{oneforteen}
\eeq
which takes the value unity in the \sm, apart from quantum corrections 
(loop effects).

The previous field-theoretical discussion of the Higgs mechanism can be
rephrased in more physical language. It is well known that a massless
vector boson such as the photon $\gamma$ or gluon $g$ has just two
polarization states: $\lambda = \pm 1$. However, a massive vector boson
such as the $\rho$ has three polarization states: $\lambda = 0, \pm 1$.
This third polarization state is provided by a spin-0 field.
In order to make $m_{W^\pm,Z^0} \not= 0$, this should
have non-zero electroweak isospin $I \not= 0$, and the simplest
possibility is a complex isodoublet $(\phi^+,\phi^0)$, as assumed above.
This has four degrees of freedom, three of which are eaten by the $W^\pm$
amd $Z^0$ as their third polarization states, leaving us with one physical
Higgs boson $H$. Once the vacuum expectation value $\vert \langle
0\vert\phi\vert 0 \rangle \vert = v/ \sqrt{2}~:~~v = \mu / 
\sqrt{2\lambda}$ is
fixed, the mass of the remaining physical Higgs boson is given by 
\beq
m^2_H = 2\mu^2 = 4 \lambda v^2, 
\label{onefifteen} 
\eeq 
which is a free parameter in the \sm.

\subsection{Precision Tests of the Standard Model}

\begin{figure}
\centerline{\includegraphics[height=3in]{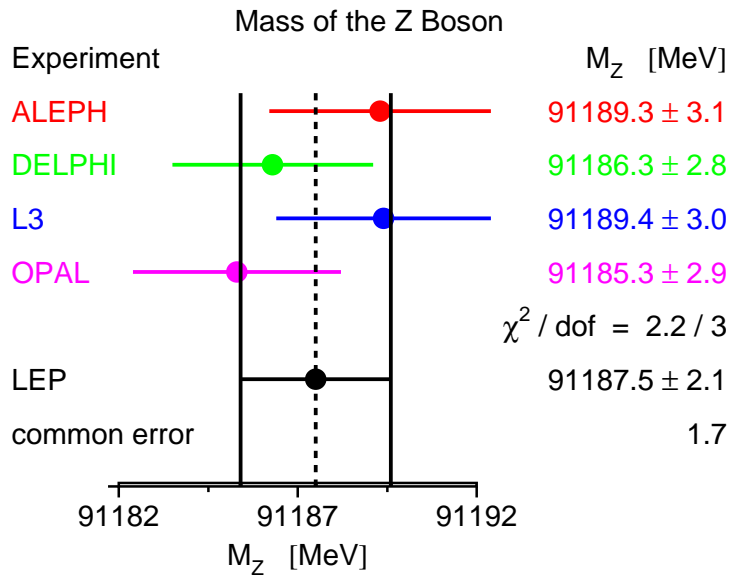}}
\caption[]{The mass of the $Z^0$ vector boson is one of the 
parameters of the \sm~ that has been measured most 
accurately~\protect\cite{LEPEWWG}.}
\label{fig:mZ}
\end{figure}

The quantity that was measured most accurately at LEP was the mass of the 
$Z^0$ boson~\cite{LEPEWWG}:
\beq
m_Z \; = \; 91,187.5 \pm 2.1~{\rm MeV},
\label{Zmass}
\eeq
as seen in Fig.~\ref{fig:mZ}.
Strikingly, $m_Z$ is now known more accurately than the muon decay
constant! Attaining this precision required understanding astrophysical
effects - those of terrestrial tides on the LEP beam energy, which
were ${\cal O}(10)$~MeV, as well as meteorological - when it rained, the
water expanded the rock in which LEP was buried, again changing the beam
energy, and seasonal - variations in the level of water in Lake
Geneva also caused the rock around LEP to expand and contract - as well as
electrical - stray currents from the nearby electric train line affected
the LEP magnets~\cite{LEPE}.

\begin{figure}
\centerline{\includegraphics[height=3in]{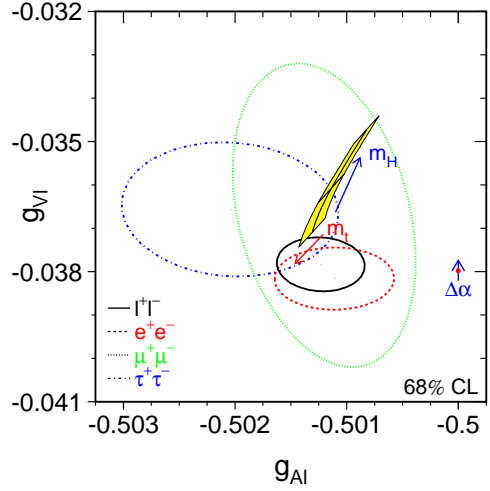}}
\caption[]{Precision measurements of the properties of the charged 
leptons $e, \mu$ and $\tau$ indicate that they have universal couplings 
to the weak vector bosons~\protect\cite{LEPEWWG}, whose value favours a 
relatively light Higgs boson.} 
\label{fig:univ}
\end{figure}

\begin{figure}
\centerline{\includegraphics[height=3in]{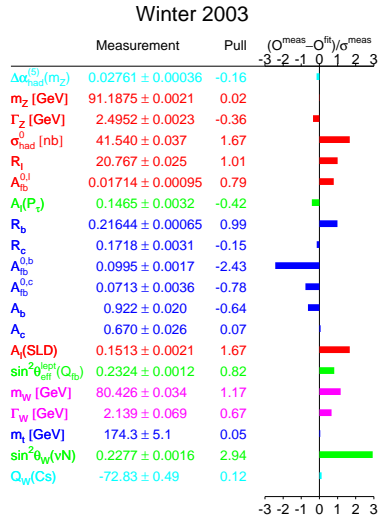}}
\caption[]{Precision electroweak measurements and the pulls they exert in a
global fit~\protect\cite{LEPEWWG}.}
\label{fig:pulls}
\end{figure}

LEP experiments also made precision measurements of many properties of the 
$Z^0$ boson~\cite{LEPEWWG}, such as the total cross section:
\beq
\sigma \; = \; {12 \pi \over m^2_Z} {\Gamma_{ee} \Gamma_{had} \over 
\Gamma^2_Z},
\label{sigmaZ}
\eeq
where $\Gamma_Z (\Gamma_{ee}, \Gamma_{had})$ is the total $Z^0$ decay rate 
(rate for decays into $e^+ e^-$, hadrons). Eq. (\ref{sigmaZ}) is the 
classical (tree-level) expression, which is reduced by about 30 \% by 
radiative corrections. The total decay rate is given by:
\beq
\Gamma_Z \; = \; \Gamma_{ee} + \Gamma_{\mu\mu} + \Gamma_{\tau\tau} + N_\nu 
\Gamma_{\nu\nu} + \Gamma_{had},
\label{GammaZ}
\eeq
where we expect $\Gamma_{ee} = \Gamma_{\mu\mu} = \Gamma_{\tau\tau}$ 
because of lepton universality, which has been verified experimentally, 
as seen in Fig.~\ref{fig:univ}~\cite{LEPEWWG}. 
Other partial decay rates have been measured via the branching ratios
\beq
R_{b,c} \; \equiv \; {\Gamma_{\bar b b, \bar c c} \over \Gamma_{had}},
\label{partials}
\eeq
as seen in Fig.~\ref{fig:pulls}.

Also measured have been various forward-backward asymmetries $A_{\ell, 
q}$ in the production of leptons and quarks, as well as the polarization 
of $\tau$ leptons produced in $Z^0$ decay, as also seen in 
Fig.~\ref{fig:pulls}. Various other measurements are also shown there, 
including the mass and decay rate of the $W^\pm$, the mass of the top 
quark, and low-energy neutral-current measurements in $\nu$-nucleon 
scattering and parity violation in atomic Cesium. The \sm~ is quite 
compatible with all these measurements, although some of them may differ 
by a couple of standard deviations: if they did not, we should be 
suspicious! Overall, the electroweak measurements tell us 
that~\cite{LEPEWWG}:
\beq
\sin^2 \theta_W \; = \; 0.23148 \pm 0.00017,
\label{sin2theta}
\eeq
providing us with a strong hint for grand unification, as we see later.

\subsection{The Search for the Higgs Boson}

\begin{figure}
\centerline{\includegraphics[height=3in]{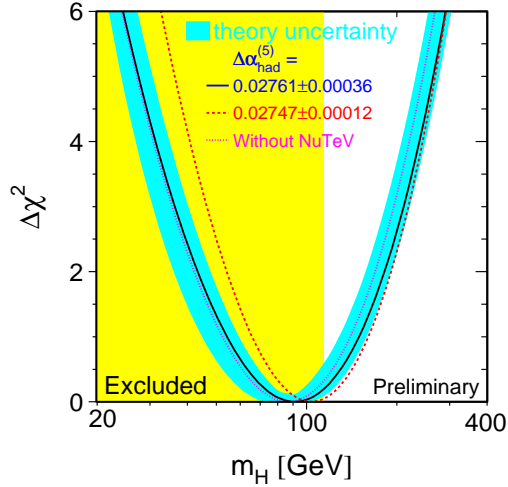}}
\caption[]{Estimate of the mass of the Higgs boson obtained 
from precision electroweak measurements. The blue band indicates 
theoretical uncertainties, and the different curves demonstrate the 
effects of different plausible estimates of the 
renormalization of the fine-structure constant at the 
$Z^0$ peak~\protect\cite{LEPEWWG}.}
\label{fig:mH} 
\end{figure}

The precision electroweak measurements at LEP and elsewhere are sensitive 
to radiative corrections via quantum loop diagrams, in particular those 
involving particles such as the top quark and the Higgs boson that are too 
heavy to be observed directly at LEP~\cite{Vt,VH}. Many of the 
electroweak observables mentioned above exhibit quadratic sensitivity to 
the mass of the top quark:
\beq 
\Delta \; \propto \; G_F m_t^2.
\label{quadtop}
\eeq
The measurements of these electroweak observables enabled the mass of the 
top quark to be predicted before it was discovered, and the measured 
value:
\beq
m_t = 174.3 \pm 5.1~{\rm GeV}
\label{onetwentyfour}
\eeq
agrees quite well with the prediction
\beq
m_t = 177.5 \pm 9.3~{\rm GeV}
\label{predictmt}
\eeq
derived from precision electroweak data~\cite{LEPEWWG}. Electroweak 
observables are also 
sensitive logarithmically to the mass of the Higgs boson:
\beq
\Delta \; \propto \; \left( {\alpha \over \pi} \right) {\rm ln} \left( 
{m_H^2 \over m_Z^2} \right),
\label{logmH}
\eeq
so their measurements can also be used to predict the mass of the Higgs 
boson. This prediction can be made more definite by combining the 
precision electroweak data with the measurement (\ref{onetwentyfour}) of 
the mass of the top quark. Making due allowance for theoretical 
uncertainties in the \sm~ calculations, as seen in Fig.~\ref{fig:mH}, one 
may estimate that~\cite{LEPEWWG}:
\beq
m_H \; = \; 91^{+58}_{-37}~{\rm GeV},
\label{predictmH}
\eeq
whereas $m_H$ is not known from first principles in the \sm.

The Higgs production and decay rates are completely fixed as
functions of the unknown mass $m_H$, enabling the search for the Higgs 
boson to be planned as a function of $m_H$~\cite{EGNH}. This search was 
one of the main 
objectives of experiments at LEP, which established the lower limit:
\beq
m_H \; > \; 114.4~{\rm GeV},
\label{LEPlimitmH}
\eeq
that is shown as the light yellow shaded region in Fig.~\ref{fig:mH}.
Combining this limit with the estimate (\ref{predictmH}), we see that 
there is 
good reason to expect that the Higgs boson may not be far away. Indeed, in 
the closing weeks of the LEP experimental programme, there was a hint for 
the discovery of the Higgs boson at LEP with a mass $\sim 115$~GeV, but 
this could not be confirmed~\cite{LEPHWG}. In the future, experiments at 
the Fermilab 
Tevatron collider and then the LHC will continue the search for the Higgs 
boson. The latter, in particular, should be able to discover it whatever 
its mass may be, up to the theoretical upper limit $m_H \lappeq 
1$~TeV~\cite{LHC}.

\subsection{Roadmap to Physics Beyond the Standard Model}

The Standard Model agrees with all confirmed experimental data from
accelerators, but is theoretically very
unsatisfactory~\cite{StAnd,CHschool}. It does not explain the particle
quantum numbers, such as the electric charge $Q$, weak isospin $I$,
hypercharge $Y$ and colour, and contains at least 19 arbitrary parameters.
These include three independent vector-boson couplings and a possible
CP-violating strong-interaction parameter, six quark and three
charged-lepton masses, three generalized Cabibbo weak mixing angles and
the CP-violating Kobayashi-Maskawa phase, as well as two independent
masses for weak bosons.

The Big Issues in physics beyond the Standard Model are conveniently
grouped into three categories~\cite{StAnd,CHschool}. These include the
problem of {\bf Mass}: what is the origin of particle masses, are they due
to a Higgs boson, and, if so, why are the masses so small; {\bf
Unification}: is there a simple group framework for unifying all the
particle interactions, a so-called Grand Unified Theory (GUT); and {\bf
Flavour}: why are there so many different types of quarks and leptons and
why do their weak interactions mix in the peculiar way observed? Solutions
to all these problems should eventually be incorporated in a Theory of
Everything (TOE) that also includes gravity, reconciles it with quantum
mechanics, explains the origin of space-time and why it has four
dimensions, makes coffee, etc. String theory, perhaps in its current
incarnation of M theory, is the best (only?) candidate we have for such a
TOE~\cite{TOE}, but we do not yet understand it well enough to make clear
experimental predictions.

As if the above 19 parameters were insufficient to appall you, at least
nine more parameters must be introduced to accommodate the neutrino
oscillations discussed in the next Lecture: 3 neutrino masses, 3 real
mixing angles, and 3 CP-violating phases, of which one is in principle
observable in neutrino-oscillation experiments and the other two in
neutrinoless double-beta decay experiments. In fact even the simplest
models for neutrino masses involve 9 further parameters, as discussed
later.

Moreover, there are many other cosmological parameters that we should also 
seek to explain. Gravity is characterized by at least two parameters, the 
Newton constant $G_N$ and the cosmological vacuum energy. We may also want 
to construct a field-theoretical model for inflation, and we certainly 
need to explain the baryon asymmetry of the Universe. So there is plenty 
of scope for physics beyond the Standard Model.

The first clear evidence for physics beyond the Standard Model of particle
physics has been provided by neutrino physics, which is also of great
interest for cosmology, so this is the subject of Lecture~2. Since there
are plenty of good reasons to study supersymmetry~\cite{CHschool},
including the possibility that it provides the cold dark matter, this is
the subject of Lecture~3. Inflation is the subject of Lecture~4, and
various further topics such as GUTs, baryo/leptogenesis and
ultra-high-energy cosmic rays are discussed in Lecture~5. As we shall see
later, neutrino physics may be the key to both inflation and baryogenesis.

\section{Neutrino Physics}

\subsection{Neutrino Masses?}

There is no good reason why either the total lepton number $L$ or the
individual lepton flavours $L_{e, \mu, \tau}$ should be
conserved. Theorists have learnt that the only conserved
quantum numbers are those associated with exact local symmetries, just as
the conservation of electromagnetic charge is associated with local U(1)
invariance. On the other hand, there is no exact local symmetry associated
with any of the lepton numbers, so we may expect non-zero neutrino masses.

However, so far we have only upper experimental limits on neutrino 
masses~\cite{PDG}.
From measurements of the end-point in Tritium $\beta$ decay, we know that:
\begin{equation}
m_{\nu_e} \; \lappeq \; 2.5~{\rm eV},
\label{mnue}
\end{equation}  
which might be improved down to about 0.5~eV with the proposed KATRIN
experiment~\cite{KATRIN}. From measurements of $\pi \to \mu \nu$ decay, we 
know that:
\begin{equation}
m_{\nu_\mu} \; < \; 190~{\rm KeV},
\label{mnumu}
\end{equation}
and there are prospects to improve this limit by a factor $\sim 
20$. Finally, from measurements of $\tau \to n \pi \nu$ decay, we know
that:
\begin{equation}
m_{\nu_\tau} \; < \; 18.2~{\rm MeV},
\label{mnutau}  
\end{equation}
and there are prospects to improve this limit to $\sim 5$~MeV.

\begin{figure}
\centerline{\includegraphics[height=3in]{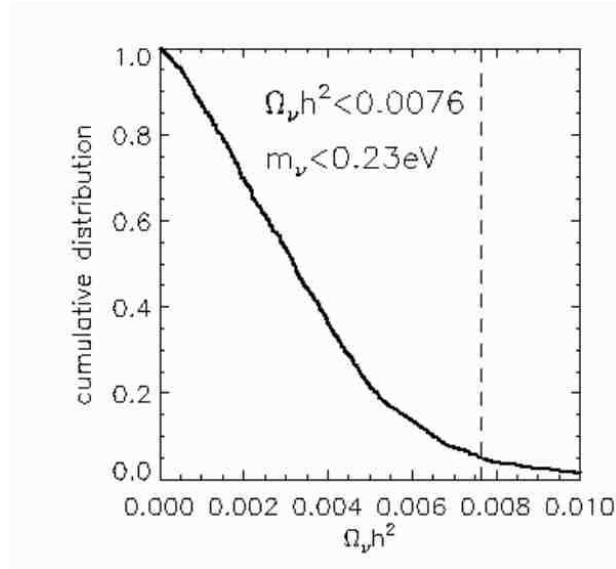}}
\caption[]{Likelihood function for the sum of neutrino masses provided by 
WMAP~\protect\cite{WMAPnu}: the quoted upper limit applies if the 3 light 
neutrino species are degenerate.}
\label{fig:WMAPnu}
\end{figure}

Astrophysical upper limits on neutrino masses are stronger than these 
laboratory limits. The 2dF data were
used to infer an upper limit on the sum of the neutrino masses of
1.8~eV~\cite{2dF}, which has recently been improved using WMAP data 
to~\cite{WMAPnu}
\begin{equation}   
\Sigma_{\nu_i} m_{\nu_i} < 0.7~{\rm eV},
\label{cosmonu}
\end{equation}
as seen in Fig.~\ref{fig:WMAPnu}.
This impressive upper limit is substantially better than
even the most stringent direct laboratory upper limit on an individual
neutrino mass.

Another interesting laboratory limit on neutrino masses comes
from searches for neutrinoless double-$\beta$ decay, which constrain the
sum of the neutrinos' Majorana masses weighted by their couplings to
electrons~\cite{KKGH}:
\begin{equation}
\langle m_\nu \rangle_e \; \equiv | \Sigma_{\nu_i} m_{\nu_i} U_{ei}^2 |
\lappeq 0.35~{\rm eV}
\label{doublebeta}
\end{equation}
which might be improved to $\sim 0.01$~eV in a future round of 
experiments. 

Neutrinos have been seen to oscillate between their different
flavours~\cite{SK,SNO}, showing that the separate lepton flavours $L_{e,
\mu, \tau}$ are indeed not conserved, though the conservation of total
lepton number $L$ is still an open question. The observation of such
oscillations strongly suggests that the neutrinos have different masses. 

\subsection{Models of Neutrino Masses and Mixing}

The conservation of lepton number is an accidental symmetry of the 
renormalizable terms in the Standard Model Lagrangian. However, one could 
easily add to the Standard Model non-renormalizable terms that would 
generate neutrino masses, even without introducing any new fields. 
For example, a non-renormalizable term of the form~\cite{BEG}
\begin{equation}
{1 \over M} \nu H \cdot \nu H,
\label{nonren}
\end{equation}
where $M$ is some large mass beyond the scale of the Standard Model, would 
generate a neutrino mass term:
\begin{equation}
m_\nu \nu \cdot \nu:
\; m_\nu \; = \; {\langle 0 \vert H \vert 0 \rangle^2 \over M}.
\label{smallm}
\end{equation}
However, a new interaction like (\ref{nonren}) seems
unlikely to be fundamental, and one should like to understand the origin 
of the large mass scale $M$.

The minimal renormalizable model of neutrino masses requires the
introduction of weak-singlet `right-handed' neutrinos $N$. These will in
general couple to the conventional weak-doublet left-handed neutrinos via
Yukawa couplings $Y_\nu$ that yield Dirac masses $m_D = Y_\nu \langle 
0 \vert H \vert 0 \rangle \sim m_W$. In
addition, these `right-handed' neutrinos $N$ can couple to themselves via
Majorana masses $M$ that may be $\gg m_W$, since they do not require 
electroweak summetry breaking. Combining the two types of
mass term, one obtains the seesaw mass matrix~\cite{seesaw}:
\begin{eqnarray}
\left( \nu_L, N\right) \left(
\begin{array}{cc}
0 & M_{D}\\ 
M_{D}^{T} & M
\end{array}
\right)
\left(
\begin{array}{c}
\nu_L \\
N
\end{array}
\right),
\label{seesaw}
\end{eqnarray}
where each of the entries should be understood as a matrix in generation 
space.

In order to provide the two measured differences in neutrino
masses-squared, there must be at least two non-zero masses, and hence at
least two heavy singlet neutrinos $N_i$~\cite{Frampton,Morozumi}.  
Presumably, all three light neutrino masses are non-zero, in which case
there must be at least three $N_i$. This is indeed what happens in simple
GUT models such as SO(10), but some models~\cite{fSU5} have more singlet
neutrinos~\cite{EGLLN}. In this Lecture, for simplicity we consider just
three $N_i$.

The effective mass matrix for light neutrinos in the seesaw model may be 
written as:
\begin{equation}
{\cal M}_\nu \; = \; Y_\nu^T {1 \over M} Y_\nu v^2,
\label{seesawmass}
\end{equation}
where we have used the relation $m_D = Y_\nu v$ with $v \equiv \langle 0 
\vert H \vert 0 \rangle$. Taking $m_D \sim m_q$ or $m_\ell$ and requiring 
light neutrino masses $\sim 10^{-1}$ to $10^{-3}$~eV, we find that heavy 
singlet neutrinos weighing $\sim 10^{10}$ to $10^{15}$~GeV seem to be 
favoured.

It is convenient to work in the field basis where the charged-lepton masses 
$m_{\ell^\pm}$ and the heavy singlet-neutrino mases $M$ are real and 
diagonal. The seesaw neutrino mass matrix ${\cal M}_\nu$ 
(\ref{seesawmass}) may then be diagonalized by a unitary transformation 
$U$:
\begin{equation}
U^T {\cal M}_\nu U \; = \; {\cal M}_\nu^d.
\label{diag}
\end{equation}
This diagonalization is reminiscent of that required for the quark mass 
matrices in the Standard Model. In that case, it is well known that one 
can redefine the phases of the quark fields~\cite{EGN} so that the mixing 
matrix 
$U_{CKM}$ has just one CP-violating phase~\cite{KM}. However, in the 
neutrino case, 
there are fewer independent field phases, and one is left with 3 
physical CP-violating parameters:
\begin{equation}
U \; = \; {\tilde P}_2 V P_0: \; P_0 \equiv {\rm Diag} \left(
e^{i\phi_1},
e^{i\phi_2}, 1 \right).
\label{MNSP}
\end{equation}   
Here ${\tilde P}_2 = {\rm Diag} \left( e^{i\alpha_1}, e^{i\alpha_2},
e^{i\alpha_3} \right)$ contains three phases that can be removed by
phase rotations and are unobservable in
light-neutrino physics, though they do play a r\^ole at high energies, 
as discussed in Lecture~5, $V$ is the light-neutrino mixing matrix 
first considered by Maki, Nakagawa and Sakata (MNS)~\cite{MNS}, and $P_0$ 
contains 2 CP-violating phases $\phi_{1,2}$ that are observable at low 
energies. The MNS matrix describes neutrino oscillations
\begin{eqnarray}
V \; = \; \left(
\begin{array}{ccc}
c_{12} & s_{12} & 0 \\
- s_{12} & c_{12} & 0 \\
0 & 0 & 1
\end{array}
\right)
\left(
\begin{array}{ccc}
1 & 0 & 0 \\
0 & c_{23} & s_{23} \\
0 & - s_{23} & c_{23}
\end{array}
\right)
\left(
\begin{array}{ccc}
c_{13} & 0 & s_{13} \\
0 & 1 & 0 \\
- s_{13} e^{- i \delta} & 0 & c_{13} e^{- i \delta}
\end{array}
\right).
\label{MNSmatrix}
\end{eqnarray}
The three real mixing angles $\theta_{12, 23, 13}$ in (\ref{MNSmatrix}) are 
analogous to the Euler angles that are familiar from the classic rotations 
of rigid mechanical bodies. The phase $\delta$ is a specific quantum 
effect that is also observable in  neutrino oscillations, and violates 
CP, as we discuss below. The other CP-violating phases $\phi_{1,2}$
are in principle observable in neutrinoless double-$\beta$ decay 
(\ref{doublebeta}).

\subsection{Neutrino Oscillations}

In quantum physics, particles such as neutrinos propagate as complex 
waves. Different mass eigenstates $m_i$ travelling with the same 
momenta $p$ oscillate with different frequencies:
\beq
e^{i E_i t}: \; \; E_i^2 \; = \; p^2 + m_i^2.
\label{frequencies}
\eeq
Now consider what happens if one produces a neutrino beam of one given 
flavour, corresponding to some specific combination of 
mass eigenstates. After propagating some distance, the different mass 
eigenstates in the beam will acquire different phase
weightings (\ref{frequencies}), so that the neutrinos in the beam will be 
detected as a 
mixture of different neutrino flavours. These oscillations will be 
proportional to the mixing $\sin^2 2 \theta$ between the different 
flavours, and also to the differences in masses-squared $\Delta 
m_{ij}^2$ between the different mass eigenstates.

\begin{figure}
\centerline{\includegraphics[height=3in]{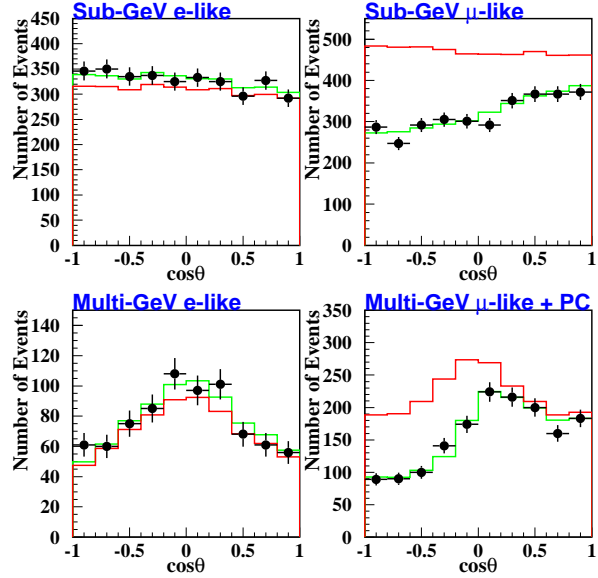}}
\caption[]{The zenith angle distributions of atmospheric neutrinos exhibit 
a deficit of downward-moving $\nu_\mu$, which is due to neutrino
oscillations~\protect\cite{SK}.}
\label{fig:atmonu}
\end{figure}

The first of the mixing angles in (\ref{MNSmatrix}) to be discovered was
$\theta_{23}$, in atmospheric neutrino experiments. Whereas the numbers of
downward-going atmospheric $\nu_\mu$ were found to agree with \sm~
predictions, a deficit of upward-going $\nu_\mu$ was observed, as seen in
Fig.~\ref{fig:atmonu}. The data from the Super-Kamiokande experiment, in
particular~\cite{SK}, favour near-maximal mixing of atmospheric neutrinos:
\beq
\theta_{23} \; \sim \; 45^o, \; \; \Delta m_{23}^2 \; \sim \; 2.4 \times 
10^{-3}~{\rm eV}^2.
\label{atmo}
\eeq
Recently, the K2K experiment using a beam of neutrinos produced by an 
accelerator has found results consistent with (\ref{atmo})~\cite{K2K}.
It seems that the atmospheric $\nu_\mu$ probably oscillate primarily into 
$\nu_\tau$, though this has yet to be established.

\begin{figure}
\centerline{\includegraphics[height=3in]{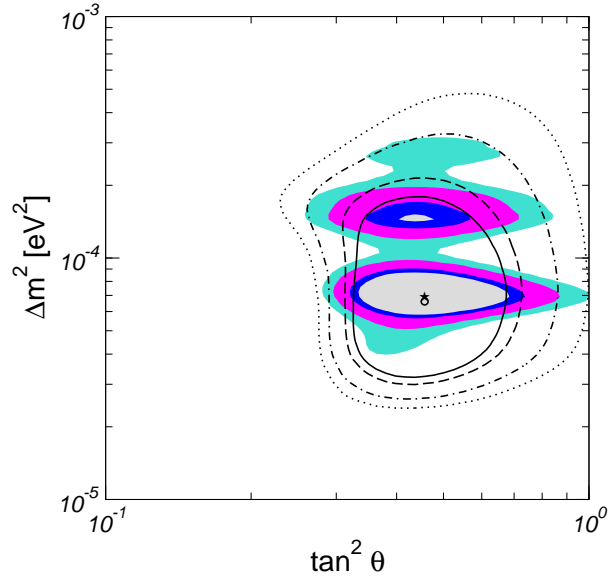}}
\caption[]{The KamLAND experiment (shadings) finds~\protect\cite{KamLAND} 
a deficit of reactor 
neutrinos that is consistent with the LMA neutrino oscillation parameters 
previously estimated (ovals) on the basis of solar neutrino 
experiments~\protect\cite{PV}.}
\label{fig:PV}
\end{figure}

More recently, the oscillation interpretation of the long-standing
solar-neutrino deficit has been established, in particular by the SNO
experiment. Solar neutrino experiments are sensitive to the mixing angle
$\theta_{12}$ in (\ref{MNSmatrix}). The
recent data from SNO~\cite{SNO} and Super-Kamiokande~\cite{SKsolar} prefer
quite strongly the large-mixing-angle (LMA) solution to the solar neutrino
problem with
\beq
\theta_{12} \; \sim \; 30^o, \; \; \Delta m_{12}^2 \; \sim \; 6 \times 
10^{-5}~{\rm eV}^2, 
\label{solar}
\eeq
though they have been unable to exclude completely the LOW solution with 
lower $\delta m^2$. However, the KamLAND experiment on reactors produced by
nuclear power reactors has recently found a deficit of $\nu_e$ that is
highly compatible with the LMA solution to the solar neutrino 
problem~\cite{KamLAND}, as seen in Fig.~\ref{fig:PV}, and excludes any 
other solution.

\begin{figure}
\hspace{-1cm}
\centerline{\includegraphics[height=3in]{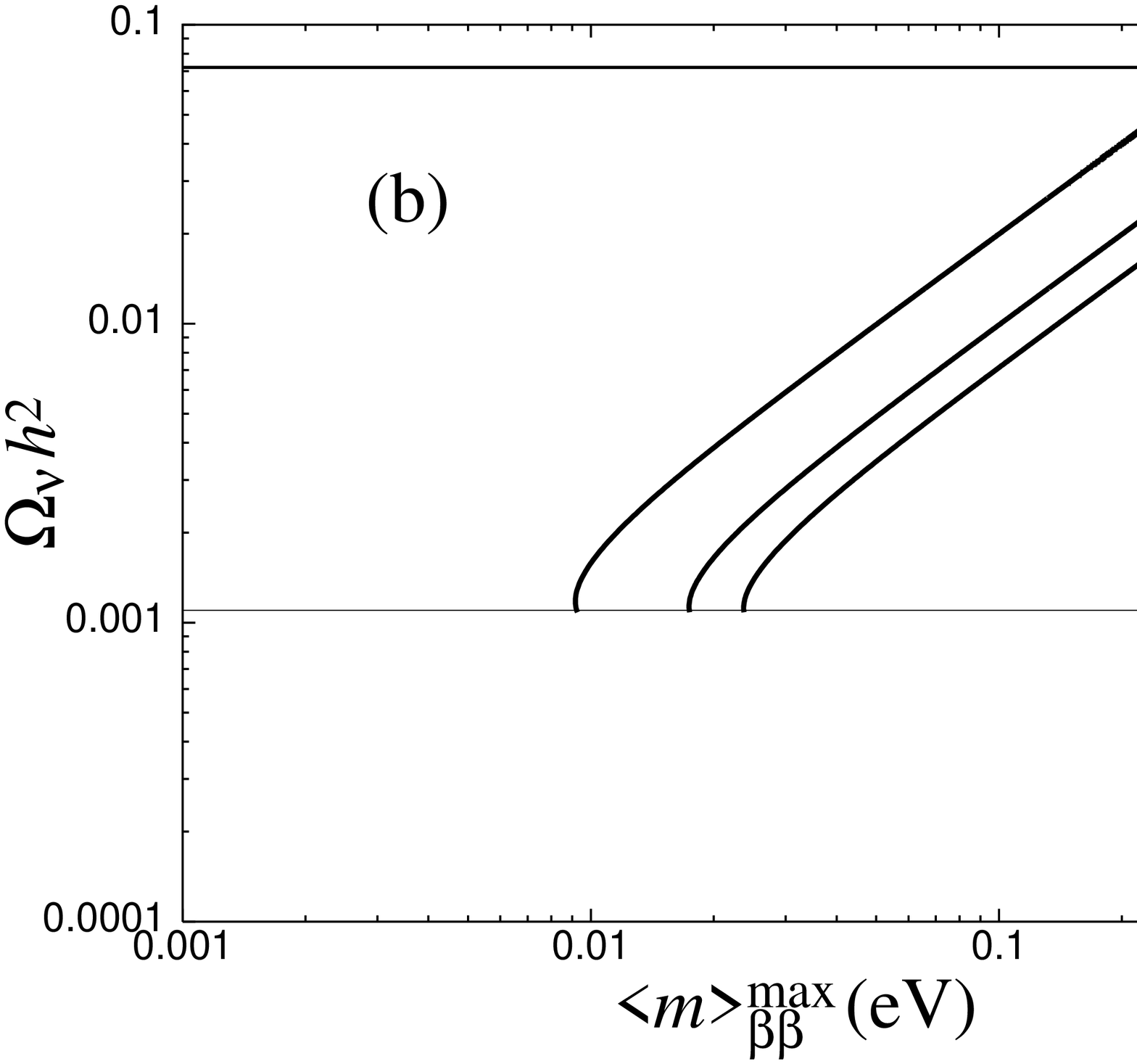}}
\caption[]{The correlation between the relic density of neutrinos 
$\Omega_\nu h^2$ and the neutrinoless double-$\beta$ decay observable: the 
different lines indicate the ranges allowed by neutrino oscillation 
experiments~\protect\cite{MS}.}
\label{fig:MS}
\end{figure}

Using the range of $\theta_{12}$ allowed by the solar and KamLAND data,
one can establish a correlation between the relic neutrino density
$\Omega_\nu h^2$ and the neutrinoless double-$\beta$ decay observable
$\langle m_\nu \rangle_e$, as seen in Fig.~\ref{fig:MS}~\cite{MS}.  
Pre-WMAP, the experimental limit on $\langle m_\nu \rangle_e$ could be
used to set the bound
\begin{equation}
10^{-3} \; \lappeq \; \Omega_\nu h^2 \; \lappeq \; 10^{-1}.
\label{MSbound}
\end{equation}
Alternatively, now that WMAP has set a tighter upper bound $\Omega_\nu
h^2 < 0.0076$ (\ref{cosmonu})~\cite{WMAPnu}, one can use this correlation 
to set an upper bound:
\begin{equation}
< m_\nu >_e \; \lappeq \; 0.1~{\rm eV},
\label{betabound}
\end{equation}
which is difficult to reconcile with the neutrinoless double-$\beta$ 
decay signal reported in~\cite{KKGH}.

The third mixing angle $\theta_{13}$ in (\ref{MNSmatrix}) is basically
unknown, with experiments such as Chooz~\cite{Chooz} and Super-Kamiokande
only establishing upper limits. {\it A fortiori}, we have no experimental
information on the CP-violating phase $\delta$.

\begin{figure}
\hspace{1cm}
\epsfig{figure=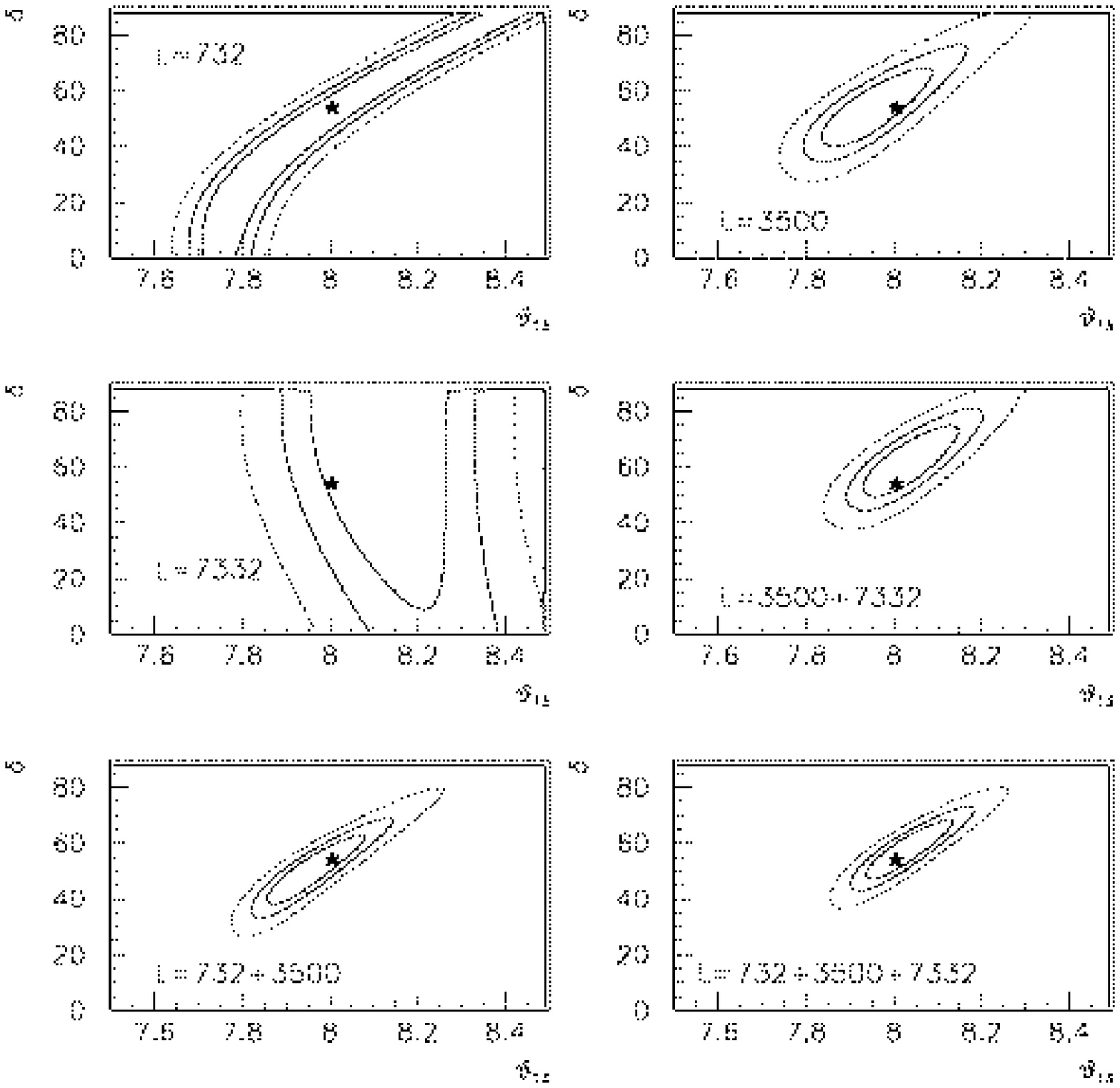,width=10cm}
\hglue3.5cm   
\caption{
Possible measurements of $\theta_{13}$ and
$\delta$ that could be made with a neutrino factory, using a neutrino 
energy threshold of about 10~GeV.
Using a single baseline correlations are very strong, but can be 
largely reduced by combining information from different baselines and 
detector techniques~\protect\cite{golden}, enabling the CP-violating phase 
$\delta$ to be extracted.}
\label{fig:cpnu}
\end{figure}

The phase $\delta$ could in principle be measured by comparing the 
oscillation 
probabilities for neutrinos and antineutrinos and computing the 
CP-violating asymmetry~\cite{DGH}:
\begin{eqnarray}
P \left( \nu_e \to \nu_\mu \right) - P \left( {\bar \nu}_e \to 
{\bar \nu}_\mu \right) \; & & = \; 
16 s_{12} c_{12} s_{13} c^2_{13} s_{23} c_{23} \sin \delta \\ \nonumber
& & \sin \left( {\Delta m_{12}^2 \over 4 E} L \right)
\sin \left( {\Delta m_{13}^2 \over 4 E} L \right)
\sin \left( {\Delta m_{23}^2 \over 4 E} L \right),
\label{CPV}
\end{eqnarray}
as seen in Fig.~\ref{fig:cpnu}~\cite{golden}.
This is possible only if $\Delta m_{12}^2$ and $s_{12}$ are large enough -
as now suggested by the success of the LMA solution to the solar neutrino 
problem, and if $s_{13}$ is large enough - which remains an open question.

A number of long-baseline neutrino experiments using beams from
accelerators are now being prepared in the United States, Europe and
Japan, with the objectives of measuring more accurately the atmospheric
neutrino oscillation parameters, $\Delta m_{23}^2, \theta_{23}$ and
$\theta_{13}$, and demonstrating the production of $\nu_\tau$ in a
$\nu_\mu$ beam. Beyond these, ideas are being proposed for intense
`super-beams' of low-energy neutrinos, produced by high-intensity,
low-energy accelerators such as the SPL~\cite{SPL} proposed at CERN. A
subsequent step could be a storage ring for unstable ions, whose decays
would produce a `$\beta$ beam' of pure $\nu_e$ or ${\bar \nu}_e$
neutrinos. These experiments might be able to measure $\delta$ via CP
and/or T violation in neutrino oscillations~\cite{betabeam}. A final step
could be a full-fledged neutrino factory based on a muon storage ring,
which would produce pure $\nu_\mu$ and ${\bar \nu}_e$ (or $\nu_e$ and
${\bar \nu}_\mu$ beams and provide a greatly enhanced capability to search
for or measure $\delta$ via CP violation in neutrino
oscillations~\cite{nufact}.

\begin{figure}[t]
\begin{center}
\hspace{-1cm}
\begin{picture}(400,300)(-200,-150)
\Oval(0,0)(30,60)(0)
\Text(-25,13)[lb]{ ${\bf Y_\nu}$  ,  ${\bf M_{N_i}}$}
\Text(-40,-2)[lb]{{\bf 15$+$3 physical}}
\Text(-30,-15)[lb]{{\bf parameters}}
\EBox(-70,90)(70,150)
\Text(-55,135)[lb]{{\bf Seesaw mechanism}}  
\Text(-8,117)[lb]{${\bf {\cal M}_\nu}$}
\Text(-65,100)[lb]{{\bf 9 effective parameters}}
\EBox(-200,-140)(-60,-80)
\Text(-167,-95)[lb]{{\bf Leptogenesis}}
\Text(-165,-113)[lb]{ ${\bf Y_\nu Y_\nu^\dagger}$ , ${\bf M_{N_i}}$}
\Text(-175,-130)[lb]{{\bf 9$+$3 parameters}}
\EBox(60,-140)(200,-80)
\Text(80,-95)[lb]{{\bf Renormalization}}
\Text(95,-113)[lb]{${\bf Y_\nu^\dagger L Y_\nu}$ , ${\bf M_{N_i}}$}
\Text(80,-130)[lb]{{\bf 13$+$3 parameters}}
\LongArrow(0,30)(0,87)
\LongArrow(-45,-20)(-130,-77)
\LongArrow(45,-20)(130,-77)
\end{picture}
\end{center}
\caption{
Roadmap for the physical observables derived from $Y_\nu$ and 
$N_i$~\protect\cite{ER}.}
\label{fig:map}
\end{figure}
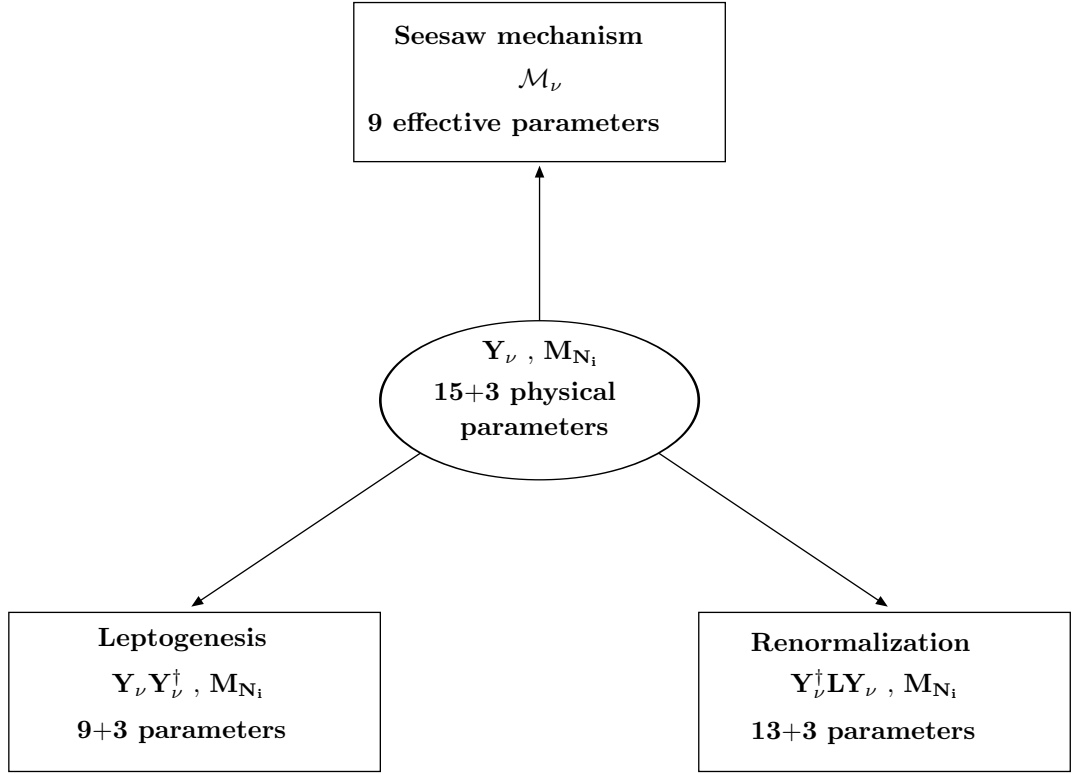

We have seen above that the effective low-energy mass matrix for the light
neutrinos contains 9 parameters, 3 mass eigenvalues, 3 real mixing angles
and 3 CP-violating phases. However, these are not all the parameters in
the minimal seesaw model. As shown in Fig.~\ref{fig:map}, this model has a
total of 18 parameters~\cite{Casas,EHLR}. The additional 9 parameters
comprise the 3 masses of the heavy singlet `right-handed' neutrinos $M_i$,
3 more real mixing angles and 3 more CP-violating phases. As illustrated
in Fig.~\ref{fig:map}, many of these may be observable via renormalization
in supersymmetric models~\cite{DI,EHLR,EHRS,EHRS2}, which may generate
observable rates for flavour-changing lepton decays such as $\mu \to e
\gamma, \tau \to \mu \gamma$ and $\tau \to e \gamma$, and CP-violating
observables such as electric dipole moments for the electron and muon.  
Some of these extra parametrs may also have controlled the generation of
matter in the Universe via leptogenesis~\cite{FY}, as discussed in
Lecture~5.

\section{Supersymmetry}

\subsection{Why?}

The main theoretical reason to expect supersymmetry at an accessible
energy scale is provided by the {\it hierarchy problem}~\cite{hierarchy}:  
why is $m_W \ll m_P$, or equivalently why is $G_F \sim 1 / m_W^2 \gg G_N =
1 / m_P^2$? Another equivalent question is why the Coulomb potential in an
atom is so much greater than the Newton potential: $e^2 \gg G_N m^2 = m^2
/ m_P^2$, where $m$ is a typical particle mass?

Your first thought might simply be to set $m_P \gg m_W$ by hand, and 
forget about the problem. Life is not so simple, because quantum 
corrections to $m_H$ and hence $m_W$ are quadratically divergent in the 
Standard Model:
\begin{equation}
\delta m_{H,W}^2 \; \simeq \; {\cal O}({\alpha \over \pi}) \Lambda^2,
\label{Qdgt}
\end{equation}
which is $\gg m_W^2$ if the cutoff $\Lambda$, which represents the scale 
where new physics beyond the Standard Model appears, is comparable to the 
GUT or Planck scale. For example, if the 
Standard Model were to hold unscathed all the way up the Planck mass $m_P 
\sim 10^{19}$~GeV, the radiative correction (\ref{Qdgt}) would be 36 
orders of magnitude greater than the physical values of $m_{H,W}^2$! 

In principle, this is not a problem from the mathematical point of view of
renormalization theory. All one has to do is postulate a tree-level value
of $m_H^2$ that is (very nearly) equal and opposite to the `correction'
(\ref{Qdgt}), and the correct physical value may be obtained by a delicate
cancellation. However, this fine tuning strikes many physicists as rather
unnatural: they would prefer a mechanism that keeps the `correction'
(\ref{Qdgt}) comparable at most to the physical value~\cite{hierarchy}.

This is possible in a supersymmetric theory, in which there are equal numbers 
of bosons and fermions with identical couplings. Since bosonic and fermionic 
loops have opposite signs, the residual one-loop correction is of the form
\begin{equation}
\delta m_{H,W}^2 \; \simeq \; {\cal O}({\alpha \over \pi}) (m_B^2 - 
m_F^2),
\label{susy}
\end{equation}
which is $\lappeq m_{H,W}^2$ and hence naturally small if the 
supersymmetric partner bosons $B$ and fermions $F$ have similar masses:
\begin{equation}
|m_B^2 - m_F^2| \; \lappeq \; 1~{\rm TeV}^2.
\label{natural}
\end{equation}
This is the best motivation we have for finding supersymmetry at 
relatively low energies~\cite{hierarchy}.
In addition to this first supersymmetric miracle of removing (\ref{susy})
the quadratic divergence (\ref{Qdgt}), many logarithmic divergences are
also absent in a supersymmetric theory~\cite{noren}, a property that also 
plays a r\^ole in the construction of supersymmetric GUTs~\cite{StAnd}. 

Supersymmetry had been around for some time before its utility for
stabilizing the hierarchy of mass scales was realized. Some theorists had
liked it because it offered the possibility of unifying fermionic matter
particles with bosonic force-carrying particles. Some had liked it because
it reduced the number of infinities found when calculating quantum
corrections - indeed, theories with enough supersymmetry can even be
completely finite~\cite{noren}. Theorists also liked the possibility of 
unifying Higgs
bosons with matter particles, though the first ideas for doing this did
not work out very well~\cite{Fayet}. Another aspect of supersymmetry, that 
made some
theorists think that its appearance should be inevitable, was that it was
the last possible symmetry of field theory not yet known to be exploited
by Nature~\cite{HLS}. Yet another asset was the observation that making 
supersymmetry
a local symmetry, like the \sm, necessarily introduced gravity, offering
the prospect of unifying {\it all} the particle interactions. Moreover,
supersymmetry seems to be an essential requirement for the consistency of
string theory, which is the best candidate we have for a Theory of
Everything, including gravity. However, none of these `beautiful'
arguments gave a clue about the scale of supersymmetric particle masses:
this was first provided by the hierarchy argument outlined above.

Could any of the known particles in the Standard Model be paired up in
supermultiplets? Unfortunately, none of the known fermions $q, \ell$ can
be paired with any of the `known' bosons $\gamma, W^\pm Z^0, g, H$,
because their internal quantum numbers do not match~\cite{Fayet}. For
example, quarks $q$ sit in triplet representations of colour, whereas the
known bosons are either singlets or octets of colour. Then again, leptons
$\ell$ have non-zero lepton number $L = 1$, whereas the known bosons have
$L = 0$. Thus, the only possibility seems to be to introduce new
supersymmetric partners (spartners) for all the known particles, as seen
in the Table below: quark $\to$ squark, lepton $\to$ slepton, photon $\to$
photino, Z $\to$ Zino, W $\to$ Wino, gluon $\to$ gluino, Higgs $\to$
Higgsino. The best that one can say for supersymmetry is that it
economizes on principle, not on particles!

\vspace{0.4cm}

\begin{center}
\begin{tabular}{lclc}\hline
&&&\\
Particle & Spin & Spartner & Spin \\ \hline
&&&\\
quark: $q$ & ${1\over 2}$ & squark: $\tilde q$ & 0 \\
&&&\\
lepton: $\ell$ & ${1\over 2}$ & slepton: $\tilde\ell$ & 0 \\
&&&\\
photon: $\gamma$ & 1 & photino: $\tilde\gamma$ & ${1\over 2}$ \\
&&&\\
$W$ & 1 & wino: $\tilde W$ & ${1\over 2}$ \\
&&&\\
$Z$ & 1 & zino: $\tilde Z$ & ${1\over 2}$ \\
&&&\\
Higgs: $H$ & 0 & higgsino: $\tilde H$ & ${1\over 2}$ \\
&&&\\ \hline
\end{tabular}
\end{center}

\vspace{0.4cm}

The minimal supersymmetric extension of the \sm~ (MSSM)~\cite{MSSM} has
the same vector interactions as the \sm, and the particle masses arise in 
much the same way. However, in addition to the \sm~ particles 
and their supersymmetric partners in the Table, the minimal supersymmetric 
extension of the \sm~ (MSSM), requires two Higgs doublets $H, \bar H$
with opposite hypercharges in order to give masses to all the matter
fermions, whereas one Higgs doublet would have sufficed in the \sm.
The two Higgs doublets couple via an extra coupling called $\mu$, and 
it should also be noted that the ratio of Higgs vacuum expectation values
\beq
\tan\beta\equiv {\langle \bar H \rangle \over \langle H \rangle}
\label{twotwentysix}
\eeq
is undetermined and should be treated as a free parameter.

\subsection{Hints of Supersymmetry}

\begin{figure}
\centerline{\hspace{10cm}\includegraphics[height=8in]{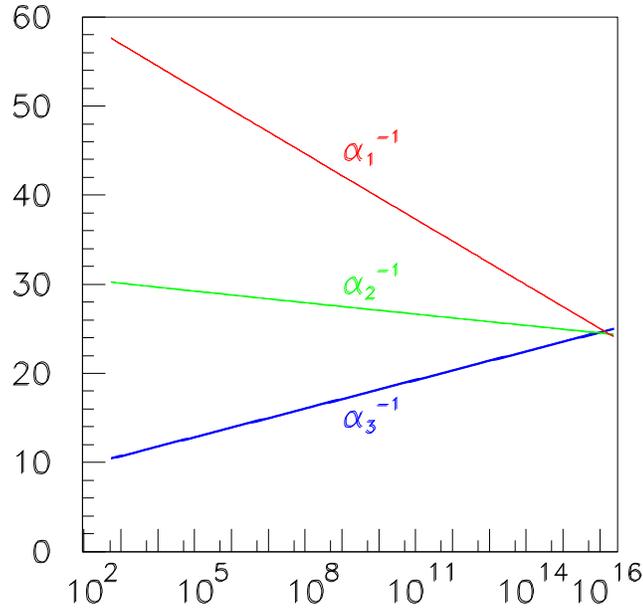}}
\vspace{-10cm}
\caption[]{The measurements of the gauge coupling strengths at LEP, 
including $\sin^2 \theta_W$ (\ref{sin2theta}),
evolve to a unified value if supersymmetry is 
included~\protect\cite{GUTs}.}
\label{fig:GUT}
\end{figure}

There are some phenomenological hints that supersymmetry may, indeed,
appear at the TeV scale.  One is provided by the strengths of the
different \sm~ interactions, as measured at LEP~\cite{GUTs}. These may be
extrapolated to high energy scales including calculable renormalization
effects~\cite{GQW}, to see whether they unify as predicted in a GUT. The 
answer is
no, if supersymmetry is not included in the calculations. In that case,
GUTs would require a ratio of the electromagnetic and weak coupling
strengths, parametrized by $\sin^2 \theta_W$, different from what is 
observed (\ref{sin2theta}), if they are to unify with the
strong interactions.  On the other hand, as seen in Fig.~\ref{fig:GUT},
minimal supersymmetric GUTs predict just the correct ratio for the weak
and electromagnetic interaction strengths, i.e., value for $\sin^2 
\theta_W$ (\ref{sin2theta}).

A second hint is the fact that precision electroweak data prefer a
relatively light Higgs boson weighing less than about
200~GeV~\cite{LEPEWWG}. This is perfectly consistent with calculations in
the minimal supersymmetric extension of the Standard Model (MSSM), in
which the lightest Higgs boson weighs less than about
130~GeV~\cite{susyHiggs}.

A third hint is provided by the astrophysical necessity of cold dark
matter. This could be provided by a neutral, weakly-interacting particle
weighing less than about 1~TeV, such as the lightest supersymmetric
particle (LSP) $\chi$~\cite{EHNOS}. This is expected to be stable in the
MSSM, and hence should be present in the Universe today as a cosmological
relic from the Big Bang~\cite{Goldberg,EHNOS}. Its stability arises
because there is a multiplicatively-conserved quantum number called $R$
parity, that takes the values +1 for all conventional particles and -1 for
all sparticles~\cite{Fayet}. The conservation of $R$ parity can be related
to that of baryon number $B$ and lepton number $L$, since
\beq
R = (-1)^{3B+L+2S}
\label{threeeighteen}
\eeq
where $S$ is the spin. There are three important consequences of $R$ 
conservation:
\begin{enumerate}
\item sparticles are always produced in pairs, e.g., $\bar
pp\rightarrow\tilde q \tilde g X$, $e^+e^-\rightarrow \tilde\mu +
\tilde\mu^-$,
\item heavier sparticles decay to lighter ones, e.g., $\tilde q 
\rightarrow
q\tilde g, \tilde\mu\rightarrow\mu\tilde\gamma$, and
\item the lightest sparticle (LSP) is stable, because it has no legal decay 
mode. 
\end{enumerate}  

This last feature constrains strongly the possible nature of the lightest
supersymmetric sparticle~\cite{EHNOS}. If it had either electric charge or
strong interactions, it would surely have dissipated its energy and
condensed into galactic disks along with conventional matter. There it
would surely have bound electromagnetically or via the strong interactions
to conventional nuclei, forming anomalous heavy isotopes that should have
been detected.

{\it A priori}, the LSP might have been a sneutrino partner of one of the
3 light neutrinos, but this possibility has been excluded by a combination
of the LEP neutrino counting and direct searches for cold dark matter.
Thus, the LSP is often thought to be the lightest neutralino $\chi$ of
spin 1/2, which naturally has a relic density of interest to
astrophysicists and cosmologists:  $\Omega_\chi h^2 = {\cal
O}(0.1)$~\cite{EHNOS}.

Finally, a fourth hint may be coming from the measured value of the muon's
anomalous magnetic moment, $g_\mu - 2$, which seems to differ slightly
from the \sm~ prediction~\cite{BNL,Davier}. If there is indeed a
significant discrepancy, this would require new physics at the TeV scale
or below, which could easily be provided by supersymmetry, as we see
later.

\subsection{Constraints on Supersymmetric Models}

Important experimental constraints on supersymmetric models have been
provided by the unsuccessful direct searches at LEP and the Tevatron 
collider. When compiling these, the supersymmetry-breaking masses of the 
different unseen scalar particles are often assumed to have a universal 
value $m_0$ at some GUT input scale, and likewise the fermionic partners 
of the vector bosons are also commonly assumed to have universal fermionic 
masses $m_{1/2}$ at the GUT scale - the so-called constrained MSSM or 
CMSSM.

\begin{figure}
\vskip 0.5in
\vspace*{-0.75in}
\begin{minipage}{8in}
\epsfig{file=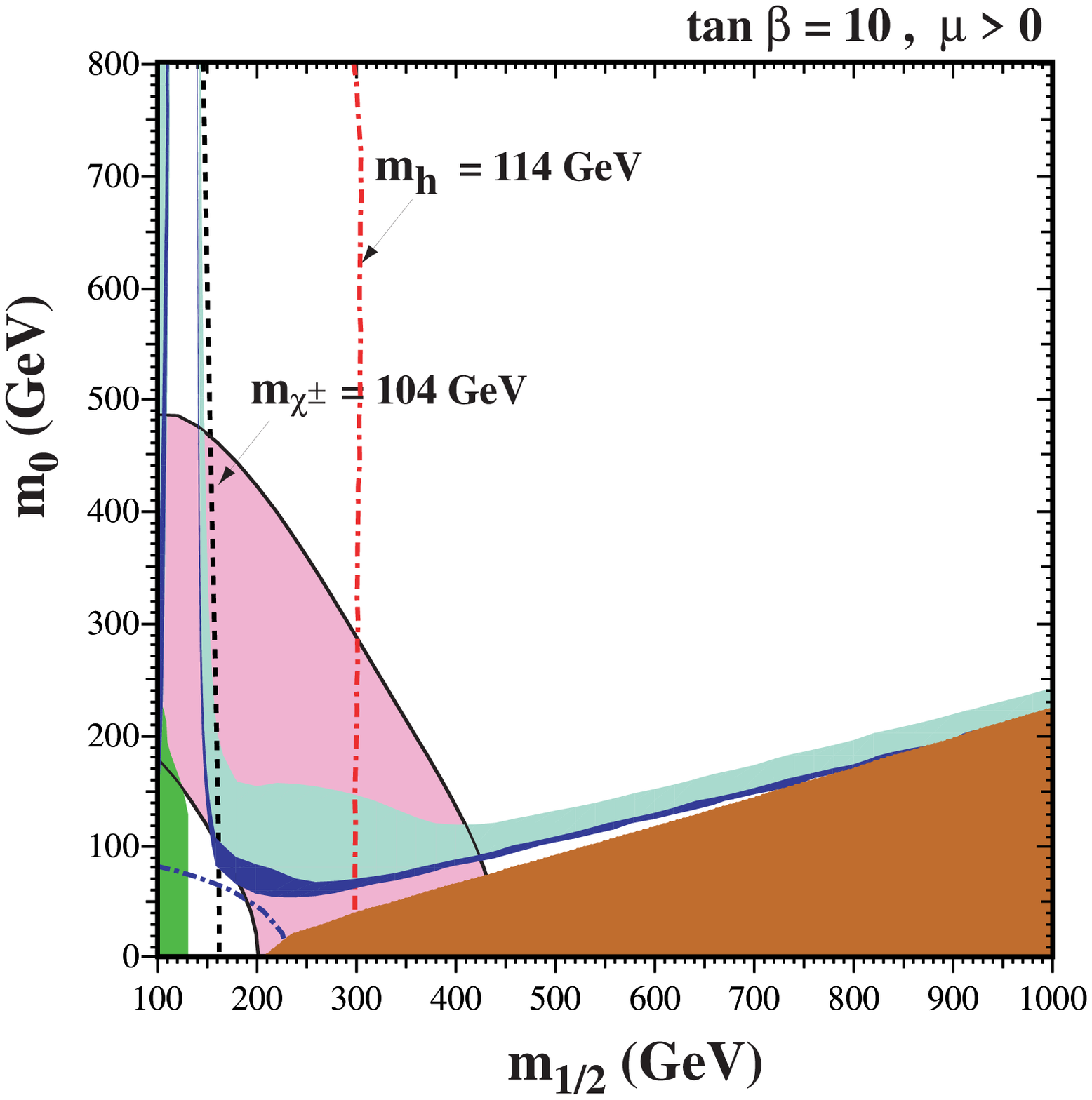,height=2.5in}
\epsfig{file=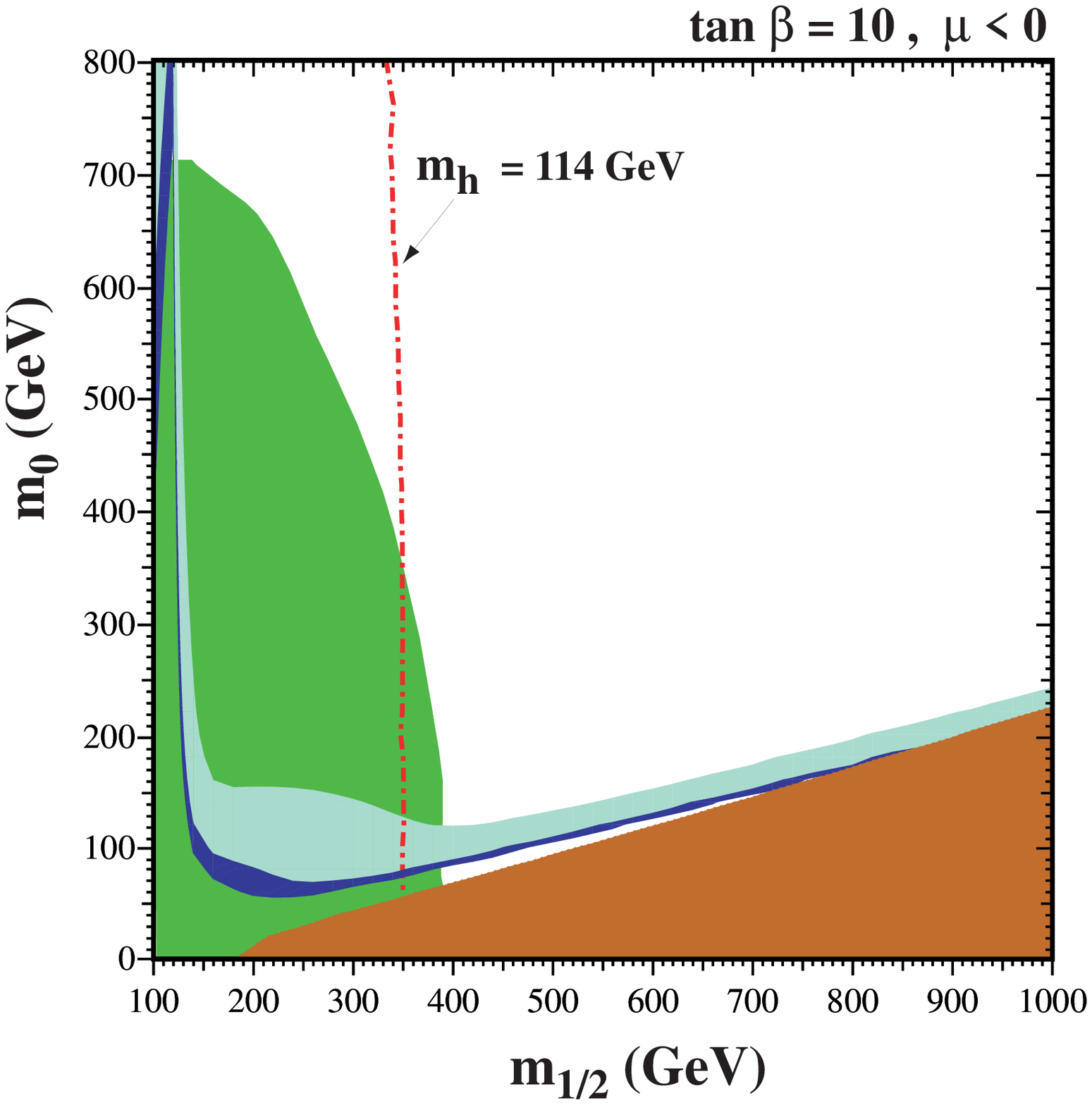,height=2.5in} \hfill
\end{minipage}
\begin{minipage}{8in}
\epsfig{file=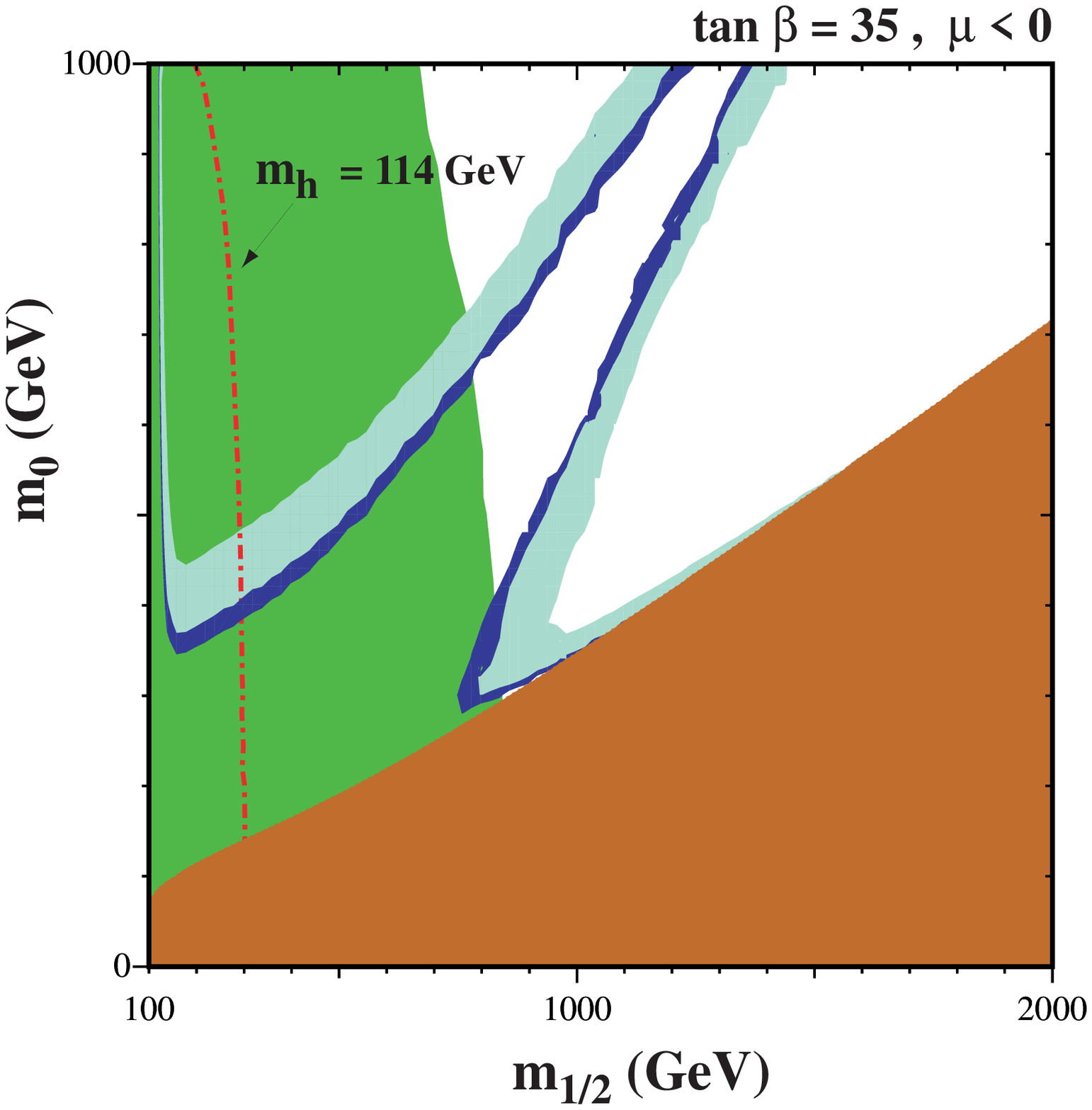,height=2.5in}
\epsfig{file=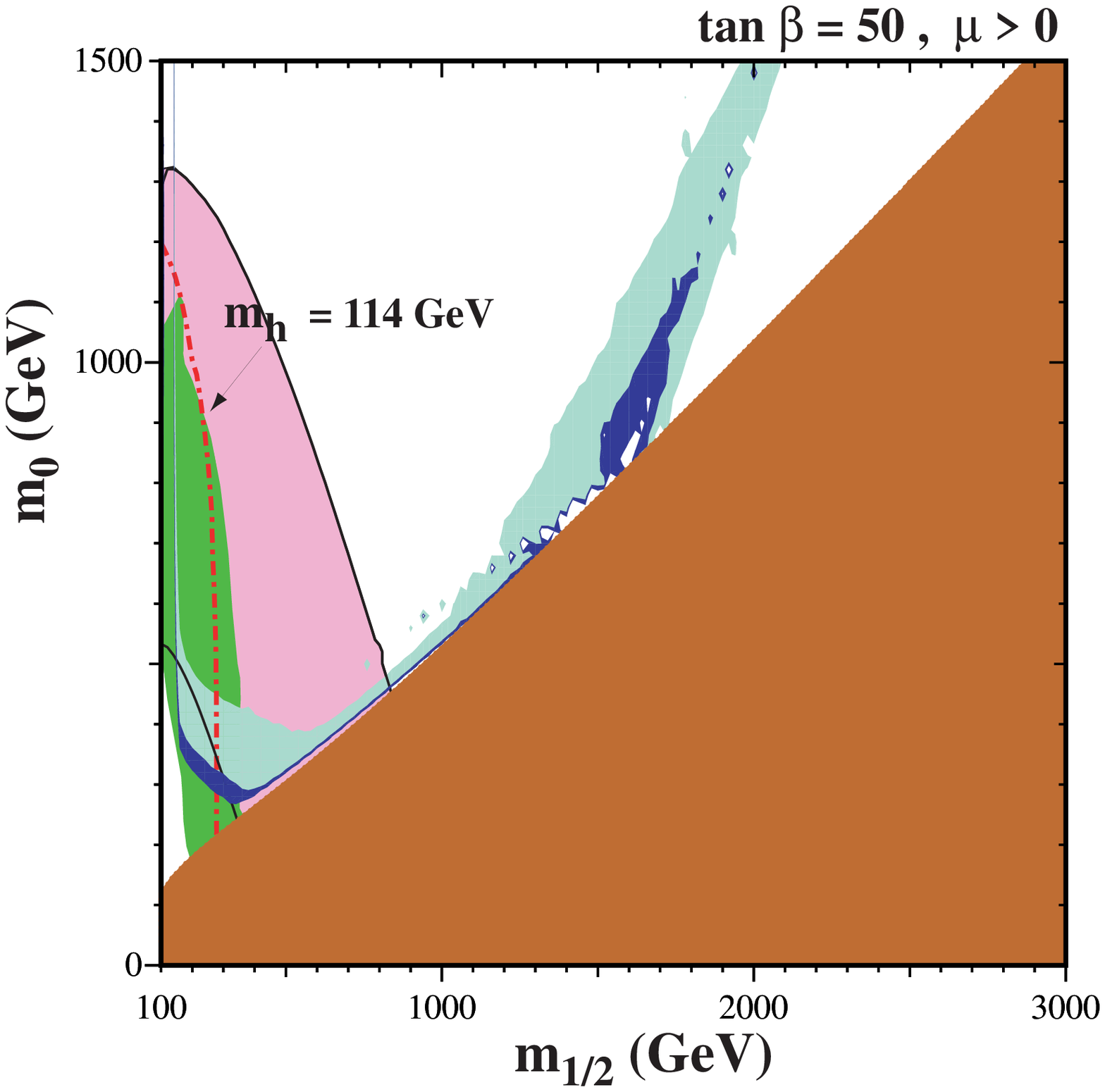,height=2.5in} \hfill
\end{minipage}
\caption{
{ 
Compilations of phenomenological constraints on the CMSSM for (a) $\tan
\beta = 10, \mu > 0$, (b) $\tan \beta = 10, \mu < 0$, (c) $\tan \beta =
35, \mu < 0$ and (d)  $\tan \beta = 50, \mu > 0$~\protect\cite{EOSS}. The
near-vertical lines are the LEP limits $m_{\chi^\pm} = 104$~GeV (dashed
and black)~\protect\cite{LEPsusy}, shown in (a) only, and $m_h = 114$~GeV
(dotted and red)~\protect\cite{LEPHWG}. Also, in the lower left corner of
(a), we show the $m_{\tilde e} = 99$ GeV 
contour~\protect\cite{LEPSUSYWG_0101}. The dark (brick red) shaded regions are
excluded because the LSP is charged. The light (turquoise) shaded areas
have \protect\mbox{$0.1\leq\ohsq\leq 0.3$}, and the smaller dark (blue)
shaded regions have \protect\mbox{$0.094\leq\ohsq\leq 0.129$}, as favoured
by WMAP~\protect\cite{EOSS}. The medium (dark green) shaded regions that
are most prominent in panels (b) and (c) are excluded by $b \to s
\gamma$~\protect\cite{bsg}. The shaded (pink) regions in panels (a) and
(d) show the $\pm 2 \, \sigma$ ranges of $g_\mu - 2$~\protect\cite{BNL}.
}}
\label{fig:CMSSM}
\end{figure}

The allowed domains in some of the $(m_{1/2}, m_0)$ planes for different
values of $\tan \beta$ and the sign of $\mu$ are shown in
Fig.~\ref{fig:CMSSM}. The various panels of this figure feature the limit
$m_{\chi^\pm} \gappeq 104$~GeV provided by chargino searches at
LEP~\cite{LEPsusy}. The LEP neutrino counting and other measurements have
also constrained the possibilities for light neutralinos, and LEP has also
provided lower limits on slepton masses, of which the strongest is
$m_{\tilde e}\gappeq$ 99 GeV~\cite{LEPSUSYWG_0101}, as illustrated in
panel (a)  of Fig.~\ref{fig:CMSSM}. The most important constraints on the
supersymmetric partners of the $u, d, s, c, b$ squarks and on the gluinos
are provided by the FNAL Tevatron collider: for equal masses $m_{\tilde q}
= m_{\tilde g} \gappeq$ 300 GeV. In the case of the $\tilde t$, LEP
provides the most stringent limit when $m_{\tilde t} - m_\chi$ is small,
and the Tevatron for larger $m_{\tilde t} - m_\chi$~\cite{LEPsusy}.

Another important constraint in Fig.~\ref{fig:CMSSM} is provided by the 
LEP lower limit on the
Higgs mass: $m_H > 114.4$~GeV~\cite{LEPHWG}. Since $m_h$ is sensitive to 
sparticle masses, particularly $m_{\tilde t}$, via loop corrections:
\begin{equation}
\delta m^2_h \propto {m^4_t\over m^2_W}~\ln\left({m^2_{\tilde t}\over
m^2_t}\right)~ + \ldots
\label{nine}
\end{equation}
the Higgs limit also imposes important constraints on the soft 
supersymmetry-breaking CMSSM parameters,
principally $m_{1/2}$~\cite{EGNO} as displayed in Fig.~\ref{fig:CMSSM}. 

Also shown in Fig.~\ref{fig:CMSSM} is the constraint imposed by
measurements of $b\rightarrow s\gamma$~\cite{bsg}. These agree with the
Standard Model, and therefore provide bounds on supersymmetric particles,
such as the chargino and charged Higgs masses, in particular.  

The final experimental constraint we consider is that due to the
measurement of the anomalous magnetic moment of the muon. Following its
first result last year~\cite{BNL1}, the BNL E821 experiment has recently
reported a new measurement~\cite{BNL} of $a_\mu\equiv {1\over 2} (g_\mu
-2)$, which deviates by about 2 standard deviations from the best
available Standard Model predictions based on low-energy $e^+ e^- \to $
hadrons data~\cite{Davier}. On the other hand, the discrepancy is more
like 0.9 standard deviations if one uses $\tau \to $ hadrons data to
calculate the Standard Model prediction. Faced with this confusion, and
remembering the chequered history of previous theoretical
calculations~\cite{lightbylight}, it is reasonable to defer judgement
whether there is a significant discrepancy with the Standard Model.
However, either way, the measurement of $a_\mu$ is a significant
constraint on the CMSSM, favouring $\mu > 0$ in general, and a specific
region of the $(m_{1/2}, m_0)$ plane if one accepts the theoretical
prediction based on $e^+ e^- \to $ hadrons data~\cite{susygmu}. The
regions preferred by the current $g-2$ experimental data and the $e^+ e^-
\to $ hadrons data are shown in Fig.~\ref{fig:CMSSM}.

Fig.~\ref{fig:CMSSM} also displays the regions where the supersymmetric 
relic density $\rho_\chi = \Omega_\chi \rho_{critical}$ falls within the 
range preferred by WMAP~\cite{WMAPnu}:
\begin{equation}
0.094 < \Omega_\chi h^2 < 0.129
\label{ten}
\end{equation}
at the 2-$\sigma$ level.
The upper limit on the relic density is rigorous, but the lower limit in 
(\ref{ten}) is optional, since there
could be other important contributions to the overall matter 
density. Smaller values of $\Omega_\chi h^2$ correspond to 
smaller values of $(m_{1/2}, m_0)$, in general.

We see in Fig.~\ref{fig:CMSSM} that there are significant regions of the 
CMSSM
parameter space where the relic density falls within the preferred range
(\ref{ten}). What goes into the calculation of the relic density? It is
controlled by the annihilation cross section~\cite{EHNOS}:
\begin{equation}
\rho_\chi = m_\chi n_\chi \, , \quad n_\chi \sim {1\over
\sigma_{ann}(\chi\chi\rightarrow\ldots)}\, ,
\label{eleven}
\end{equation}
where the typical annihilation cross section $\sigma_{ann} \sim 1/m_\chi^2$.
For this reason, the relic density typically increases with the relic
mass, and this combined with the upper bound in (\ref{ten}) then leads to
the common expectation that $m_\chi \lappeq {\cal O}(1)$~GeV. 

However, there are various ways in which the generic upper bound on
$m_\chi$ can be increased along filaments in the $(m_{1/2},m_0)$ plane.
For example, if the next-to-lightest sparticle (NLSP) is not much heavier
than $\chi$: $\Delta m/m_\chi \lappeq 0.1$, the relic density may be
suppressed by coannihilation: $\sigma (\chi + $NLSP$ \rightarrow \ldots
)$~\cite{coann}.  In this way, the allowed CMSSM region may acquire a
`tail' extending to larger sparticle masses. An example of this
possibility is the case where the NLSP is the lighter stau: $\tilde\tau_1$
and $m_{\tilde\tau_1} \sim m_\chi$, as seen in Figs.~\ref{fig:CMSSM}(a)
and (b)~\cite{ourcoann}. 

Another mechanism for extending the allowed CMSSM region to large $m_\chi$
is rapid annihilation via a direct-channel pole when $m_\chi \sim {1\over
2} m_{Higgs}$~\cite{funnel,EFGOSi}. This may yield a `funnel' extending
to large $m_{1/2}$ and $m_0$ at large $\tan\beta$, as seen in panels (c)
and (d) of Fig.~\ref{fig:CMSSM}~\cite{EFGOSi}. Yet another allowed region
at large $m_{1/2}$ and $m_0$ is the `focus-point' region~\cite{focus},
which is adjacent to the boundary of the region where electroweak symmetry
breaking is possible. The lightest
supersymmetric particle is relatively light in this region.

\subsection{Benchmark Supersymmetric Scenarios}

\begin{figure}
\begin{centering}
\hspace{1cm}
\epsfig{figure=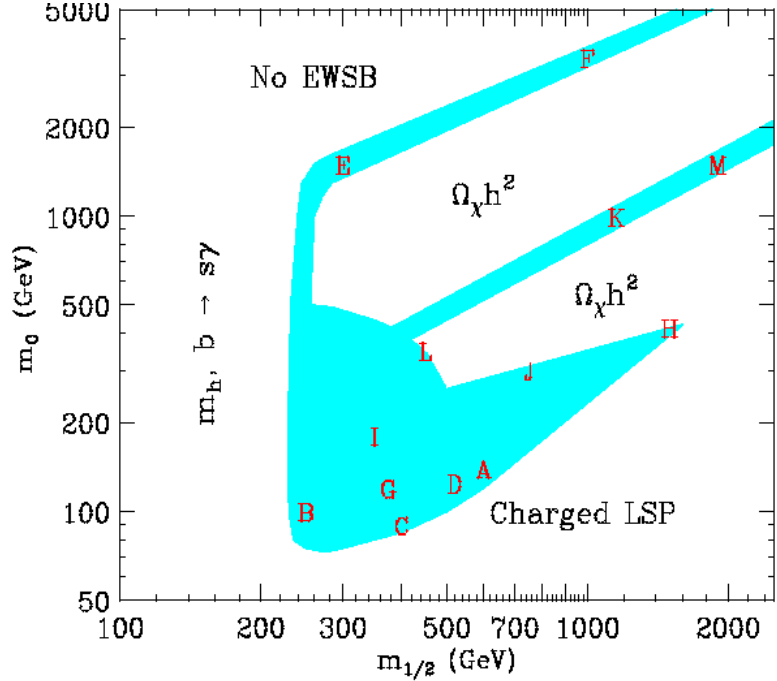,width=4in}
\end{centering}
\hglue3.5cm   
\caption[]{Sketch of the locations of the benchmark points proposed 
in~\cite{Bench} in the region of the
$(m_{1/2}, m_0)$ plane where $\ohsq$ falls within the range preferred by 
cosmology (shaded blue). Note that the filaments of the allowed parameter 
space extending to large $m_{1/2}$ and/or $m_0$ are sampled.}
\label{fig:Bench}
\end{figure}

As seen in Fig.~\ref{fig:CMSSM}, all the experimental, cosmological and
theoretical constraints on the MSSM are mutually compatible. As an aid to
understanding better the physics capabilities of the LHC and various 
other accelerators, as well as non-accelerator experiments, a set of
benchmark supersymmetric scenarios have been proposed~\cite{Bench}. 
Their distribution in the $(m_{1/2}, m_0)$ plane is sketched in 
Fig.~\ref{fig:Bench}. These benchmark scenarios
are compatible with all the accelerator constraints mentioned above,
including the LEP searches and $b \to s \gamma$, and yield relic densities
of LSPs in the range suggested by cosmology and astrophysics. The
benchmarks are not intended to sample `fairly' the allowed parameter
space, but rather to illustrate the range of possibilities currently
allowed.

In addition to a number of benchmark points falling in the `bulk' region
of parameter space at relatively low values of the supersymmetric particle
masses, as see in Fig.~\ref{fig:Bench}, we also proposed~\cite{Bench} some
points out along the `tails' of parameter space extending out to larger
masses. These clearly require some degree of fine-tuning to obtain the
required relic density~\cite{EO} and/or the correct $W^\pm$
mass~\cite{EENZ}, and some are also disfavoured by the supersymmetric
interpretation of the $g_\mu - 2$ anomaly, but all are logically
consistent possibilities.

\subsection{Prospects for Discovering Supersymmetry at Accelerators}

\begin{figure}
\begin{centering}
\hspace{0.5cm}
\epsfig{figure=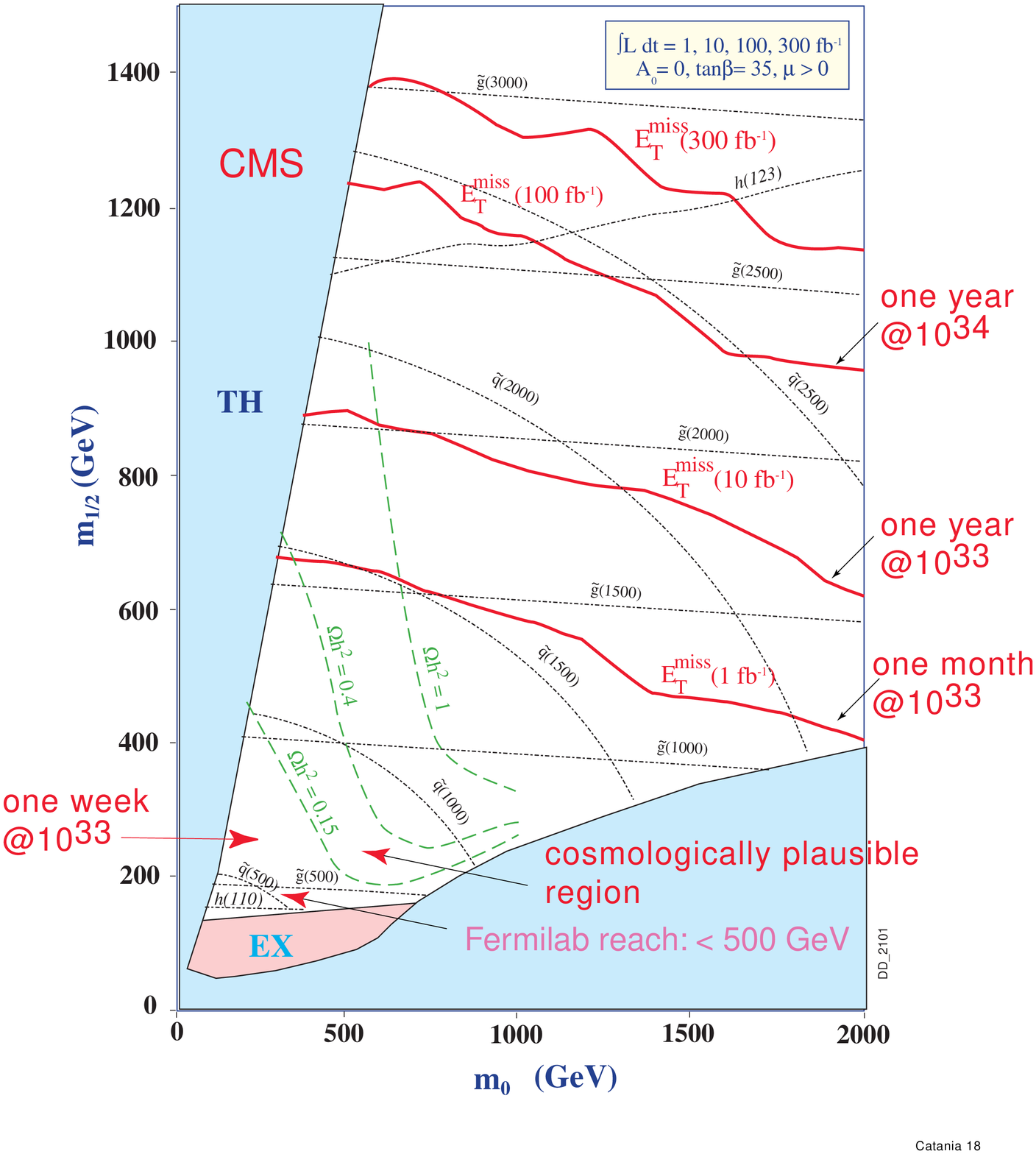,width=5in}
\end{centering}
\caption[]{\it The regions of the $(m_0, m_{1/2})$ plane that can be 
explored by the LHC with various integrated 
luminosities~\protect\cite{CMS}, using 
the missing energy + jets signature~\protect\cite{Paige}.}
\label{fig:Paige1}
\end{figure}

\begin{figure}
\begin{centering}
\hspace{2cm}
\epsfig{figure=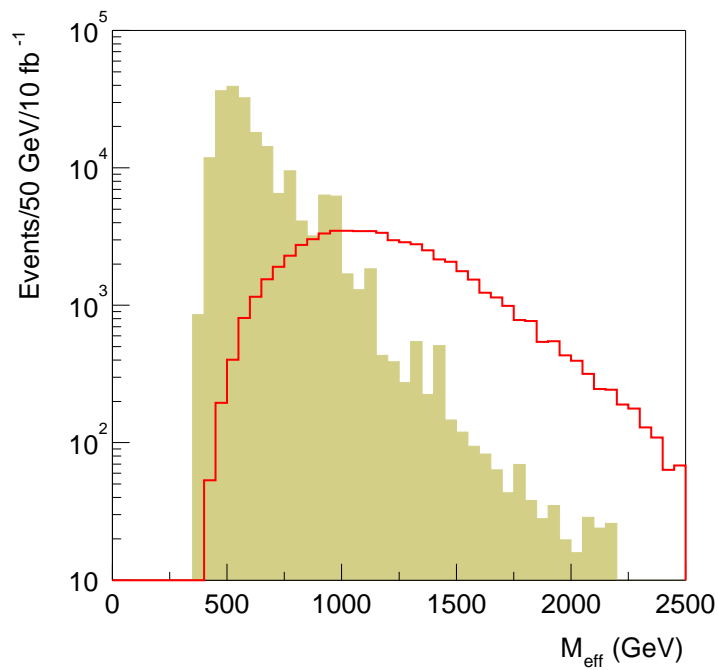,width=4in}
\end{centering}
\hglue3.5cm   
\caption[]{The distribution expected at the LHC in the variable 
$M_{\rm eff}$ that combines the jet energies with the missing 
energy~\protect\cite{HP,Tovey,Paige}.}
\label{fig:Paige3}
\end{figure}

\begin{figure}
\begin{centering}
\hspace{2cm}
\epsfig{figure=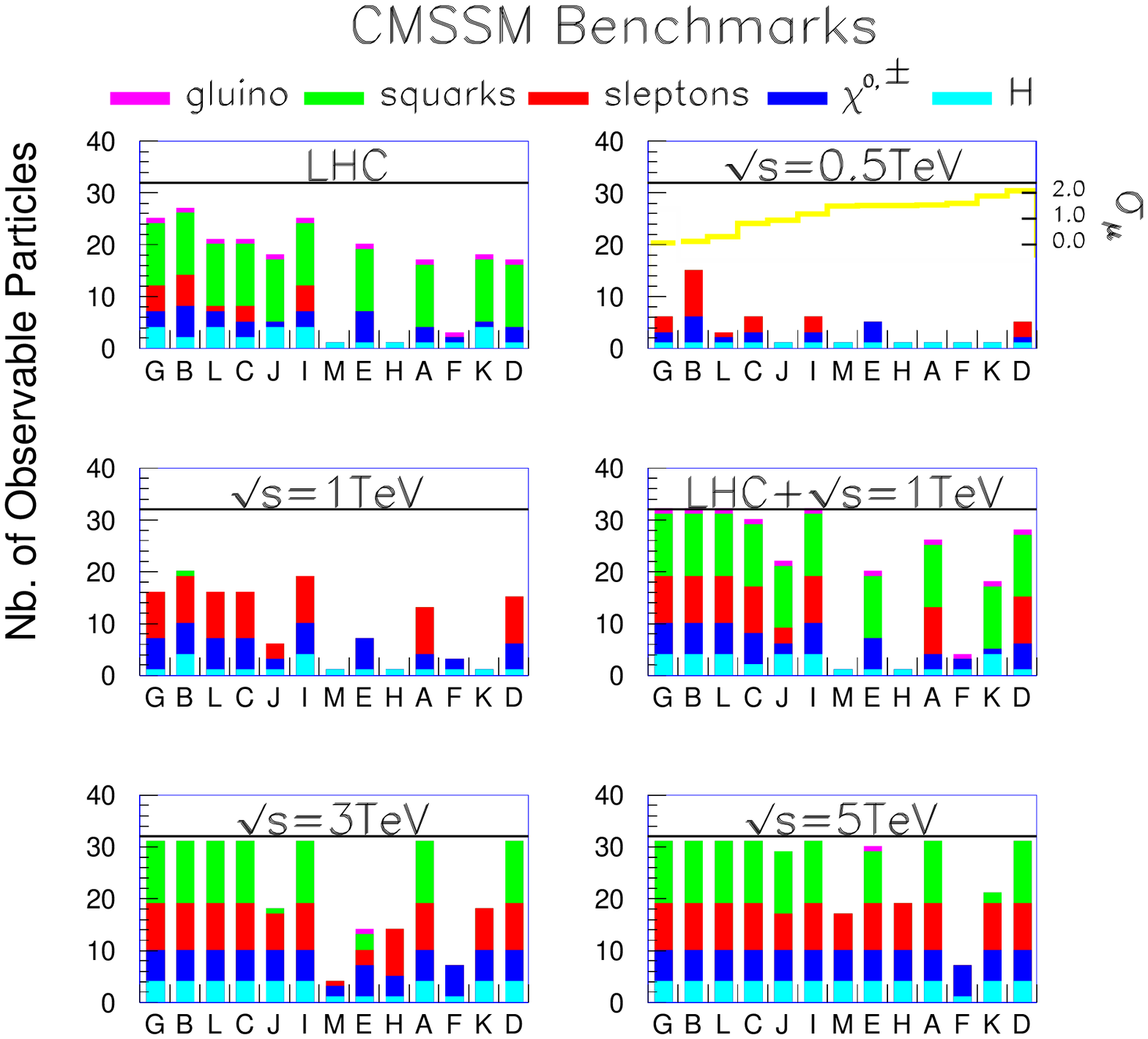,width=4in}
\end{centering}
\hglue3.5cm   
\caption[]{The numbers of different sparticles expected to be 
observable at the LHC and/or linear $e^+ e^-$ colliders with various 
energies, in each of the proposed benchmark 
scenarios~\protect\cite{Bench}, 
ordered by their 
difference from the present central experimental value of $g_\mu - 
2$~\protect\cite{BNL}.} 
\label{fig:Manhattan} 
\end{figure}

In the CMSSM discussed here, there are just a few prospects for
discovering supersymmetry at the FNAL {\it Tevatron
collider}~\cite{Bench}, but these could be increased in other
supersymmetric models~\cite{BenchKane}. On the other 
hand, there are good prospects for discovering 
supersymmetry at the {\it LHC}, and Fig.~\ref{fig:Paige1} shows its
physics reach for observing pairs of supersymmetric particles. The 
signature for supersymmetry - multiple jets (and/or leptons)  
with a large amount of missing energy - is quite distinctive, as seen in
Fig.~\ref{fig:Paige3}~\cite{Tovey,Paige}. Therefore, the detection of the
supersymmetric partners of quarks and gluons at the LHC is expected to be
quite easy if they weigh less than about 2.5~TeV~\cite{CMS}. Moreover, in
many scenarios one should be able to observe their cascade decays into
lighter supersymmetric particles. As seen in Fig.~\ref{fig:Manhattan},
large fractions of the supersymmetric spectrum should be seen in most of
the benchmark scenarios, although there are a couple where only the
lightest supersymmetric Higgs boson would be seen~\cite{Bench}, as seen in
Fig.~\ref{fig:Manhattan}.

{\it Electron-positron colliders} provide very clean experimental
environments, with egalitarian production of all the new particles that
are kinematically accessible, including those that have only weak
interactions, and hence are potentially complementary to the LHC, as
illustrated in Fig.~\ref{fig:Manhattan}. Moreover, polarized beams provide
a useful analysis tool, and $e \gamma$, $\gamma \gamma$ and $e^- e^-$
colliders are readily available at relatively low marginal costs. However,
the direct production of supersymmetric particles at such a collider
cannot be guaranteed~\cite{EGO}. We do not yet know what the
supersymmetric threshold energy may be (or even if there is one!). We may
well not know before the operation of the LHC, although $g_\mu - 2$ might
provide an indication~\cite{susygmu}, if the uncertainties in the Standard
Model calculation can be reduced. However, if an $e^+ e^-$ collider is
above the supersymmetric threshold, it will be able to measure very
accurately the sparticle masses. By combining its measurements with those
made at the LHC, it may be possible to calculate accurately from first
principle the supersymmetric relic density and compare it with the
astrophysical value.

\subsection{Searches for Dark Matter Particles}

In the above discussion, we have paid particular attention to the region
of parameter space where the lightest supersymmetric particle could
constitute the cold dark matter in the Universe~\cite{EHNOS}. How easy
would this be to detect? 

$\bullet$ One strategy is to look for relic annihilations in the {\it
galactic halo}, which might produce detectable antiprotons or positrons in
the cosmic rays~\cite{SS}. Unfortunately, the rates for
their production are not very promising in the benchmark scenarios we
studied~\cite{EFFMO}.

\begin{figure}[htb]
\centerline{\epsfxsize = 0.5\textwidth \epsffile{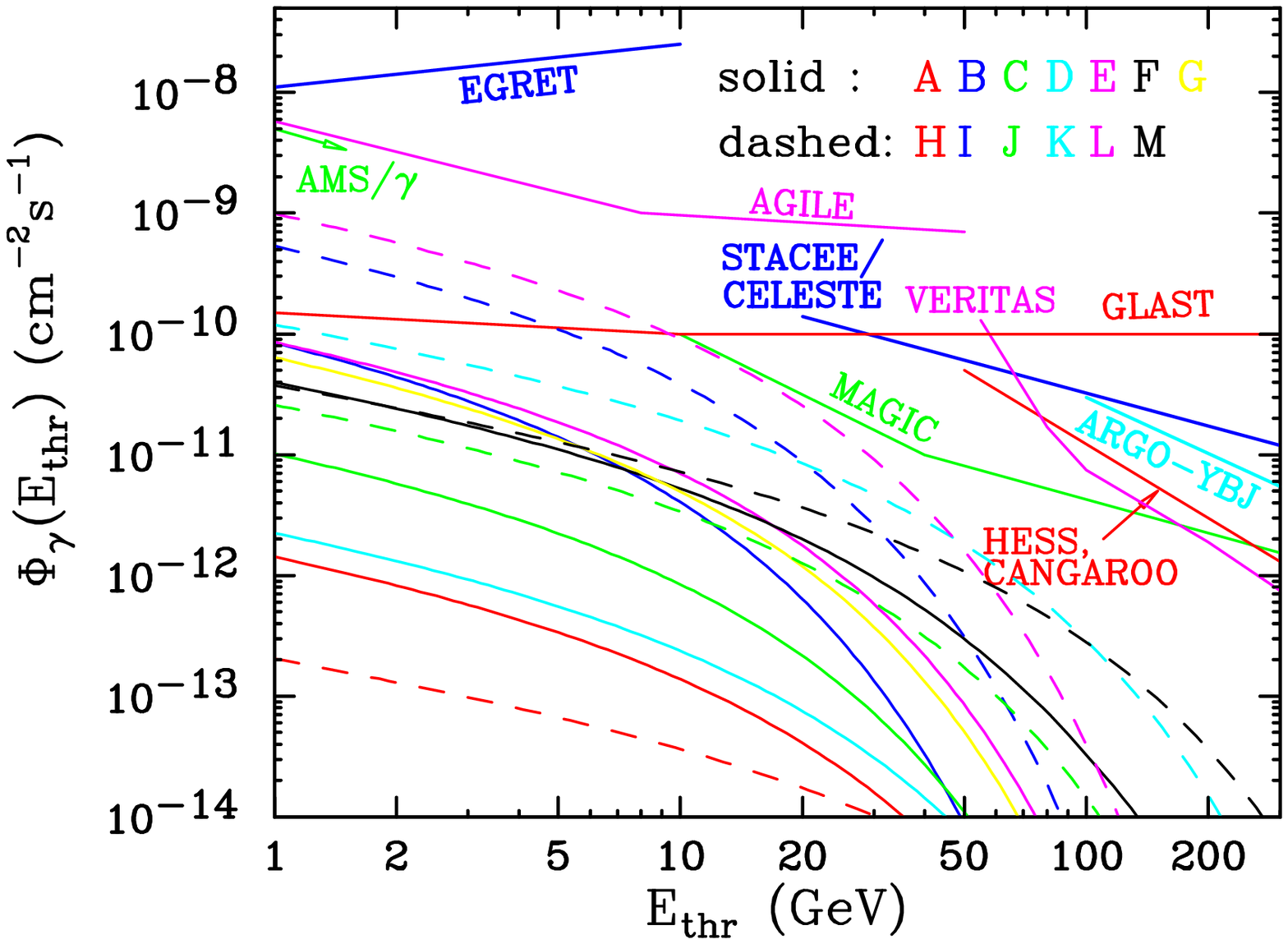}
\hfill \epsfxsize = 0.5\textwidth \epsffile{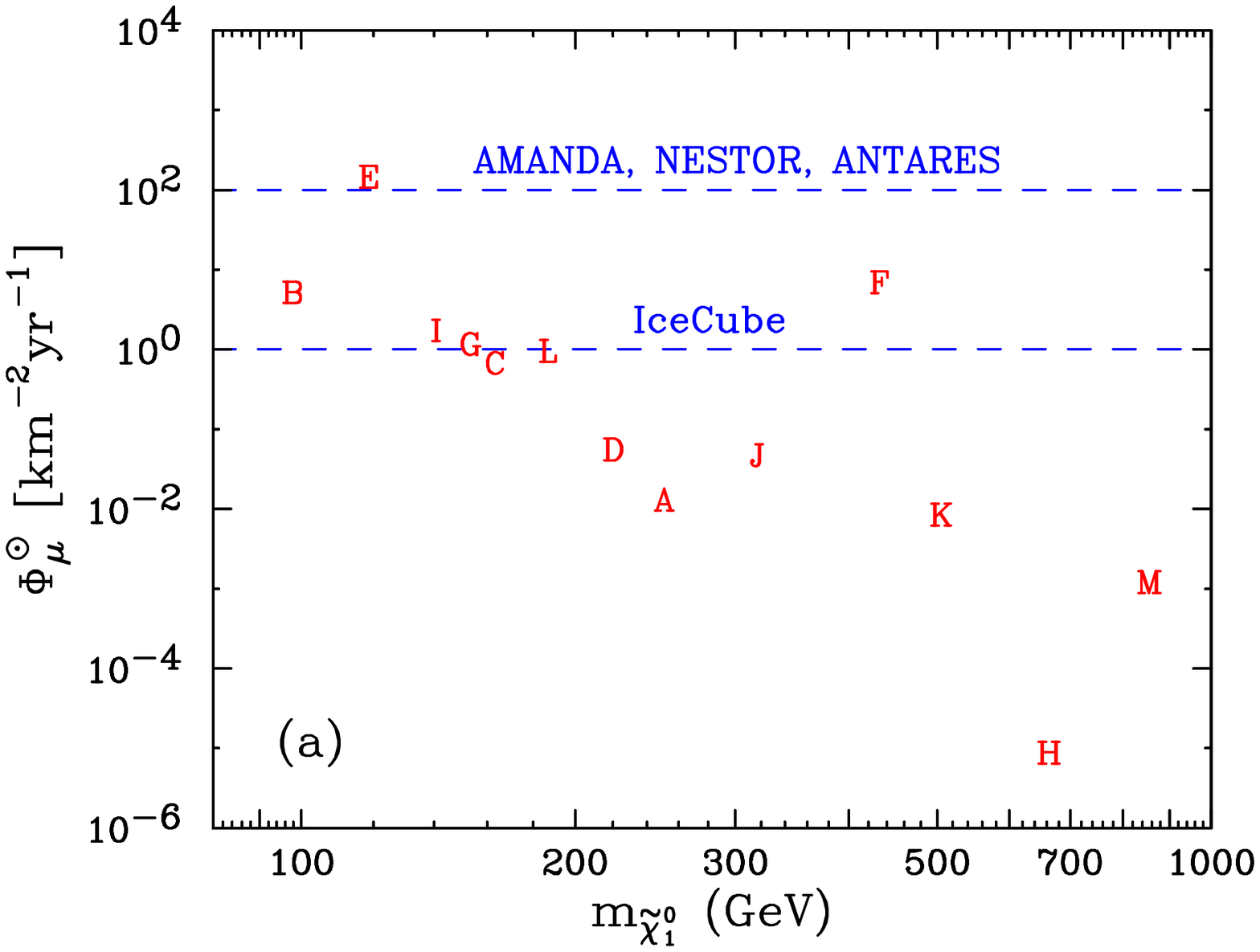}
}
\caption{
Left panel: Spectra of photons from the annihilations of
dark matter particles in the core of our galaxy, in different benchmark   
supersymmetric models~\protect\cite{EFFMO}. Right panel: Signals for
muons produced by energetic neutrinos originating from annihilations of 
dark matter particles in the core of the Sun, in the same benchmark
supersymmetric models~\protect\cite{EFFMO}.}
\vspace*{0.5cm}
\label{fig:RS13}
\end{figure}

$\bullet$ Alternatively, one might look for annihilations in the core of
our galaxy, which might produce detectable {\it gamma rays}. As seen in
the left panel of Fig.~\ref{fig:RS13}, this may be possible in certain
benchmark scenarios~\cite{EFFMO}, though the rate is rather uncertain 
because of the unknown enhancement of relic particles in our galactic core.

$\bullet$ A third strategy is to look for {\it annihilations inside the
Sun or Earth}, where the local density of relic particles is enhanced in a
calculable way by scattering off matter, which causes them to lose energy
and become gravitationally bound~\cite{SOS}. The signature would then be
energetic neutrinos that might produce detectable muons. Several
underwater and ice experiments are underway or planned to look for this
signature, and this strategy looks promising for several benchmark
scenarios, as seen in the right panel of Fig.~\ref{fig:RS13}~\cite{EFFMO}.
It will be interesting to have such neutrino telescopes in different
hemispheres, which will be able to scan different regions of the sky for
astrophysical high-energy neutrino sources.

\begin{figure}[htb]
\centerline{\epsfxsize = 0.5\textwidth \epsffile{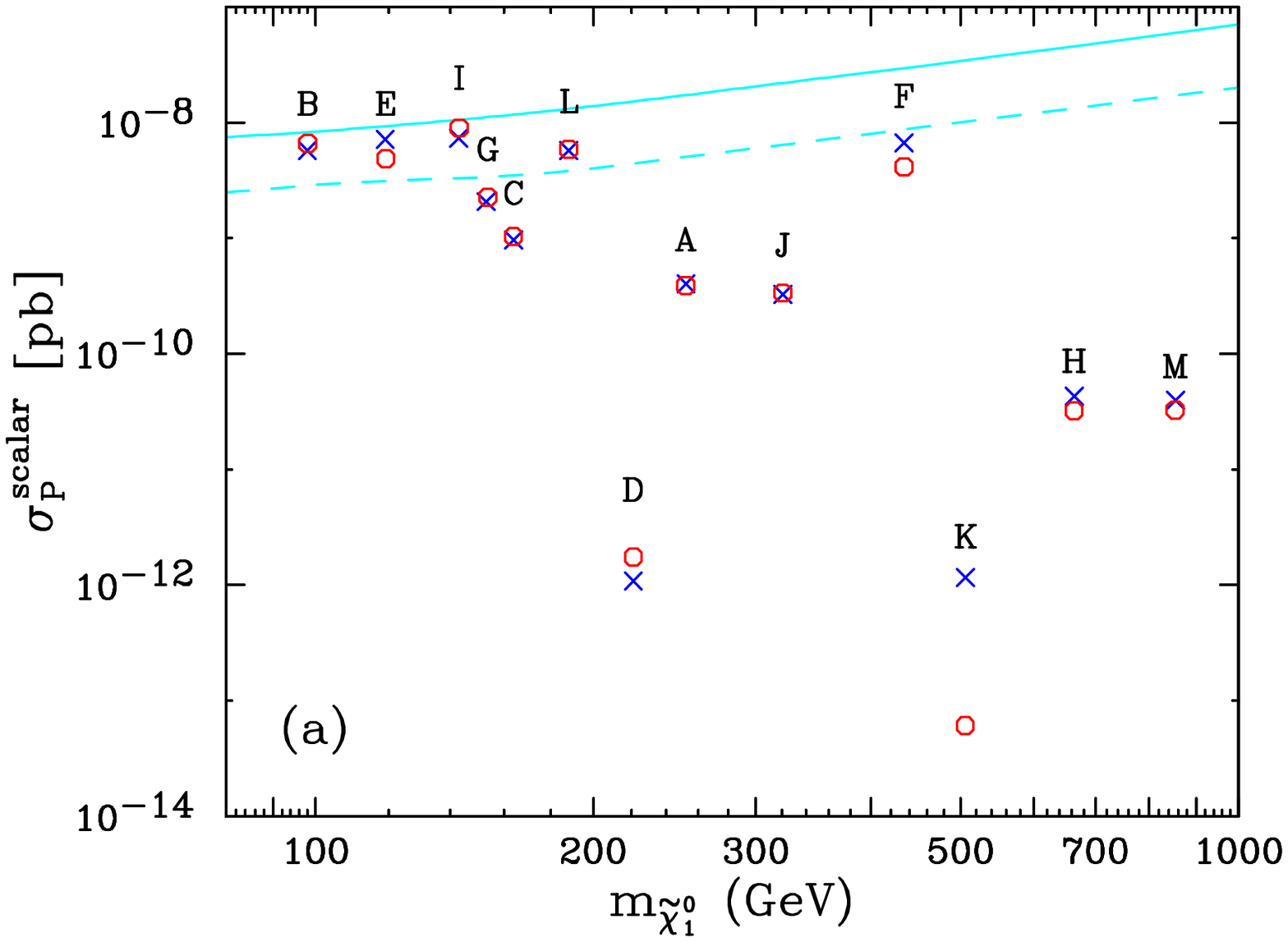}
\hfill \epsfxsize = 0.5\textwidth \epsffile{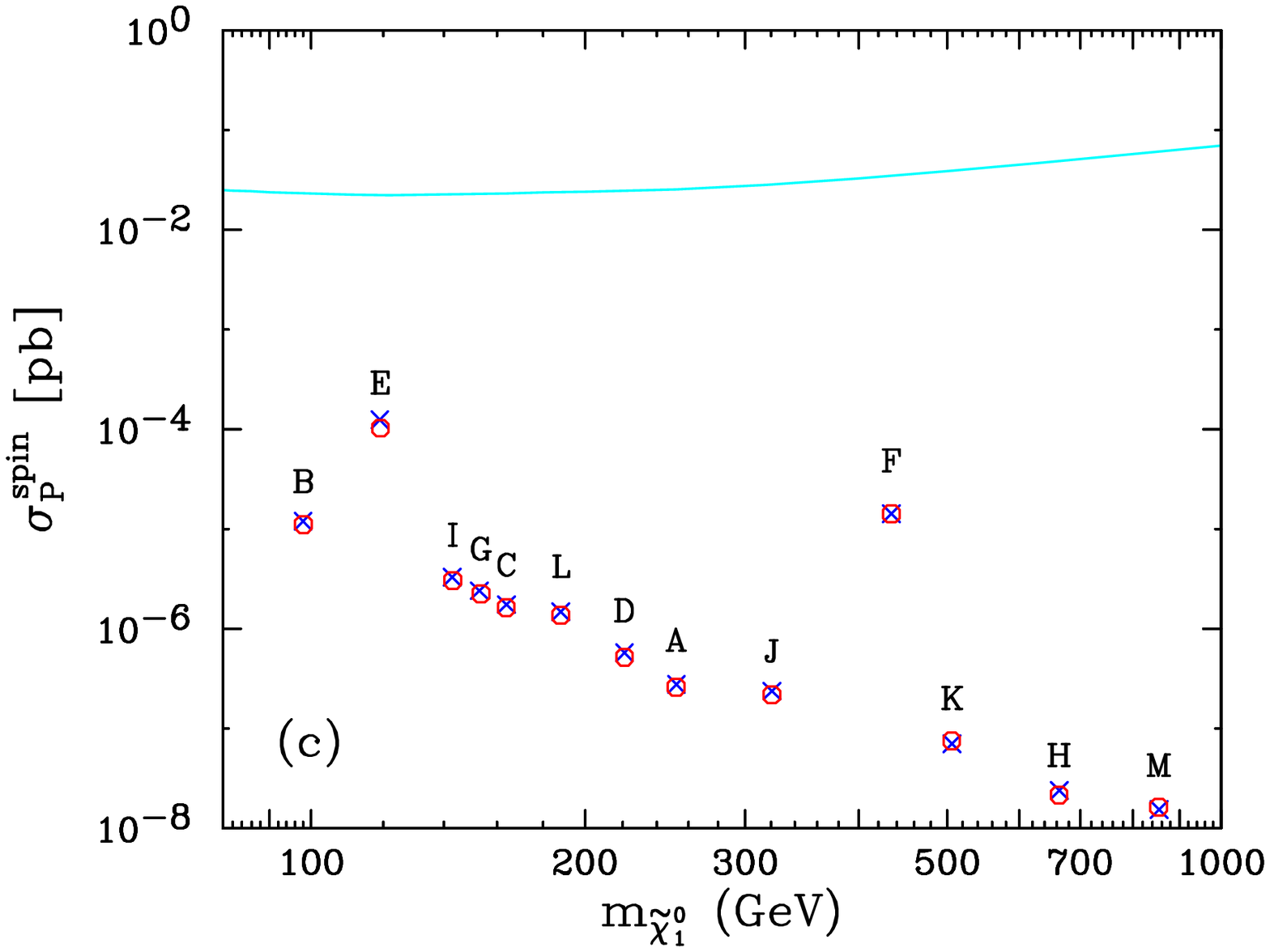}
}
\caption[]{
Left panel: elastic spin-independent scattering of supersymmetric relics
on protons calculated in benchmark scenarios~\protect\cite{EFFMO},
compared with the projected sensitivities for CDMS
II~\protect\cite{Schnee:1998gf} and CRESST~\protect\cite{Bravin:1999fc}
(solid) and GENIUS~\protect\cite{GENIUS} (dashed). The predictions of the
{\tt SSARD} code (blue crosses) and {\tt Neutdriver}\cite{neut} (red
circles) for neutralino-nucleon scattering are
compared~\protect\cite{EFFMO}. The labels A, B, ...,L correspond to the
benchmark points as shown in Fig.~\protect\ref{fig:Bench}. Right panel:
prospects for detecting elastic spin-dependent scattering in the benchmark
scenarios, which are less bright~\protect\cite{EFFMO}.}
\label{fig:DM}
\end{figure}

$\bullet$ The most satisfactory way to look for supersymmetric relic
particles is directly via their {\it elastic scattering on nuclei} in a
low-background laboratory experiment~\cite{GW}. There are two types of
scattering matrix elements, spin-independent - which are normally dominant
for heavier nuclei, and spin-dependent - which could be interesting for
lighter elements such as fluorine. The best experimental sensitivities so
far are for spin-independent scattering, and one experiment has claimed a
positive signal~\cite{DAMA}. However, this has not been confirmed by a
number of other experiments~\cite{unDAMA}. In the benchmark scenarios the
rates are considerably below the present experimental
sensitivities~\cite{EFFMO}, but there are prospects for improving the
sensitivity into the interesting range, as seen in Fig.~\ref{fig:DM}.

\section{Inflation}

\subsection{Motivations}

One of the main motivations for inflation~\cite{inflation} is the {\it 
horizon} or {\it 
homogeneity} problem: why are distant parts of the Universe so similar:
\beq
\left( { \delta T \over T } \right)_{CMB} \; \sim \; 10^{-5} \; ?
\label{similar}
\eeq
In conventional Big Bang cosmology, the largest patch of the CMB sky which 
could have been causally connected, i.e., across which a signal could have 
travelled at the speed of light since the initial singularity, is about 2 
degrees. So how did opposite parts of the Universe, 180 degrees apart, 
`know' how to coordinate their temperatures and densities?

Another problem of conventional Big bang cosmology is the {\it size} or 
{\it age} problem. The Hubble expansion rate in conventional Big bang 
cosmology is given by:
\beq
H^2 \; \equiv \; \left( {{\dot a} \over a} \right)^2 \; = \; {8 \pi G_N 
\rho 
\over 3} - {k \over a^2},
\label{FRW}
\eeq
where $k = 0$ or $\pm 1$ is the curvature. The only dimensionful 
coefficient in (\ref{FRW}) is the Newton constant, $G_N \equiv 1/M_P^2: 
M_P \simeq 1.2 \times 10^{19}$~GeV. A generic solution of (\ref{FRW}) 
would have a characteristic scale size $a \sim \ell_P \equiv 1 / M_P \sim 
10^{-33}$~s and live to the ripe old age of $t \sim t_P \equiv \ell_P /c 
\simeq 10^{-43}$~s. Why is our Universe so long-lived and big? Clearly, we 
live in an atypical solution of (\ref{FRW})!

A related issue is the {\it flatness} problem. Defining, as usual
\beq
\Omega \; \equiv \; {\rho \over \rho_c}: \; \rho_c \equiv {3 H^2 \over 8 
\pi G_N},
\label{Omega}
\eeq
we have
\beq
\Omega (t) \; = \; {1 \over 1 - { \left( {k / a^2} \right) \over \left( 8 
\pi G_N 
\rho / 3 \right) } }.
\label{notequal1}
\eeq
Since $\rho \sim a^{-4}$ during the radiation-dominated era and $\sim 
a^{-3}$ during the matter-dominated era, it is clear from 
(\ref{notequal1}) that $\Omega (t) \to 0$ rapidly: for 
$\Omega$ to be ${\cal O}(1)$ as it is today, $|\Omega - 1|$ must have been 
${\cal O}(10^{-60})$ at the Planck epoch when $t_P \sim 10^{-43}$~s. The 
density of the very early Universe must have been very finely tuned in 
order for its geometry to be almost flat today.

Then there is the {\it entropy} problem: why are there so many particles 
in the visible Universe: $S \sim 10^{90}$? A `typical' Universe would have 
contained ${\cal O}(1)$ particles in its size $\sim \ell_P^3$.

All these particles have diluted what might have been the primordial 
density of {\it unwanted massive particles} such as magnetic monopoles and 
gravitinos. Where did they go?

The basic idea of inflation~\cite{Guth} is that, at some early epoch in 
the history of 
the Universe, its energy density may have been dominated by an almost 
constant term:
\beq
\left( { {\dot a} \over a} \right)^2 \; = \; {8 \pi G_N \rho \over 3} - {k 
\over a^2}: \; \; \rho \; = \; V,
\label{infV}
\eeq
leading to a phase of almost de Sitter expansion.
It is easy to see that the second (curvature) term in (\ref{infV}) rapidly 
becomes negligible, and that
\beq
a \; \simeq \; a_I e^{Ht}: \; \; H \; = \; \sqrt{ {8 \pi G_N \over 3} V }
\label{expV}
\eeq
during this inflationary expansion.

It is then apparent that the {\it horizon} would have expanded (near-) 
exponentially, so that the entire visible Universe might have been within 
our pre-inflationary horizon. This would have enabled initial homogeneity 
to have been established. The trick is not somehow to impose connections 
beyond the horizon, but rather to make the horizon much larger than 
naively expected in conventional Big Bang cosmology:
\beq
a_H \; \simeq \; a_I e^{H \tau} \; \gg \; c \tau,
\label{newhorizon}
\eeq
where $H \tau$ is the number of e-foldings during inflation.
It is also apparent that the $- {k \over a^2}$ term in (\ref{infV})  
becomes negligible, so that the Universe is almost {\it flat} with
$\Omega_{tot} \simeq 1$. However, as we see later, perturbations during
inflation generate a small deviation from unity: $|\Omega_{tot} - 1 |
\simeq 10^{-5}$. Following inflation, the conversion of the inflationary 
vacuum energy into 
particles reheats the Universe, filling it with the required {\it 
entropy}. Finally, the closest pre-inflationary {\it monopole} or {\it 
gravitino} is 
pushed away, further than the origin of the CMB, by the exponential 
expansion of the Universe.

{}From the point of view of general relativity, the (near-) 
constant inflationary vacuum 
energy is equivalent to a cosmological constant $\Lambda$:
\beq
R_{\mu\nu} - {1 \over 2} g_{\mu\nu} R \; = \; 8 \pi G_N T_{\mu\nu} + 
\Lambda g_{\mu\nu}.
\label{cosmoconst}
\eeq
We may compare the right-hand side of (\ref{cosmoconst}) with  the 
energy-momentum tensor of a standard fluid:
\beq
T_{\mu\nu} \; = \; - p g_{\mu \nu} + ( \rho + p ) U_\mu U_\nu
\label{fluid}
\eeq
where $U_\mu = (1, 0, 0, 0)$ is the four-momentum vector for a comoving 
fluid. We can therefore write
\beq
R_{\mu\nu} - {1 \over 2} g_{\mu\nu} R \; = \; 8 \pi G_N 
T^\Lambda_{\mu\nu},
\label{RLambda}
\eeq
where 
\beq
\rho_\Lambda \; \equiv \; {\Lambda \over 8 \pi G_N} \; \equiv \; - 
p_\Lambda.
\label{rhoLambda}
\eeq
Thus, we see that inflation has negative pressure. The value of the 
cosmological constant today, as suggested by recent 
observations~\cite{SN,Bahcall}, is {\it 
many} orders of magnitude smaller than would have been required during 
inflation: $\rho_\Lambda \sim 10^{-50}$~GeV$^4$ compared with the density 
$V \sim 10^{+65}$~GeV$^4$ required during inflation, as we see later.

Such a small value of the cosmological energy density is also {\it much 
smaller} than many contributions to it from identifiable physics sources: 
$\rho (QCD) \sim 10^{-4}$~GeV$^4$, $\rho (Electroweak) \sim 10^9$~GeV$^4$,
$\rho (GUT) \sim 10^{64} (?)$~GeV$^4$ and $\rho (Quantum Gravity) \sim 
10^{74} (?)$~GeV$^4$. Particle physics offers no reason to expect the 
present-day vacuum energy to lie within the range suggested by cosmology, 
and raises the question why it is not many orders of magnitude larger.

\subsection{Some Inflationary Models}

The first inflationary potential $V$ to be proposed was one with a
`double-dip' structure \`a la Higgs~\cite{Guth}. The {\it old inflation} 
idea was that 
the Universe would
have started in the false vacuum with $V \ne 0$, where it would have
undergone many e-foldings of de Sitter expansion. Then, the Universe was 
supposed to
have tunnelled through the potential barrier to the true vacuum with $V
\simeq 0$, and subsequently thermalized. The inflation required before 
this tunnelling was
\beq
H \tau \; \gappeq \; 60: \; \; H \; = \; \sqrt{ {\Lambda \over 3}}, \; 
\Lambda \; = \; 8 \pi G_N V.
\label{oldinf}
\eeq
The problem with this old inflationary scenario was that the phase 
transition to the new vacuum would never have been completed. The Universe 
would look like a `Swiss cheese' in which the bubbles of true vacuum would 
be expanding as $t^{1/2}$ or $t^{2/3}$, while the `cheese' between them 
would still have been expanding exponentially as $e^{H t}$. Thus, the 
fraction of space in the false vacuum would be
\beq
f \; \sim \; exp \left[ H t \left( 3 - {4 \pi \over 3} {\Gamma \over H^4} 
\right) \right],
\label{cheese}
\eeq
where $\Gamma$ is the bubble nucleation rate per unit four-volume. The 
fraction $f \to 0$ only if $\Gamma / H^4 \simeq {\cal O}(1)$, but in this 
case there would not have been sufficient e-foldings for adequate 
inflation.

One of the fixes for this problem trades under the name of {\it new
inflation}~\cite{Linde}. The idea is that the near-exponential expansion 
of the
Universe took place in a flat region of the potential $V(\phi)$ that is
not separated from the true vacuum by any barrier. It might have been
reached after a first-order transition of the type postulated in old
inflation, in which case one can regard our Universe as part of a bubble
that expanded near-exponentially inside the `cheese' of old vacuum, and
there could be regions beyond our bubble that are still expanding (near-)  
exponentially. For the Universe to roll eventually downhill into the true
vacuum, $V(\phi)$ could not quite be constant, and hence the Hubble
expansion rate $H$ during inflation was also not constant during new
inflation.

An example of such a scenario is {\it chaotic inflation}~\cite{chaos},
according to which there is no `bump' in the effective potential
$V(\phi)$, and hence no phase transition between old and new vacua.
Instead, any given region of the Universe is assumed to start with some
random value of the inflaton field $\phi$ and hence the potential
$V(\phi)$, which decreases monotonically to zero. If the initial value of
$V(\phi)$ is large enough, and the potential flat enough, (our part of)
the Universe will undergo sufficient expansion.

Another fix for old inflation trades under the name of {\it extended
inflation}~\cite{extended}. Here the idea is that the tunnelling rate
$\Gamma$ depends on some other scalar field $\chi$ that varies while the
inflaton $\phi$ is still stuck in the old vacuum. If $\Gamma (\chi)$ is
initially small, but $\chi$ then changes so that $\Gamma (\chi)$ becomes
large, the problem of completing the transition in the `Swiss cheese'
Universe is solved.

All these variants of inflation rely on some type of elementary scalar
inflaton field. Therefore, the discovery of a Higgs boson would be a
psychological boost for inflation, even though the electroweak Higgs boson
cannot be responsible for it directly. Moreover, just as supersymmetry is
well suited for stabilizing the mass scale of the electroweak Higgs boson,
it may also be needed to keep the inflationary potential under
control~\cite{susyinf}. Later in this Lecture, I discuss a specific
supersymmetric inflationary model.

\subsection{Density Perturbations}

The above description is quite classical. In fact, one should expect 
quantum fluctuations in the initial value of the inflaton field $\phi$, 
which would cause the roll-over into the true vacuum to take place 
inhomogeneously, and different parts of the Universe to expand 
differently. As we discuss below in more detail, these quantum 
fluctuations would give rise to a Gaussian random field of perturbations 
with similar magnitudes on different scale sizes, just as the 
astrophysicists have long wanted. The magnitudes of these perturbations 
would be linked to the value of the effective potential during inflation, 
and would be visible in the CMB as adiabatic temperature fluctuations:
\beq
{\delta T \over T } \; \sim \; {\delta \rho \over \rho} \; \sim \; \mu^2 
G_N,
\label{deltaT}
\eeq
where $\mu \equiv V^{1/4}$ is a typical vacuum energy scale during 
inflation. As we discuss later in more detail, consistency with the CMB 
data from COBE {\it et al}., that find $\delta T / T \simeq 10^{-5}$, is 
obtained if
\beq
\mu \; \simeq \; 10^{16}~{\rm GeV},
\label{iscale}
\eeq
comparable with the GUT scale.

Each density perturbation can be regarded as an embryonic potential well,
into which non-relativistic cold dark matter particles may fall,
increasing the local contrast in the mass-energy density. On the other 
hand, relativistic hot dark matter particles will escape from small-scale 
density perturbations, modifying their rate of growth. This also depends 
on the expansion rate of the Universe and hence the cosmological constant. 
Present-day data are able to distinguish the effects of different 
categories of dark matter. In particular, as we already discussed, the 
WMAP and other data tell us that the 
density of hot dark matter neutrinos is relatively small~\cite{WMAPnu}:
\beq
\Omega_\nu h^2 \; < \; 0.0076,
\label{smallhot}
\eeq
whereas the density of cold dark matter is relatively 
large~\cite{WMAPnu}:
\beq
\Omega_{CDM} h^2 = 0.1126^{+0.0081}_{-0.0091},
\label{bigcold}
\eeq
and the cosmological constant is even larger: $\Omega_\Lambda \simeq 
0.73$.

The cold dark matter amplifies primordial perturbations already while the
conventional baryonic matter is coupled to radiation before
(re)combination. Once this epoch is passed and the CMB decouples from the
conventional baryonic matter, the baryons become free to fall into the
`holes' prepared for them by the cold dark matter that has fallen into the
overdense primordial perturbations. In this way, structures in the
Universe, such as galaxies and their clusters, may be formed earlier than
they would have appeared in the absence of cold dark matter. 

All this theory is predicated on the presence of primordial perturbations
laid down by inflation~\cite{perts}, which we now explore in more detail.

There are in fact two types of perturbations, namely density fluctuations 
and gravity waves. To describe the first, we consider the density field 
$\rho ( {\mathbf x} )$ and its perturbations $\delta ( {\mathbf x} ) 
\equiv ( \rho ( {\mathbf x} ) - <\rho> ) / <\rho>$, which we can decompose 
into Fourier modes:
\beq
\delta ( {\mathbf x} ) \; = \; \int d^3 {\mathbf x} \delta_{\mathbf k} 
e^{-i {\mathbf k} \cdot {\mathbf x}}.
\label{deltak}
\eeq
The density perturbation on a given scale $\lambda$ is then given by
\beq
\left( {\delta \rho \over \rho} \right)^2_\lambda \; = \; \left( { k^3 
|\delta_{\mathbf k}|^2 \over 2 \pi^2 } \right)_{k^{-1} = \lambda},
\label{deltarho}
\eeq
whose evolution depends on the ratio $\lambda / a_H$, where $a_H \equiv c 
\cdot t$ is the naive horizon size.

The evolution of small-scale perturbations with $\lambda / a_H < 1$ 
depends on the astrophysical dynamics, such as the quation of state, 
dissipation, the Jeans instability, etc.:
\beq
{\ddot {\delta}}_{\mathbf k} + 2 H {\dot \delta}_{\mathbf k} + v_s^2 {k^2 
\over a^2} 
\delta_{\mathbf k} \; = \; 4 \pi G_N <\rho> \delta_{\mathbf k},
\label{evdeltak}
\eeq
where $v_s$ is the sound speed: $v_s^2 = dp / d \rho$. If the wave number 
$k$ is larger than the characteristic Jeans value
\beq
k_J \; = \; \sqrt{ {4 \pi G_N a^2 <\rho> \over v_s^2}},
\label{Jeans}
\eeq
the density perturbation $\delta_{\mathbf k}$ oscillates, whereas it 
grows if $k < k_J$. Cold dark matter effectively provides $v_s \to 0$, in 
which case $k_J \to \infty$ and perturbations with all wave numbers grow.

In order to describe the evolution of large-scale perturbations with 
$\lambda / a_H > 1$, we use the gauge-invariant ratio $\delta \rho / \rho 
+ p$, which remains constant outside the horizon $a_H$. Hence, the value 
when such a density perturbation comes back within the horizon is 
identical with its value when it was inflated beyond the horizon. During 
inflation, one had $\rho + p \simeq < {\dot \phi}^{2} >$, and
\beq
\delta \rho \; = \; \delta \phi \times {\partial V \over \partial \phi} \; 
= \; \delta \phi \times V^\prime (\phi).
\label{deltas}
\eeq
During roll-over, one has ${\ddot \phi} + 3 H {\dot \phi} + V^\prime 
(\phi) 
= 0$, and, if the roll-over is slow, one has
\beq
{\dot \phi} \; \simeq \; - {V^\prime (\phi) \over 3 H},
\label{phidot}
\eeq
where the Hubble expansion rate
\beq
H^2 \; = \; {8 \pi \over 3} G_N \left( V(\phi) + {1\over2} {\dot \phi}^{2} 
\right) \; \simeq \; {8 \pi G_N \over 3} V(\phi).
\label{H2}
\eeq
The quantum fluctuations of the inflaton field in de Sitter space are 
given by:
\beq
\delta \phi \; \simeq \; {H \over 2 \pi},
\label{deSitterphi}
\eeq
so initially
\beq
{\delta \rho \over \rho + p} \; \simeq \; {\delta \phi \cdot V^\prime 
\over \phi^{.2} } \; \simeq \; { H^3 \cdot V^\prime \over (V^\prime)^2} \; 
\simeq \; {V^{3/2} \over V^\prime}.
\label{invariant}
\eeq
This is therefore also the value when the perturbation comes back within 
the horizon:
\beq
\left( {\delta \rho \over \rho} \right)_{\lambda = a_H} \; \equiv \; A_S 
(\phi) \; = \; { \sqrt{2} \kappa^2 \over 8 \pi^{3/2} } { H^2 \over | 
H^\prime |}: \; \; \kappa^2 \; \equiv \; 8 \pi G_N,
\label{AS}
\eeq
assuming that $\rho \gg p$ at this epoch.

Gravity-wave perturbations obey an equation analogous to (\ref{evdeltak}):
\beq
{\ddot {h}}^{1,2}_{\mathbf k} + 2 H {\dot h}^{1,2}_{\mathbf k} + v_s^2 
{k^2 
\over 
a^2} h^{1,2}_{\mathbf k} \; = \; 0,
\label{evdeltah}
\eeq
for each of the two graviton polarization states $h^{1,2}_{\mu\nu}$, where
\beq
g_{\mu\nu} \; = \; g_{\mu\nu}^{FRW} + h_{\mu\nu}.
\label{nonFRW}
\eeq
The $h^{1,2}_{\mathbf k}$ also remain unchanged outside the horizon $a_H$, 
and have initial values
\beq
h^{1,2}_{\mathbf k} \; \simeq \; {H \over 2 \pi},
\label{h12I}
\eeq
yielding
\beq
{ k^3 |h_{\mathbf k}|_{\lambda = a_H} \; \equiv \; A_T (\phi) \; = \; 
{\kappa \over 4 \pi^{3/2}} H: \; \; H \; \simeq \; \sqrt{{8 \pi G_N \over 
3} V}}.
\label{AT}
\eeq
Comparing (\ref{AS}, \ref{AT}), we see that
\beq
r \; \equiv \; {A_S (\phi) \over A_T (\phi)} \; = \; {\kappa \over 
\sqrt{2}} {H \over |H^\prime|}: \; \; \kappa^2 \; = \; 8 \pi G_N.
\label{defr}
\eeq
Hence, if the roll-over is very slow, so that $|H^\prime|$ is very small, 
the density waves dominate over the tensor gravity waves. However, in the 
real world, also the gravity waves may be observable, furnishing a 
possible signature of inflation~\cite{Kinney}.

\subsection{Inflation in Scalar Field Theories}

Let now consider in more detail chaotic inflation in a generic scalar 
field theory~\cite{Kinney}, described by a Lagrangian
\beq
{\cal L}(\phi) \; = \; {1 \over 2} \partial^\mu \phi \partial_\mu \phi 
\; - \; V(\phi),
\label{generic}
\eeq
where the first term yields the kinetic energy of the inflaton field 
$\phi$ and the second term is the inflaton potential. One may treat the 
inflaton field as a fluid with density
\beq
\rho \; = \; {1 \over 2} {\dot \phi}^2 \; + \; V(\phi),
\label{density}
\eeq
and pressure
\beq
p =  \; {1 \over 2} {\dot \phi}^2 \; - \; V(\phi).
\label{pressure}
\eeq
Inserting these expressions into the standard FRW equations, we find that 
the Hubble expansion rate is given by
\beq
H^2 \; = \; {8 \pi \over 3 \pi^2} \left[ {1 \over 2} {\dot \phi}^2 + 
V(\phi) \right],
\label{Hubble}
\eeq
as discussed above, the deceleration rate is given by
\beq
\left( {{\ddot a} \over a} \right) \; = \; {8 \pi \over 3 \pi^2} \left[ 
V(\phi) - {1 \over 2} {\dot \phi}^2 \right],
\label{decel}
\eeq
and the equation of motion of the inflaton field is
\beq
{\ddot {\phi}} \; + \; 3 H {\dot \phi} \; + \; V^\prime(\phi) \; = \; 0.
\label{motion}
\eeq
The first term in (\ref{motion}) is assumed to be negligible, in which 
case the equation of motion is dominated by the second (Hubble drag) term, 
and one has
\beq
{\dot \phi} \; \simeq \; - {V^\prime \over 3 H},
\label{roll}
\eeq
as assumed above.
In this slow-roll approximation, when the kinetic term in (\ref{Hubble}) 
is 
negligible, and the Hubble expansion rate is dominated by the potential 
term:
\beq
H \; \simeq \; \sqrt{ {1 \over 3 M_P^2} V(\phi) }.
\label{slowroll}
\eeq
where $M_P \equiv 1/\sqrt{8 \pi G_N} \simeq 2.4 \times 10^{18}$~GeV. It is 
convenient to introduce the following slow-roll parameters:
\beq
\epsilon \; \equiv \; {1 \over 2} M_P^2 \left( {V^\prime \over V} 
\right)^2, \; \eta \; \equiv M_P^2 \left( {V^{\prime\prime} \over 
V} \right), \; \xi \; \equiv \;  M_P^4 \left( {V V^{\prime\prime\prime} 
\over V^2} \right).
\label{epseta}
\eeq
Various observable quantities can then be expressed in terms of 
$\epsilon, \eta$ 
and $\xi$, including the spectral index for scalar density perturbations:
\beq
n_s \; = \; 1 \; - \; 6 \epsilon \; + \; 2 \eta,
\label{index}
\eeq
the ratio of scalar and tensor perturbations at the quadrupole scale:
\beq
r \; \equiv {A_T \over A_S} = 16 \epsilon,
\label{scalartensor}
\eeq
the spectral index of the tensor perturbations:
\beq
n_T \; = \; -2 \epsilon,
\label{tensoridex}
\eeq
and the running parameter for the scalar spectral index:
\beq
{d n_s \over d {\rm ln} k} \; = \; {2 \over 3} \left[ \left( n_s - 1 
\right)^2 -
4 \eta^2 \right] + 2 \xi.
\label{running}
\eeq
The amount $e^N$ by which the Universe expanded during inflation is also 
controlled by the slow-roll parameter $\epsilon$:
\beq
e^N : \; N \; = \; \int H dt \; = \; {2 \sqrt{\pi} \over m_P } 
\int^{\phi_{final}}_{\phi_{initial}} {d \phi \over \sqrt{\epsilon (\phi)} 
}.
\label{expansion}
\eeq
In order to explain the size of a feature in the observed Universe, one 
needs:
\beq
N \; = \; 62 - {\rm ln} {k \over a_0 H_0} - {\rm ln} {10^{16} {\rm GeV} 
\over V_k^{1/4}} + {1 \over 4} {\rm ln} {V_k \over V_e} - {1 \over 3} 
{\rm ln} {V_e^{1/4} \over \rho_{reheating}^{1/4}},
\label{amount}
\eeq
where $k$ characterizes the size of the feature, $V_k$ is the magnitude of 
the inflaton potential when the feature left the horizon, $V_e$ is the 
magnitude of the inflaton potential at the end of inflation, and 
$\rho_{reheating}$ is the density of the Universe immediately following 
reheating after inflation.

As an example of the above general slow-roll theory, let us consider
chaotic inflation~\cite{chaos} with a $V = {1 \over 2} m^2
\phi^2$ potential~\footnote{This is motivated by the sneutrino inflation 
model~\cite{ERY} discussed later.}, and
compare its predictions with the WMAP data~\cite{WMAPnu}. In this model,
the conventional slow-roll inflationary parameters are
\beq
\epsilon \; = \; {2 M_P^2 \over \phi_I^2}, \;
\eta \; = \; {2 M_P^2 \over \phi_I^2}, \;
\xi \; = \; 0,
\label{slowroll2}
\eeq
where $\phi_I$ denotes the {\it a priori} unknown inflaton field value
during inflation at a typical CMB scale $k$. The overall scale of the
inflationary potential is normalized by the WMAP data on density
fluctuations:
\beq
\Delta_R^2 = {V \over 24 \pi^2 M_P^2 \epsilon} = 2.95 \times 10^{-9} A 
~~~:~~~
A = 0.77 \pm 0.07,
\label{normn} 
\eeq
yielding
\beq
V^{1 \over 4} = M_P ^4\sqrt{\epsilon \times 24 \pi^2 \times 2.27 \times
10^{-9}} = 0.027 M_P \times \epsilon^{1 \over 4},
\label{WMAP}
\eeq
corresponding to
\beq
m^{1 \over 2} \phi_I \; = \; 0.038 \times M_P^{3 \over 2}
\label{onecombn}
\eeq
in any simple chaotic $\phi^2$ inflationary model.
The above expression (\ref{amount}) for the number of 
e-foldings after the
generation of the CMB density fluctuations observed by COBE 
could be as low as $N \simeq 50$ for a reheating temperature $T_{RH}$ as 
low as $10^6$~GeV. In the $\phi^2$ inflationary model, this value of $N$ 
would imply
\beq
N \; = {1 \over 4} {\phi^2_I \over M_P^2} \simeq \; 50,
\label{fixN}
\eeq
corresponding to
\beq
\phi^2_I \simeq 200 \times M_P^2.
\label{fixphi}
\eeq
Inserting this requirement into the WMAP normalization condition
(\ref{WMAP}), we find the following required mass for any quadratic
inflaton:
\beq
m \; \simeq 1.8 \times 10^{13}~{\rm GeV}.
\label{phimass}
\eeq
This is comfortably within the range of heavy
singlet (s)neutrino masses usually considered, namely $m_N \sim 10^{10}$
to $10^{15}$~GeV, motivating the sneutrino inflation model~\cite{ERY} 
discussed below.

Is this simple $\phi^2$ model compatible with the WMAP data? 
It predicts the following values for the
primary CMB observables~\cite{ERY}: the scalar spectral index
\beq
n_s \; = \; 1 - {8 M_P^2 \over \phi^2_I} \simeq 0.96,
\label{ns}
\eeq
the tensor-to scalar ratio
\beq
r \; = \; {32 M_P^2 \over \phi^2_I} \; \simeq \; 0.16,
\label{r}
\eeq
and the running parameter for the scalar spectral index:
\beq
{d n_s \over d {\rm ln} k} \; = \; 
{32 M_P^4 \over \phi^4_I}  \simeq  8 \times 
10^{-4}.
\label{values}
\eeq
The value of $n_s$ extracted from WMAP data depends whether, for example,
one combines them with other CMB and/or large-scale structure data.
However, the $\phi^2$ model value $n_s \simeq 0.96$ appears to
be compatible with the data at the 1-$\sigma$ level. The $\phi^2$
model value $r\simeq 0.16$ for the relative tensor strength is
also compatible with the WMAP data. In fact, we note that the favoured
individual values for $n_s, r$ and ${d n_s / d {\rm ln} k}$ reported in an
independent analysis~\cite{realbarger} {\it all coincide} with the
$\phi^2$ model values, within the latter's errors!

One of the most interesting features of the WMAP analysis is the
possibility that ${d n_s / d {\rm ln} k}$ might differ from zero. The
$\phi^2$ model value ${d n_s / d {\rm ln} k} \simeq 8 \times
10^{-4}$ derived above is negligible compared with the WMAP preferred
value and its uncertainties. However, ${d n_s / d {\rm ln} k} = 0$ appears
to be compatible with the WMAP analysis at the 2-$\sigma$ level or better,
so we do not regard this as a death-knell for the $\phi^2$ model.

\subsection{Could the Inflaton be a Sneutrino?}

This `old' idea~\cite{MY} has recently been resurrected~\cite{ERY}. We
recall that seesaw models~\cite{seesaw} of neutrino masses involve three
heavy singlet right-handed neutrinos weighing around $10^{10}$ to
$10^{15}$~GeV, which certainly includes the preferred inflaton mass found
above (\ref{phimass}).  Moreover, supersymmetry requires each of the heavy
neutrinos to be accompanied by scalar sneutrino partners. In addition,
singlet (s)neutrinos have no interactions with vector bosons, so their
effective potential may be as flat as one could wish.  Moreover,
supersymmetry safeguards the flatness of this potential against radiative
corrections. Thus, singlet sneutrinos have no problem in meeting the
slow-roll requirements of inflation.

\begin{figure}[t]
\centerline{\epsfxsize = 0.7\textwidth \epsffile{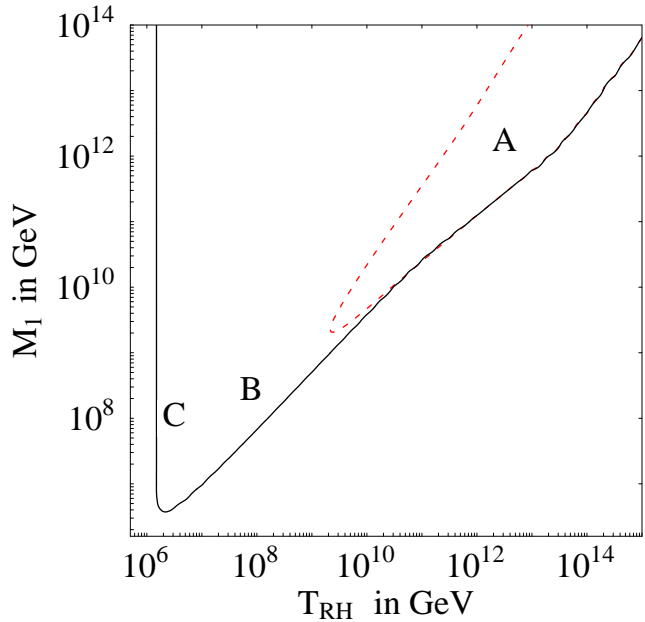}
}
\caption{
The solid curve bounds the region allowed for leptogenesis in the
$(T_{RH},\,M_{N_1})$ plane, assuming a baryon-to-entropy ratio $Y_B > 
7.8\times
10^{-11}$ and the maximal CP asymmetry $\epsilon_1^{max}(M_{N_1}).$ In the
area bounded by the red dashed curve leptogenesis is entirely 
thermal~\protect\cite{ERY}.
\vspace*{0.5cm}}
\label{fig:reheat}
\end{figure}

On the other hand, their Yukawa interactions $Y_D$ are eminently suitable
for converting the inflaton energy density into particles via $N \to H +
\ell$ decays and their supersymmetric variants. Since the magnitudes of
these Yukawa interactions are not completely determined, there is
flexibility in the reheating temperature after inflation, as we see in 
Fig.~\ref{fig:reheat}~\cite{ERY}. Thus the answer to the question in the 
title of this
Section seems to be `yes', so far.

\section{Further Beyond}

Some key cosmological and astrophysical problems may be resolved only by
appeal to particle physics beyond the ideas we have discussed so far. One
of the greatest successes of Big Bang cosmology has been an explanation of
the observed abundances of light elements, ascribed to cosmological
nucleosynthesis when the temperature $T \sim 1$ to $0.1$~MeV. This
requires a small baryon-to-entropy ratio $n_B / s \simeq 10^{-10}$. How
did this small baryon density originate?

Looking back to the previous quark epoch, there must have been a small
excess of quarks over antiquarks. All the antiquarks would then have
annihilated with quarks when the temperature of the Universe was $\sim
200$~MeV, producing radiation and leaving the small excess of quarks to
survive to form baryons. So how did the small excess of quarks originate?

Sakharov~\cite{Sakharov} pointed out that microphysics, in the form of
particle interactions, could generate a small excess of quarks if the
following three conditions were satisfied:

$\bullet$ {\it The interactions of matter and antimatter particles should
differ}, in the sense that both charge conjugation C and its combination
CP with mirror reflection should be broken, as discovered in the weak 
interactions.

$\bullet$ {\it There should exist interactions capable of changing the net
quark number}. Such interactions do exist in the \sm, mediated by unstable
field configurations called sphalerons. They have not been observed at low
temperatures, where they would be mediated by heavy states called
sphalerons and are expected to be very weak, but they are thought to have
been important when the temperature of the Universe was $\gappeq 100$~GeV.
Alternatively, one may appeal to interactions in Grand Unified Theories
(GUTs) that are thought to change quarks into leptons and {\it vice versa}
when their energies $\sim 10^{15}$~GeV.

$\bullet$ {\it There should have been a breakdown of thermal equilibrium}.  
This could have occurred during a phase transition in the early Universe,
for example during the electroweak phase transition when $T \sim 100$~GeV,
during inflation, or during a GUT phase transition when $T \sim
10^{15}$~GeV.

The great hope in the business of cosmological baryogenesis is to find a 
connection with physics accessible to accelerator experiments, and some 
examples will be mentioned later in this Lecture.

Another example of observable phenomena related to GUT physics may be
ultra-high-energy cosmic rays (UHECRs)~\cite{UHECR}, which have energies
$\gappeq 10^{11}$~GeV. The UHECRs might either have originated from some
astrophysical source, such as an active galactic nuclei (AGNs) or
gamma-ray bursters (GRBs), or they might be due to the decays of
metastable GUT-scale particles, a possibility discussed in the last part
of this Lecture.

\subsection{Grand Unified Theories}

The philosophy of grand unification~\cite{GG} is to seek a simple
group that includes the untidy separate interactions of the \sm, QCD and
the electroweak sector. The hope is that this Grand Unification can be
achieved while neglecting gravity, at least as a first approximation. If
the grand unification scale turns out to be significantly less than the
Planck mass, this is not obviously a false hope. The Grand Unification
scale is indeed expected to be exponentially large:
\beq
{m_{GUT}\over m_W} = \exp \left( {\cal
O}\left({1\over\alpha_{em}}\right)\right)
\label{fourone}
\eeq
and typical estimates are that $m_{GUT} = {\cal O}(10^{16}$ GeV). 
Such a calculation involves an extrapolation of known physics by many
orders of magnitude further than, e.g., the extrapolation that Newton
made from the apple to the Solar System.

If the grand unification scale is indeed so large, most tests of it are
likely to be indirect, such as relations between
\sm~ vector couplings and between particle masses. Any new interactions,
such as those that might cause protons to decay or give masses to
neutrinos, are likely to be very strongly suppressed.

To examine the indirect GUT predictions for the \sm~ 
vector interactions in more detail,
one needs to study their variations with the energy scale~\cite{GQW}, 
which are described by the following two-loop renormalization equations:
\beq
Q~~{\partial\alpha_i(Q)\over\partial Q} = -{1\over 2\pi}~~\left( b_i +
{b_{ij}\over 4\pi}~~\alpha_j(Q)\right)~~\left[\alpha_i(Q)\right]^2
\label{fourseven1}
\eeq
where the $b_i$ receive the one-loop contributions
\beq
b_i = \left(\matrix{ 0 \cr -{22\over 3} \cr -11}\right) + N_g
\left(\matrix{{4\over 3} \cr\cr {4\over 3} \cr\cr {4\over 3}}\right) + N_H
\left(\matrix{{1\over 10} \cr\cr {1\over 6} \cr \cr 0}\right)
\label{foureight}
\eeq
from vector bosons, $N_g$ matter generations and $N_H$ Higgs doublets,
respectively, and at two loops
\beq
b_{ij} = \left(\matrix{0&0&0\cr\cr 0&-{136\over 3} & 0 \cr\cr
0&0&-102}\right) + N_g
\left(\matrix{{19\over 15} & {3\over 5} & {44\over 15} \cr\cr {1\over 5} & {49\over 3} & 4
\cr\cr {4\over 30} & {3\over 2} & {76\over 3}}\right) +  N_H \left(
\matrix{{9\over 50} &
{9\over 10} & 0 \cr\cr {3\over 10} & {13\over 6} & 0 \cr\cr 0 & 0 & 
0}\right).
\label{fournine}
\eeq
These coefficients are all independent of any specific GUT model,
depending only on the light particles contributing to the
renormalization. 

Including supersymmetric particles as in the MSSM, one finds~\cite{DRW}
\beq
b_i = \left(\matrix{0 \cr\cr -6 \cr\cr -9}\right) + N_g
\left(\matrix{2\cr\cr 2 \cr\cr 2}\right) + N_H \left(\matrix{{3\over 10}
\cr\cr {1\over 2}\cr\cr 0}\right)
\label{fourten}
\eeq
and
\beq
b_{ij} = \left(\matrix{0&0&0\cr\cr 0&-24 & 0 \cr\cr 0&0&-54}\right) + N_g
\left(\matrix{{38\over 15} & {6\over 5} & {88\over 15} \cr\cr {2\over 5} & 14 & 8
\cr\cr {11\over 5} & 3 & {68\over 3}}\right) +  N_H \left( \matrix{{9\over 50} &
{9\over 10} & 0 \cr\cr {3\over 10} & {7\over 2} & 0 \cr\cr 0 & 0 & 
0}\right),
\label{foureleven}
\eeq
again independent of any specific supersymmetric GUT.

Calculations with these equations show that non-supersymmetric models are
not consistent with the measurements of the \sm~ interactions at LEP and
elsewhere.  However, although extrapolating the experimental
determinations of the interaction strengths using the non-supersymmetric
renormalization-group equations (\ref{foureight}), (\ref{fournine}) does
not lead to a common value at any renormalization scale, we saw in
Fig.~\ref{fig:GUT} that extrapolation using the supersymmetric equations
(\ref{fourten}), (\ref{foureleven}) {\bf does} lead to possible
unification at $m_{GUT} \sim 10^{16}$ GeV~\cite{GUTs}.

The simplest GUT model is based on the group SU(5)~\cite{GG}, whose 
most useful representations are the complex 
vector \underline{5} representation
denoted by $F_\alpha$, its conjugate $\underline{\bar 5}$ 
denoted by $\bar F^\alpha$, the
complex two-index antisymmetric tensor \underline{10} 
representation $T_{[\alpha\beta]}$,
and the adjoint \underline{24} representation 
$A^\alpha_\beta$. The latter is used to
accommodate the vector bosons of SU(5):
\beq
\left( \matrix{
&&& \vdots & \bar X \;\;\bar Y \cr
&g_{1,\ldots ,8} && \vdots & \bar X \;\;\bar Y \cr
&&& \vdots & \bar X \;\;\bar Y \cr
\multispan5 \dotfill \cr
X & X & X & \vdots & \cr
&&& \vdots& W_{1,2,3} \cr
Y & Y & Y & \vdots &}
\right)
\label{fourseventeen}
\eeq
where the $g_{1,\ldots,8}$ are the gluons of QCD, the $W_{1,2,3}$ are weak 
bosons, and the $(X,Y)$ are new vector bosons, whose interactions we
discuss in the next section.

The quarks and leptons of each generation are 
accommodated in $\underline{\bar 5}$ and 
$\underline{10}$ representations of SU(5):
\beq
\bar F = \left(\matrix{d^c_R\cr\cr d^c_Y \cr\cr d^c_B \cr \dotfill \cr -e^- \cr
\nu_e}\right)_L~,~~~T = 
\left(
\matrix{0  & u^c_B & -u^c_Y & \vdots & -u_R & -d_R \cr
-u^c_B & 0 & u^c_R & \vdots & -u_Y & -d_Y \cr
u^c_Y & -u^c_R & 0 & \vdots & -u_B & -d_B \cr 
\multispan6 \dotfill \cr
u_R & u_Y & u_B & \vdots & 0 & -e^c \cr
d_R & d_Y & d_B & \vdots & e^c & 0} \right)_L
\label{foureighteen}
\eeq
The particle assignments are unique up to the 
effects of mixing between generations, which we do
not discuss in detail here~\cite{SU5mix}.

\subsection{Baryon Decay and Baryogenesis}

Baryon instability is to be expected on general grounds, 
since there is no exact 
symmetry to guarantee that baryon number $B$ is conserved, just as we 
discussed previously for lepton number.
Indeed, baryon decay is a
generic prediction of GUTs, which we illustrate with the 
simplest SU(5) model.
We see in (\ref{fourseventeen}) that
there are two species of vector bosons in SU(5)
that couple the colour indices
(1,2,3) to the electroweak indices (4,5), 
called $X$ and $Y$. As we can see from the
matter representations (\ref{foureighteen}), 
these may enable two quarks or a quark and
lepton to annihilate.
Combining these possibilities leads to
interactions with $\Delta B  =
\Delta L = 1$. The forms of effective 
four-fermion interactions mediated by the exchanges of
massive $Z$ and $Y$ bosons, respectively, are~\cite{BEGN}:
\bea
&&\left(\epsilon_{ijk} u_{R_k} \gamma_\mu u_{L_j}\right)~~{g^2_X\over 8 m^2_X} ~~ \left(2 e_R
~\gamma^\mu ~d_{L_i} + e_L~\gamma^\mu~d_{R_i} \right)~, \nonumber \\
&&\left(\epsilon_{ijk} u_{R_k} \gamma_\mu d_{L_j}\right)~~{g^2_Y\over 8 m^2_X} ~~ \left(\nu_L
~\gamma^\mu ~d_{R_i}\right),
\label{fourtwentyfive}
\eea
up to generation mixing factors.

Since the couplings $g_X = g_Y$ 
in an SU(5) GUT, and $m_X \simeq m_Y$,
we expect that
\beq
G_X \equiv {g^2_X\over 8m^2_X}\simeq G_Y \equiv {g^2_Y\over 8m^2_Y}.
\label{fourtwentysix}
\eeq
It is clear from (\ref{fourtwentyfive}) that the baryon 
decay amplitude $A\propto G_X$, and
hence the baryon $B\rightarrow \ell +$ meson decay rate
\beq
\Gamma_B = c G^2_X m^5_p,
\label{fourtwentyseven}
\eeq
where the factor of $m^5_p$ comes from dimensional 
analysis, and $c$ is a coefficient that
depends on the GUT model and the non-perturbative 
properties of the baryon and meson.

The decay rate (\ref{fourtwentyseven}) corresponds to a proton lifetime
\beq
\tau_p = {1\over c} ~{m^4_X\over m^5_p}.
\label{fourtwentyeight}
\eeq
It is clear from (\ref{fourtwentyeight}) that 
the proton lifetime is very sensitive to
$m_X$, which must therefore be calculated very 
precisely. In minimal SU(5), the best
estimate was
\beq
m_X \simeq (1~{\rm to}~2) \times 10^{15} \times \Lambda_{QCD}
\label{fourtwentynine}
\eeq
where $\Lambda_{QCD}$ is the characteristic 
QCD scale.
Making an analysis of the generation mixing factors~\cite{SU5mix}, one
finds that the preferred proton (and bound neutron) decay
modes in minimal
SU(5) are
\begin{eqnarray}
&&p \rightarrow e^+\pi^0~,~~e^+\omega~,~~\bar\nu \pi^+~,
~~\mu^+K^0~,~~\ldots \nonumber \\
&& n \rightarrow e^+\pi^-~,~~ e^+\rho^-~,~~\bar\nu \pi^0~,~~\ldots
\label{fourthirty}
\end{eqnarray}
and the best numerical estimate of the lifetime is
\beq
\tau (p\rightarrow e^+\pi^0) \simeq 2\times 10^{31\pm 1} \times
\left({\Lambda_{QCD}\over 400~{\rm MeV}}\right)^4~~y
\label{fourthirtyone}
\eeq
This is in {\it prima facie} conflict with the latest experimental lower limit
\beq
\tau (p \rightarrow e^+\pi^0) > 1.6 \times 10^{33}~y
\label{fourthirtytwo}
\eeq
from super-Kamiokande~\cite{SKpdk}.

We saw earlier that supersymmetric GUTs, including SU(5), fare better with
coupling unification. They also predict a larger GUT scale~\cite{DRW}:
\beq
m_X \simeq 10^{16}~{\rm GeV},
\label{fourthirtythree}
\eeq
so that $\tau (p\rightarrow e^+\pi^0)$ is considerably 
longer than the experimental lower
limit. However, this is not the dominant proton 
decay mode in supersymmetric SU(5)~\cite{susySU5pdk}. In
this model, there are important $\Delta B = \Delta L = 1$ 
interactions mediated by the
exchange of colour-triplet Higgsinos $\tilde H_3$, 
dressed by gaugino exchange~\cite{WSY}:
\beq
G_X\rightarrow {\cal O}~\left({\lambda^2g^2\over 
16\pi^2}\right)~{1\over m_{\tilde
H_3}\tilde m}
\label{fourthirtyfour}
\eeq
where $\lambda$ is a Yukawa coupling. 
Taking into account colour factors and the increase
in $\lambda$ for more massive particles, 
it was found~\cite{susySU5pdk} that decays into neutrinos and
strange particles should dominate:
\beq
p\rightarrow \bar\nu K^+~,~~n\rightarrow\bar\nu K^0~,~~\ldots
\label{fourthirtyfive}
\eeq
Because there is only one factor of a heavy mass 
$m_{\tilde H_3}$ in the denominator of
(\ref{fourthirtyfour}), these decay modes are 
expected to dominate over $p\rightarrow
e^+\pi^0$, etc.,  in minimal supersymmetric SU(5). 
Calculating carefully the other factors
in (\ref{fourthirtyfour})~\cite{susySU5pdk}, it seems that the modes
(\ref{fourthirtyfive}) may now be close to exclusion at rates compatible 
with this model. The current experimental 
limit is $\tau(p\rightarrow \bar\nu
K^+) > 6.7 \times 10^{32} y$. However, there are other GUT 
models~\cite{fSU5} that remain compatible with the baryon decay limits.

The presence of baryon-number-violating interactions opens the way to 
cosmological baryogenesis via the out-of-equilibrium decays of GUT 
bosons~\cite{Yoshimura}:
\beq
X \; \to \; q + {\bar \ell} \; \; \; vs \; \; \; {\bar X} \; \to {\bar q} 
+ \ell.
\label{GUTdecays}
\eeq
In the presence of C and CP violation, the branching ratios for $X \; \to 
\; q + {\bar \ell}$ and ${\bar X} \; \to {\bar q} + \ell$ may differ. Such 
a difference may in principle be generated by quantum (loop) corrections 
to the leading-order interactions of GUT bosons. This effect is 
too small in the minimal SU(5) GUT described above~\cite{EGNbnotu}, but 
could be larger in 
some more complicated GUT. One snag is that, with GUT bosons as heavy as 
suggested above, the CP-violating decay asymmetry may tend to get washed 
out by thermal effects. This difficulty may in principle be avoided by 
appealing to the decays of GUT Higgs bosons, which might weigh $\ll 
10^{15}$~GeV, though this possibility is not strongly motivated. 

Although neutrino masses might arise without a GUT framework, they appear 
very naturally in most GUTs, and this framework helps motivated the 
mass scale $\sim 10^{10}$ to $10^{15}$~GeV required for the heavy 
singlet neutrinos. Their decays provide an alternative mechanism for
generating the baryon asymmetry of the Universe, namely
leptogenesis~\cite{FY}. In the presence of C and CP violation, the
branching ratios for $N \to {\rm Higgs} + \ell$ may differ from that for
$N \to {\rm Higgs} + {\bar \ell}$, producing a net lepton asymmetry. The
likely masses for heavy singlet neutrinos could be significantly lower
than the GUT scale, so it may be easier to avoid thermal washout effects.  
However, you may ask what is the point of generating a lepton asymmetry,
since we want a quark asymmetry? The answer is provided by the weak
sphaleron interactions that are present in the \sm, and would have
converted part of the lepton asymmetry into the desired quark asymmetry. 
We now discuss how this scenario might have operated in the minimal seesaw 
model for neutrino masses discussed in Lecture 2.

\subsection{Leptogenesis in the Seesaw Model}

As mentioned in the second Lecture, the minimal seesaw neutrino model
contains 18 parameters~\cite{Casas}, of which only 9 are observable in
low-energy neutrino interactions: 3 light neutrino masses, 3 real mixing
angles $\theta_{12,23,31}$, the oscillation phase $\delta$ and the 2
Majorana phases $\phi_{1,2}$.

To see how the extra 9 parameters appear~\cite{EHLR}, we reconsider the
full lepton sector, assuming that we have diagonalized the charged-lepton 
mass
matrix:
\begin{equation}
\left( Y_\ell \right)_{ij} \; = \; Y^d_{\ell_i} \delta_{ij},
\end{equation}
as well as that of the heavy singlet neutrinos:
\begin{equation}
M_{ij} \; = M^d_i \delta_{ij}.
\label{diagM}
\end{equation}
We can then parametrize the neutrino Dirac coupling matrix $Y_\nu$ in
terms of its real and diagonal eigenvalues and unitary rotation matrices:
\begin{equation}
Y_\nu \; = \; Z^* Y^d_{\nu_k} X^\dagger,
\label{diagYnu}
\end{equation}
where $X$ has 3 mixing angles and one CP-violating phase, just like the
CKM matrix, and we can write $Z$ in the form
\begin{equation}
Z \; = \; P_1 {\bar Z} P_2,
\label{PZP}
\end{equation}
where ${\bar Z}$ also resembles the
CKM matrix, with  3 mixing angles and one CP-violating phase, and the
diagonal matrices $P_{1,2}$ each have two CP-violating phases:
\begin{equation}
P_{1,2} \; = \; {\rm Diag} \left( e^{i\theta_{1,3}}, e^{i\theta_{2,4}}, 1
\right).
\label{PP}
\end{equation}  
In this parametrization, we see explicitly that the neutrino
sector has 18 parameters~\cite{Casas}: the 3 heavy-neutrino mass 
eigenvalues
$M^d_i$, the 3 real eigenvalues of $Y^D_{\nu_i}$, the $6 = 3 + 3$ real
mixing angles in $X$ and ${\bar Z}$, and the $6 = 1 + 5$ CP-violating  
phases in $X$ and ${\bar Z}$~\cite{EHLR}.   

The total decay rate of a heavy neutrino $N_i$ may be written in the
form
\begin{equation}
\Gamma_i \; = \; {1 \over 8 \pi} \left( Y_\nu Y^\dagger_\nu \right)_{ii}
M_i.
\label{gammai}
\end{equation}
One-loop CP-violating diagrams involving the exchange of heavy
neutrino $N_j$ would generate an asymmetry in $N_i$ decay of the form: 
\begin{equation}
\epsilon_{ij} \; = \; {1 \over 8 \pi} {1 \over \left( Y_\nu Y^\dagger_\nu
\right)_{ii}} {\rm Im} \left( \left( Y_\nu Y^\dagger_\nu \right)_{ij}
\right)^2 f \left( {M_j \over M_i} \right),
\label{epsilon}
\end{equation}
where $f ( M_j / M_i )$ is a known kinematic function.

Thus we see that leptogenesis~\cite{FY} is proportional to the product
\begin{equation}
Y_\nu Y^\dagger_\nu \; = \; P_1^* {\bar Z}^* \left( Y^d_\nu \right)^2
{\bar Z}^T P_1,
\label{leptog}
\end{equation}  
which depends on 13 of the real parameters and 3 CP-violating phases.
As mentioned in Lecture 2, the extra seesaw parameters also contribute to
the renormalization of soft supersymmetry-breaking masses, in leading 
order via  the combination
\begin{equation}
Y_\nu^\dagger Y_\nu \; = \; X \left( Y^d_\nu \right)^2 X^\dagger,
\label{renn}
\end{equation}
which depends on just 1 CP-violating phase, with two
more phases appearing in higher orders, when one allows the heavy singlet
neutrinos to be non-degenerate~\cite{EHRS}.

In order to see how the low-energy sector is embedded in the full
parametrization of the seesaw model, and hence its (lack of) relation to 
leptogenesis~\cite{ER}, we first recall that the 3 phases in ${\tilde 
P}_2$ (\ref{MNSP}) become observable when one also considers high-energy   
quantities. Next, we introduce a complex orthogonal matrix
\begin{equation}
R \; \equiv \; \sqrt{M^d}^{-1} Y_\nu U \sqrt{M^d}^{-1} \left[ v \sin \beta
\right],
\label{defR}
\end{equation}
which has 3 real mixing angles and 3 phases: $R^T R = 1$. These 6
additional parameters may be used to characterize $Y_\nu$, by inverting
\begin{equation}
Y_\nu \; = \; {\sqrt{M^d} R \sqrt{M^d} U^\dagger \over \left[ v \sin \beta
\right]},
\label{invertY}
\end{equation}
giving us the grand total of $18 = 9 + 3 + 6$ parameters~\cite{EHLR}.
The leptogenesis observable (\ref{leptog}) may now be written in the form
\begin{equation}
Y_\nu Y^\dagger_\nu \; = \; { \sqrt{M^d} R {\cal M}^d_\nu R^\dagger
\sqrt{M^d} \over \left[ v^2 \sin^2 \beta \right]},
\label{newleptog}
\end{equation}
which depends on the 3 phases in $R$, but {\it not} the 3 low-energy
phases $\delta, \phi_{1,2}$, {\it nor} the 3 real MNS mixing
angles~\cite{EHLR}!

The basic reason for this is that one makes a unitary
sum over all the light lepton species in evaluating the
asymmetry $\epsilon_{ij}$. It is easy to derive a compact expression for
$\epsilon_{ij}$ in terms of the heavy neutrino masses and the complex
orthogonal matrix $R$:
\begin{equation}
\epsilon_{ij} \; = \; {1 \over 8 \pi} M_j f \left( {M_j \over M_i} \right)
{ {\rm Im} \left( \left( R {\cal M}_\nu^d R^\dagger \right)_{ij} \right)^2
\over \left( R {\cal M}_\nu^d R^\dagger \right)_{ii} },
\label{epsilonR}
\end{equation}
which depends explicitly on the extra phases in $R$. How can we measure
them?

\begin{figure}[htb]
\centerline{\epsfxsize = 0.5\textwidth \epsffile{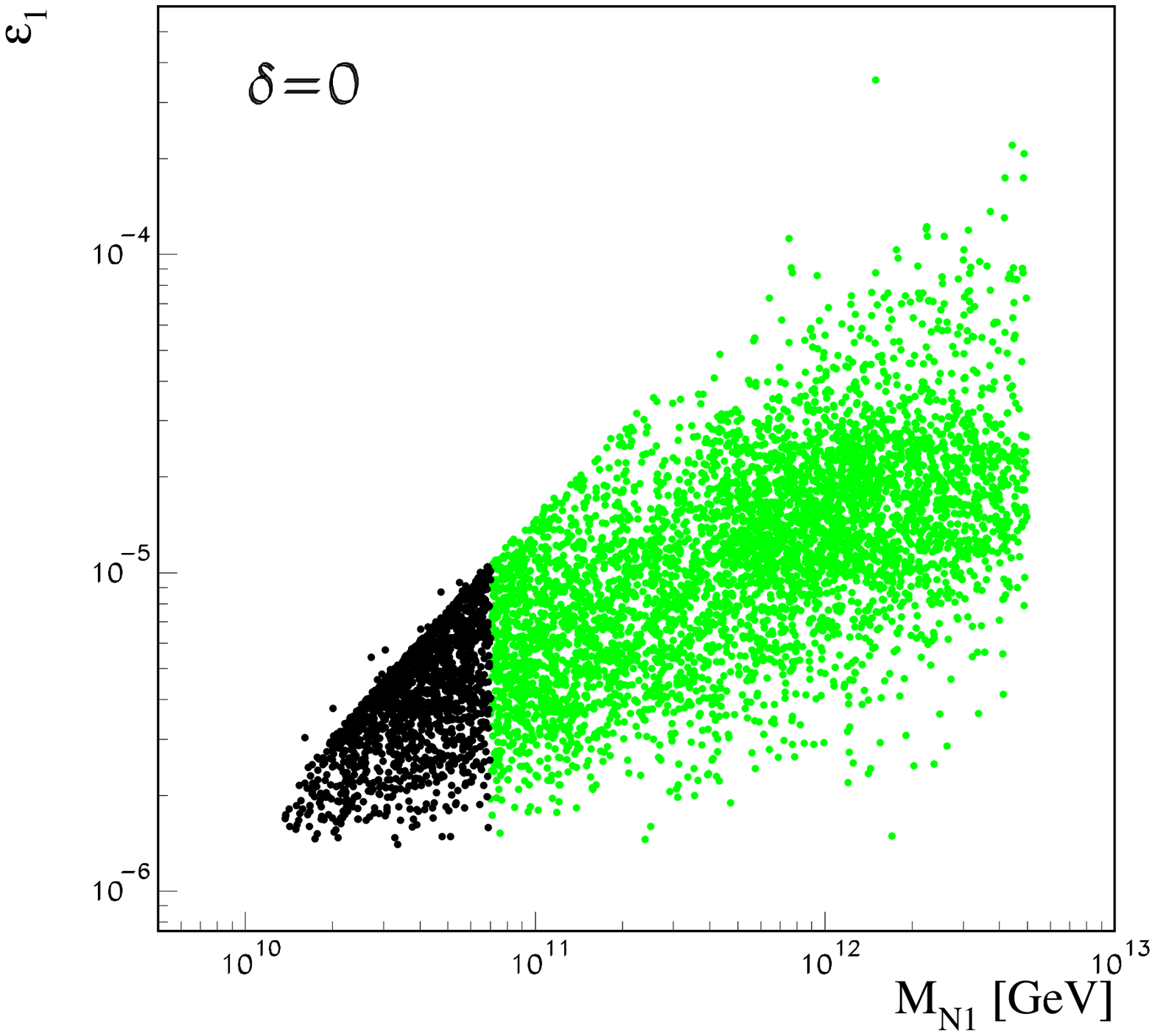}
\hfill \epsfxsize = 0.5\textwidth \epsffile{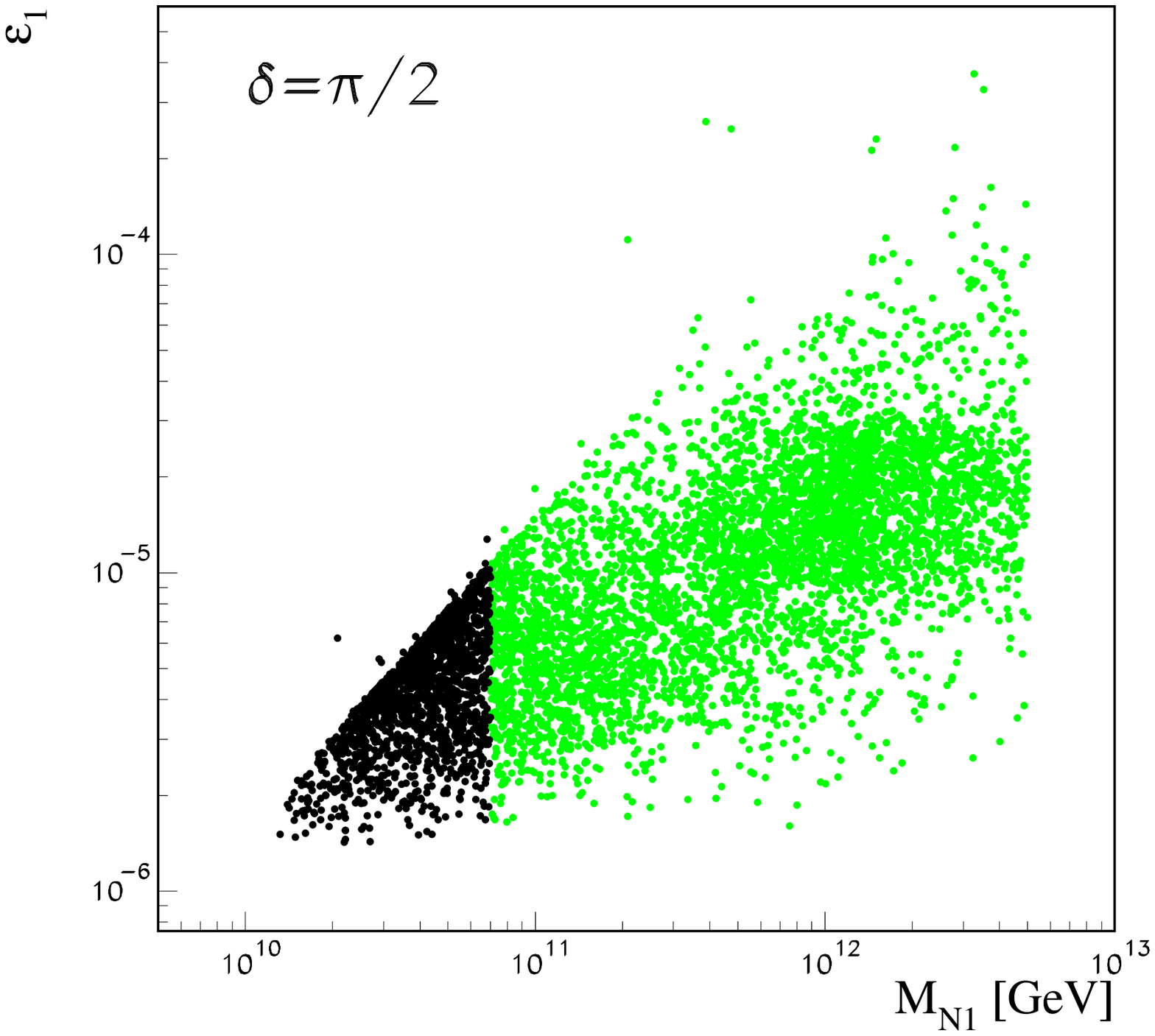}}
\caption{
Comparison of the CP-violating asymmetries in the decays of heavy singlet
neutrinos giving rise to the cosmological baryon asymmetry via
leptogenesis (left panel) without and (right panel) with maximal
CP violation in neutrino oscillations~\protect\cite{ER}. They are 
indistinguishable.}
\vspace*{0.5cm} 
\label{fig:nodelta}
\end{figure}

In general, one may formulate the following strategy for calculating
leptogenesis in terms of laboratory observables~\cite{EHLR,ER}:
\begin{itemize}
\item{Measure the neutrino oscillation phase $\delta$ and the Majorana 
phases $\phi_{1,2}$,}
\item{Measure observables related to the renormalization of soft
supersymmetry-breaking parameters, that are functions of $\delta,
\phi_{1,2}$ and the leptogenesis phases,} 
\item{Extract the effects of the known values of $\delta$ and
$\phi_{1,2}$, and isolate the leptogenesis parameters.}
\end{itemize}
In the absence of complete information on the first two steps above, we
are currently at the stage of preliminary explorations of the
multi-dimensional parameter space. As seen in Fig.~\ref{fig:nodelta}, the
amount of the leptogenesis asymmetry is explicitly independent of
$\delta$~\cite{ER}. However, in order to make more definite 
predictions, one must make extra hypotheses.

\begin{figure}[t]
\centerline{\epsfxsize = 0.7\textwidth \epsffile{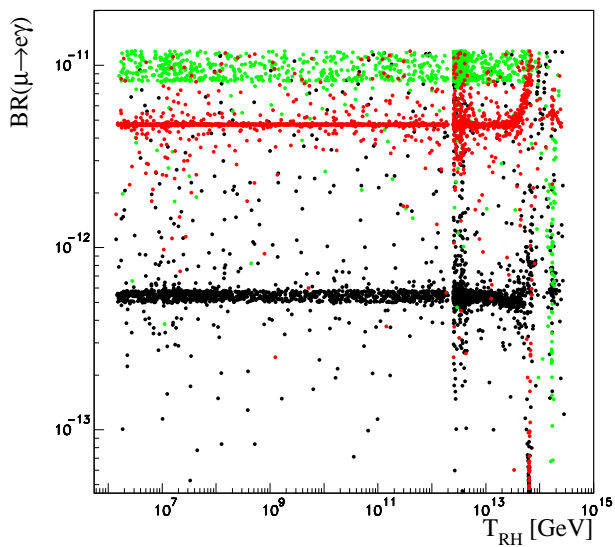}
}
\caption{
Calculations of BR$(\mu \to e \gamma)$ in the sneutrino inflation 
model. Black points correspond to
$\sin \theta_{13} = 0.0$,  $M_2 = 10^{14}$~GeV,
and $5 \times 10^{14}$~GeV $< M_3 < 5 \times 10^{15}$~GeV.
Red points correspond to $\sin \theta_{13} = 0.0$,  $M_2 = 5 \times 
10^{14}$~GeV and  $M_3 = 5 \times 10^{15}$~GeV, while green points 
correspond to $\sin \theta_{13} = 0.1$,
$M_2 = 10^{14}$~GeV and $M_3  = 5 \times 10^{14}$~GeV~\protect\cite{ERY}. 
We 
assume for illustration that $(m_{1/2}, m_0) = (800, 170)$~GeV and $\tan 
\beta = 10$.}
\vspace*{0.5cm}
\label{fig:ERY}
\end{figure}

One possibility is that the inflaton might be a heavy singlet sneutrino,
as discussed in the previous Lecture~\cite{ERY}. As shown there, this
hypothesis would require a mass $\simeq 1.8 \times 10^{13}$~GeV for the
lightest sneutrino, which is well within the range favoured by seesaw
models. As also discussed in the previous Lecture, this sneutrino inflaton
model predicts values of the spectral index of scalar perturbations, the
fraction of tensor perturbations and other CMB observables that are
consistent with the WMAP data. The sneutrino inflaton model is quite
compatible with a low reheating temperature, as seen in
Fig.~\ref{fig:reheat}. Moreover, because of this and the other constraints
on the seesaw model parameters in this model, it makes predictions for the
branching ratio for $\mu \to e \gamma$ that are more precise than in the
generic seesaw model. As seen in Fig.~\ref{fig:ERY}, it predicts that this
decay should appear within a couple of orders of magnitude of the present
experimental upper limit~\cite{ERY}.

\subsection{Ultra-High-Energy Cosmic Rays}

\begin{figure}
\centerline{\includegraphics[height=3in]{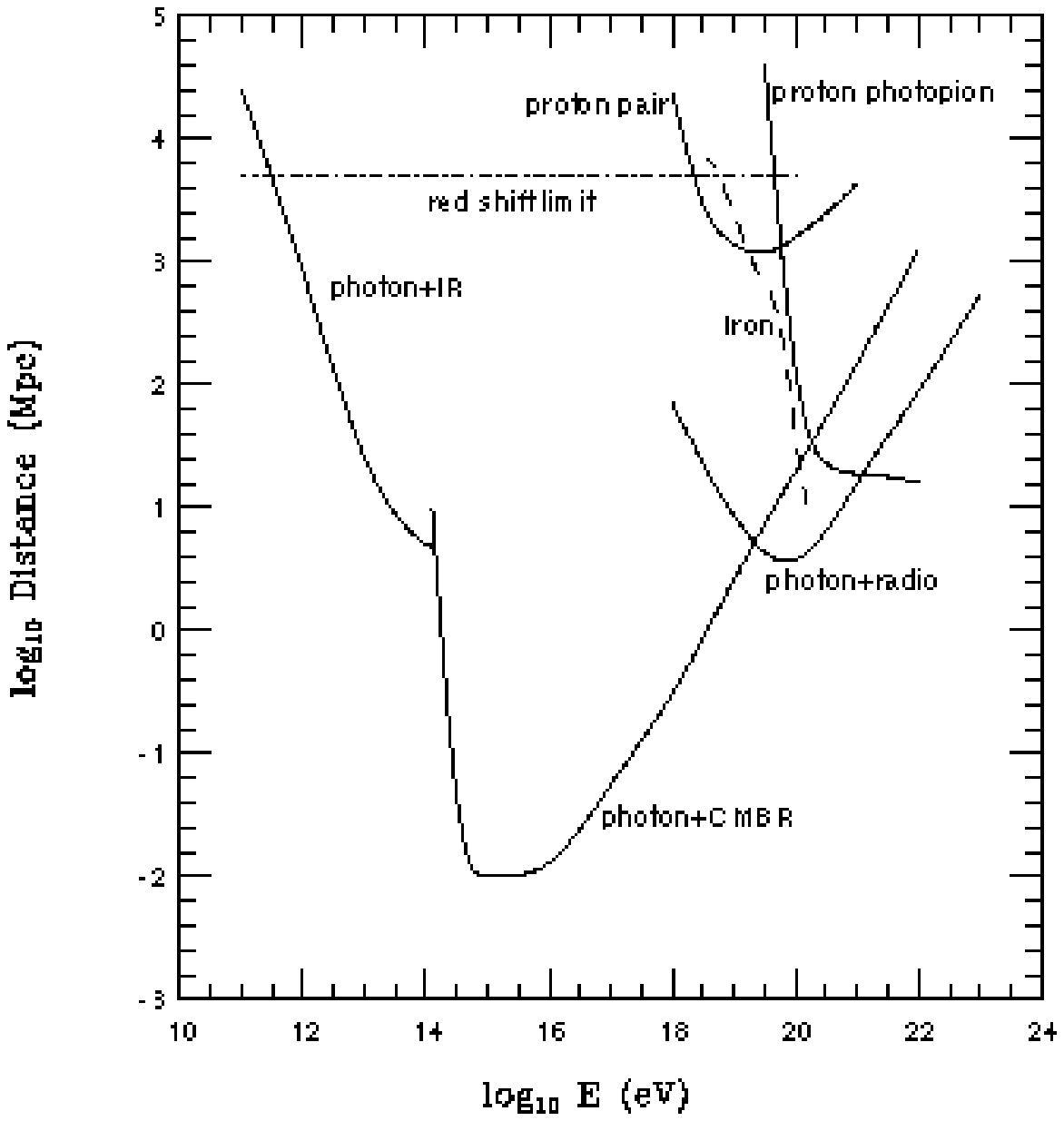}}
\caption[]{Energetic particles propagating through the Universe scatter 
on relic photons, imposing a cutoff on the maximum distance over which 
they can propagate~\protect\cite{GKZ}.}
\label{fig:GKZ}
\end{figure}

The flux of cosmic rays falls approximately as $E^{-3}$ from $E \sim$ 1
GeV, through $E\sim 10^6$ GeV where there is a small change in slope
called the `knee', continuing to about $10^{10}$ GeV, the `ankle'. Beyond
about $5\times 10^{10}$ GeV, as seen in Fig.~\ref{fig:GKZ}, one expects a
cutoff~\cite{GKZ} due to the photopion reaction $p +
\gamma_{CMB}\rightarrow\Delta^+\rightarrow p + \bar \pi^0, n + \pi^+$, for
all primary cosmic rays that originate from more than about 50~Mpc away.
However, some experiments report cosmic-ray events with higher energies of
$10^{11}$ GeV or more~\cite{UHECR}. If this excess flux beyond the GKZ
cutoff is confirmed, conventional physics would require it to originate
from distances $\lappeq$ 100 Mpc, in which case one would expect to see
some discrete sources. Analogous cutoffs are expected for primary
cosmic-ray photons or nuclei, as also seen in Fig.~\ref{fig:GKZ}.

There are two general categories of sources considered for such
ultra-high-energy cosmic rays (UHECRs): {\it bottom-up} and {\it top-down}
scenarios~\cite{UHECR}.

Astrophysical sources capable of accelerating high-energy cosmic rays 
in some bottom-up scenario must be larger than the
gyromagnetic radius $R$ corresponding to their internal magnetic field 
$B$:
\beq
R \sim \left({100\over Z}\right)~~\left({E\over 10^{11}{\rm
GeV}}\right)~~\left({\mu G\over B}\right) ~~{\rm kpc},
\label{six}
\eeq
where $Z$ is the atomic number of the cosmic ray particle.
Candidate astrophysical sources include gamma-ray busters (GRBs) and
active galactic nuclei (AGNs).

If UHECRs are produced by such localized sources, one would expect to see
a clustering in their arrival directions. Such clustering has been claimed
in both the AGASA and Yakutsk data~\cite{Igor}, but I personally do not
find the evidence overwhelming. A correlation has also been claimed with
BL Lac objects~\cite{Igor2}, which are AGNs emitting relativistic jets
pointing towards us, but this is also a claim that I should like to see
confirmed by more data, as will be provided soon by the HiRes and Auger
experiments.

Favoured top-down scenarios involve physics at the GUT scale $\gappeq
10^{15}$ GeV that produces UHECRs with energies $\sim 10^{12}$ GeV via
some `trickle-down' mechanism. Suggestions have included topological
structures, such as cosmic strings, that are present in some GUTs and
would radiate energetic particles, and the decays of metastable superheavy
relic particles.

In the latter case, one would expect most of the observed UHECRs to come
from the decays of relics in our own galactic halo. In this case, one
would expect the UHECRs to exhibit an anisotropy correlated with the
orientation of the galaxy~\cite{aniso}. The present data are insufficient
to confirm or exclude an isotropy of the magnitude predicted in different
halo models, but the Auger experiment should be able to decide the issue.
One might naively expect that superheavy relic particles would be spread
smoothly through the halo, and hence that they would not cause clustering
in the UHECRs.  However, this is not necessarily the case, as many cold
dark matter models predict clumps within the halo~\cite{Navarro}, which
could contribute a clustered component on top of an apparently smooth
background.

How might suitable metastable superheavy relic particles
arise~\cite{cryptons}? The proton is a prototype for a metastable
particle. As discussed earlier in this Lecture, we know that its lifetime
must exceed about $10^{33}$~y or so, much longer than if it decayed via
conventional weak interactions. On the other hand, there is no known exact
symmetry principle capable of preventing the proton from decaying.
Therefore, we believe that it is only metastable, decaying very slowly via
some higher-dimensional non-renormalizable interaction that violates
baryon number. For example, as we saw earlier, in many GUT models there is
a dimension-6 $qqq\ell$ interaction with a coefficient $\propto 1 /M^2$,
where $M$ is some superheavy mass scale. This would yield a decay
amplitude $A\sim 1/M^2$, and hence a long lifetime $\tau \sim {M^4\over
m^5_p}$.

We must work harder in the case of a superheavy relic weighing $\gappeq
10^{12}$ GeV, but the principle is the same. For an interaction of dimension
$4 + n$, we expect
\beq
\tau\sim {M^{2n}\over m^{2n+1}_{\rm relic}}
\label{eight}
\eeq
This could yield a lifetime greater than the age of the Universe, even for
$m_{\rm relic} \sim 10^{12}$ GeV, if $M$ and/or $n$ are large enough, for
example if $M \sim 10^{17}$ GeV and $n \geq 9$~\cite{BEN}.

Phenomenological constraints on such metastable relic particles were
considered some time ago for reasons other than explaining
UHECRs~\cite{EGLNS}. Constraints from the abundances of light elements,
from the CMB and from the high-energy $\nu$ flux have been considered.
They provide no obstacle to postulating a superheavy relic particle with
$\Omega h^2\sim 0.1$ if $\tau \gappeq 10^{15}$ y. Hence, metastable
superheavy relic particles could in principle constitute most of the cold
dark matter.

\begin{figure}
\centerline{\includegraphics[height=3in]{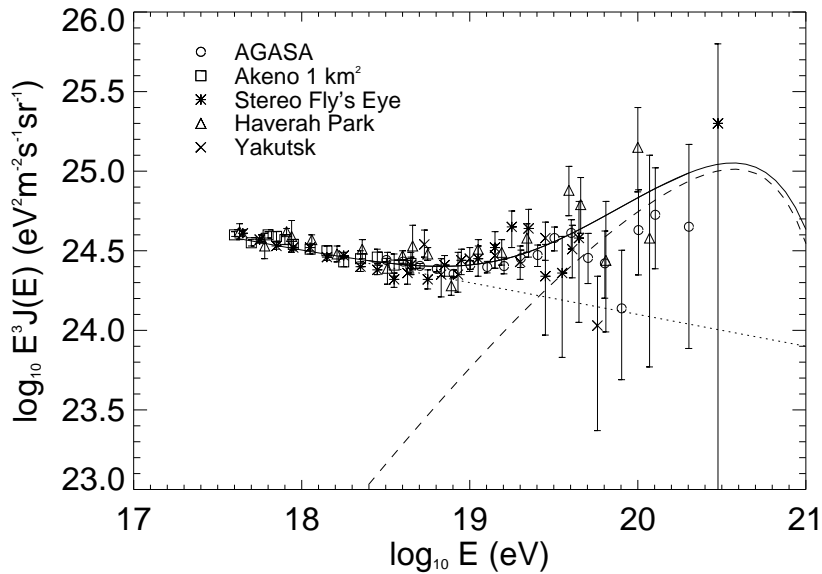}}
\caption[]{The spectrum of UHECRs can be explained by the decays of 
superheavy metastable particles such as cryptons~\protect\cite{Sarkar}.}
\label{fig:Subir}
\end{figure}

Possible theoretical candidates within a general framework of string
and/or M theory have been considered~\cite{cryptons,BEN}. These models
have the generic feature that, in addition to the interactions of the
Standard Model, there are others that act on a different set of `hidden'
matter particles, which communicate with the Standard Model only via
higher-order interactions scaled by some inverse power of a large mass
scale $M$. Just as the strong nuclear interactions bind quarks to form
metastable massive particles, the protons, so some `hidden-sector'
interactions might become strong at some higher energy scale, and form
analogous, but supermassive, metastable particles. Just like the proton,
these massive `cryptons' generally decay through high-dimension
interactions into multiple quarks and leptons. The energetic quarks
hadronize via QCD in a way that can be modelled using information from
$Z^0$ decays at LEP. Several simulations have shown that the resulting
spectrum of UHECRs is compatible with the available data, whether
supersymmetry is included in the jet fragmentation process, or not, as
shown in Fig.~\ref{fig:Subir}~\cite{Sarkar}.

A crucial issue is whether there is a mechanism that might produce a relic
density of superheavy particles that is large enough to be of interest for
cosmology, without being excessive. As was discussed in Lecture~3, the
plausible upper limit on the mass of a relic particle that was initially
in thermal equilibrium is of the order of a TeV. However, equilibrium
might have been violated in the early Universe, around the epoch of
inflation, and various non-thermal production mechanisms have been
proposed~\cite{Chung}. These include out-of-equilibrium processes at the
end of the inflationary epoch, such as parametric resonance effects, and
gravitational production as the scale factor of the Universe changes
rapidly. It is certainly possible that superheavy relic particles might be
produced with a significant fraction of the critical density.

We have seen that UHECRs could perhaps be due to the decays of metastable
superheavy relic particles. They might have the appropriate abundance,
their lifetimes might be long on a cosmological time-scale, and the decay
spectrum might be compatible with the events seen. Pressure points on this
interpretation of UHECRs include the composition of the UHECRs -- there
should be photons and possibly neutrinos, as well as protons, and no
heavier nuclei; their isotropy -- UHECRs from relic decays would exhibit a
detectable galactic anisotropy; and clustering -- this would certainly be
expected in astrophysical source models, but is not excluded in the
superheavy relic interpretation. 

The Auger project currently under construction in Argentina should provide
much greater statistics on UHECRs and be able to address many of these
issues~\cite{Auger}. In the longer term, the EUSO project now being
considered by ESA for installation on the International Space Station
would provide even greater sensitivity to UHECRs~\cite{EUSO}. Thus an
experimental programme exists in outline that is capable of clarifying
their nature and origin, telling us whether they are indeed due to new
fundamental physics.

\subsection{Summary}

We have seen in these lectures that the \sm~ must underlie any description
of the physics of the early Universe. Its extensions may provide the
answers to many of the outstanding issues in cosmology, such as the nature
of dark matter, the origin of the matter in the Universe, the size and age
of the Universe, and the origins of the structures within it. Theories
capable of resolving these issues abound, and include many new options not
stressed in these lectures. Continued progress in understanding these
issues will involve a complex interplay between particle physics and
cosmology, involving experiments at new accelerators such as the LHC, as
well as new observations.


\begin{thebibliography}{99}

\bibitem{CMB}
C. Lineweaver, Lectures at this School.

\bibitem{BBN}
For a recent review, see:
K.~A.~Olive, arXiv:astro-ph/0202486.

\bibitem{KT}
E.~W.~Kolb and M.~S.~Turner, {\it The Early Universe}
(Addison-Wesley, Redwood City, USA, 1990).

\bibitem{LHC}
See the LHC home page: \\
{\tt http://lhc-new-homepage.web.cern.ch/lhc-new-homepage/}.

\bibitem{SM}
S.L. Glashow, {\it Nucl. Phys.} {\bf 22}, 579 (1961);
S. Weinberg, {\it Phys. Rev. Lett.} {\bf 19}, 1264 (1967);
A. Salam, Proc. 8th Nobel Symposium, Stockholm 1968, ed. N. Svartholm 
(Almqvist and Wiksells, Stockholm, 1968), p. 367.

\bibitem{LEPEWWG}
LEP Electroweak Working Group, \\
{\tt http://lepewwg.web.cern.ch/LEPEWWG/Welcome.html}.

\bibitem{BIM}
C. Bouchiat, J. Iliopoulos and Ph. Meyer, {\it Phys. Lett.} B {\bf
138}, 652 (1972) and references therein.

\bibitem{SSB}
C. Quigg, {\it Gauge Theories of the Strong, Weak and
Electromagnetic
Interactions} (Benjamin-Cummings, Reading, 1983).

\bibitem{LEPE}
D.~Brandt, H.~Burkhardt, M.~Lamont, S.~Myers and J.~Wenninger,
{\it Rept. Prog. Phys.} {\bf 63}, 939 (2000).

\bibitem{Vt}
M. Veltman, {\it Nucl. Phys.} B {\bf 123}, 89 (1977);
M.S. Chanowitz, M. Furman and I. Hinchliffe, {\it Phys. Lett.} B {\bf 78}, 
285 (1978).

\bibitem{VH}
M. Veltman, {\it Acta Phys.Pol.} {\bf 8}, 475 (1977).

\bibitem{EGNH}
J.~R.~Ellis, M.~K.~Gaillard and D.~V.~Nanopoulos,
{\it Nucl. Phys.} B {\bf 106}, 292 (1976).

\bibitem{LEPHWG}
LEP Higgs Working Group for Higgs boson searches, OPAL Collaboration,
ALEPH Collaboration, DELPHI Collaboration and L3
Collaboration,
{\it Search for the Standard Model Higgs Boson at LEP},
CERN-EP/2003-011.

\bibitem{StAnd}
J.~R.~Ellis, Lectures at 1998 CERN Summer School, St. Andrews, {\it Beyond 
the Standard Model for Hillwalkers},
arXiv:hep-ph/9812235.

\bibitem{CHschool}
J.~R.~Ellis, Lectures at 2001 CERN Summer School, Beatenberg,
{\it Supersymmetry for Alp hikers},
arXiv:hep-ph/0203114.

\bibitem{TOE}
J. Scherk and J. H. Schwarz, {\it Nucl. Phys.} {\bf B81}, 118 (1974);
M.~B.~Green and J.~H. Schwarz, {\it Phys. Lett.} {\bf 149B}, 117 (1984)
and {\bf 151B}, 21 (1985);
J.~R.~Ellis,
{\it The Superstring: Theory Of Everything, Or Of Nothing?},
Nature {\bf 323}, 595 (1986).

\bibitem{PDG}
K.~Hagiwara {\it et al.}  [Particle Data Group Collaboration],
Phys.\ Rev.\ D {\bf 66}, 010001 (2002).

\bibitem{KATRIN}
A.~Osipowicz {\it et al.}  [KATRIN Collaboration],
arXiv:hep-ex/0109033.

\bibitem{2dF}
O.~Elgaroy {\it et al.},
{\it Phys. Rev. Lett.} {\bf 89}, 061301 (2002)
[arXiv:astro-ph/0204152].

\bibitem{WMAPnu}
C.~L.~Bennett {\it et al.},
arXiv:astro-ph/0302207;
D.~N.~Spergel {\it et al.},
arXiv:astro-ph/0302209;
H.~V.~Peiris {\it et al.},
arXiv:astro-ph/0302225.

\bibitem{KKGH}
H.~V.~Klapdor-Kleingrothaus {\it et al.}, {\it Eur.\ Phys.\ J.} 
A {\bf 12}, 147 (2001) [arXiv:hep-ph/0103062]; see, however,
H.~V.~Klapdor-Kleingrothaus {\it et al.},
{\it Mod.\ Phys.\ Lett.} A {\bf 16}, 2409 (2002)
[arXiv:hep-ph/0201231].

\bibitem{SK}
Y.~Fukuda {\it et al.}  [Super-Kamiokande Collaboration],
{\it Phys. Rev. Lett.}  {\bf 81}, 1562 (1998)
[arXiv:hep-ex/9807003].

\bibitem{SNO}
Q.~R.~Ahmad {\it et al.}  [SNO Collaboration],
{\it Phys. Rev. Lett.}  {\bf 89}, 011301 (2002) 
[arXiv:nucl-ex/0204008];
{\it Phys. Rev. Lett.}  {\bf 89}, 011302 (2002) 
[arXiv:nucl-ex/0204009].

\bibitem{BEG}
R.~Barbieri, J.~R.~Ellis and M.~K.~Gaillard,
{\it Phys. Lett.} B {\bf 90}, 249 (1980).

\bibitem{seesaw}
M. Gell-Mann, P. Ramond and R. Slansky, Proceedings of
the Supergravity Stony Brook Workshop, New York, 1979, eds. P. Van
Nieuwenhuizen and D. Freedman (North-Holland, Amsterdam);
T. Yanagida, Proceedings of
the  Workshop  on Unified  Theories  and  Baryon  Number in the
Universe,  Tsukuba,  Japan 1979 (edited by A.  Sawada and A.
Sugamoto, KEK Report No.  79-18, Tsukuba);
R.~Mohapatra and G.~Senjanovic,
{\it Phys. Rev. Lett.} {\bf 44}, 912 (1980).

\bibitem{Frampton}
P.~H.~Frampton, S.~L.~Glashow and T.~Yanagida,
arXiv:hep-ph/0208157.

\bibitem{Morozumi}
T.~Endoh, S.~Kaneko, S.~K.~Kang, T.~Morozumi and M.~Tanimoto,
arXiv:hep-ph/0209020.

\bibitem{fSU5}
J.~R.~Ellis, J.~S.~Hagelin, S.~Kelley and D.~V.~Nanopoulos,
{\it Nucl. Phys.} B {\bf 311}, 1 (1988).

\bibitem{EGLLN}
J.~R.~Ellis, M.~E.~G\'omez, G.~K.~Leontaris, S.~Lola and D.~V.~Nanopoulos,
{\it Eur. Phys. J.} C {\bf 14}, 319 (2000).

\bibitem{EGN}
J.~R.~Ellis, M.~K.~Gaillard and D.~V.~Nanopoulos,
{\it Nucl. Phys.} B {\bf 109}, 213 (1976).

\bibitem{KM}
M.~Kobayashi and T.~Maskawa,
{\it Prog. Theor. Phys.} {\bf 49}, 652 (1973).

\bibitem{MNS}
Z.~Maki, M.~Nakagawa and S.~Sakata,
{\it Prog. Theor. Phys.} {\bf 28}, 870 (1962).

\bibitem{K2K}
Y.~Oyama, arXiv:hep-ex/0210030.

\bibitem{SKsolar}
S.~Fukuda {\it et al.}  [Super-Kamiokande Collaboration],
{\it Phys. Lett.} B {\bf 539}, 179 (2002)
[arXiv:hep-ex/0205075].

\bibitem{KamLAND}
S.~A.~Dazeley  [KamLAND Collaboration], arXiv:hep-ex/0205041.

\bibitem{PV}
S.~Pakvasa and J.~W.~Valle, arXiv:hep-ph/0301061.

\bibitem{MS}
H.~Minakata and H.~Sugiyama, arXiv:hep-ph/0212240.

\bibitem{Chooz}
Chooz Collaboration,
{\it Phys. Lett.} B {\bf 420}, 397 (1998).

\bibitem{DGH}
A. De R\'ujula, M.B. Gavela and P. Hern\'andez,
{\it Nucl. Phys.} B {\bf 547}, 21 (1999) [arXive:hep-ph/9811390].

\bibitem{golden}
A.~Cervera {\it et al.},
{\it Nucl. Phys.} B {\bf 579}, 17 (2000)
[Erratum-ibid. B {\bf 593}, 731 (2001)].

\bibitem{SPL}
B.~Autin {\it et al.},
{\it Conceptual design of the SPL, a high-power superconducting H$^-$ 
linac at CERN}, CERN-2000-012.

\bibitem{betabeam}
P.~Zucchelli,
{\it Phys. Lett.} B {\bf 532}, 166 (2002).

\bibitem{nufact}
M. Apollonio {\it et al}., {\it Oscillation physics with a neutrino 
factory}, arXiv:hep-ph/0210192;
and references therein.

\bibitem{Casas}
J.~A.~Casas and A.~Ibarra,
{\it Nucl. Phys.} B {\bf 618}, 171 (2001) [arXiv:hep-ph/0103065].

\bibitem{EHLR}
J.~R.~Ellis, J.~Hisano, S.~Lola and M.~Raidal,
{\it Nucl. Phys.} B {\bf 621}, 208 (2002)
[arXiv:hep-ph/0109125].

\bibitem{DI}
S.~Davidson and A.~Ibarra, JHEP {\bf 0109}, 013 (2001).

\bibitem{EHRS}
J.~R.~Ellis, J.~Hisano, M.~Raidal and Y.~Shimizu,
{\it Phys. Lett.} B {\bf 528}, 86 (2002)
[arXiv:hep-ph/0111324].

\bibitem{EHRS2}
J.~R.~Ellis, J.~Hisano, M.~Raidal and Y.~Shimizu,
{\it Phys. Rev.} D {\bf 66}, 115013 (2002)
[arXiv:hep-ph/0206110].

\bibitem{FY}
M.~Fukugita and T.~Yanagida,
{\it Phys. Lett.} B {\bf 174}, 45 (1986).

\bibitem{ER}
J.~R.~Ellis and M.~Raidal,
{\it Nucl. Phys.} B {\bf 643}, 229 (2002)
[arXiv:hep-ph/0206174].   

\bibitem{hierarchy}
L.~Maiani, {\it Proceedings of the 1979 Gif-sur-Yvette Summer School On
Particle
Physics}, 1;
G.~'t Hooft, in {\it Recent Developments in Gauge Theories, Proceedings
of the Nato Advanced Study
Institute, Cargese, 1979}, eds. G.~'t Hooft {\it et al.}, (Plenum Press,
NY, 1980); E.~Witten,
{\it Phys. Lett.}  B {\bf 105}, 267 (1981).

\bibitem{noren}
S.~Ferrara, J.~Wess and B.~Zumino,
{\it Phys. Lett.} B {\bf 51}, 239 (1974);
S.~Ferrara, J.~Iliopoulos and B.~Zumino,
{\it Nucl. Phys.} B {\bf 77}, 413 (1974).

\bibitem{Fayet}
P.~Fayet, as reviewed in
{\it Supersymmetry, Particle Physics And Gravitation},
CERN-TH-2864,
published in {\it Proc. of Europhysics Study Conf. on Unification of
Fundamental Interactions}, Erice, Italy, Mar 17-24, 1980, eds. S.
Ferrara, J. Ellis, P. van Nieuwenhuizen (Plenum Press, 1980).

\bibitem{HLS}
R. Haag, J. Lopusz\'anski and M. Sohnius, {\it Nucl. Phys.} B {\bf 88},
257 (1975).

\bibitem{MSSM}
H.~E. Haber and G.~L. Kane, {\it Phys. Rep.} {\bf 117}, 75 (1985).

\bibitem{GUTs}
J.~Ellis, S.~Kelley and D.~V.~Nanopoulos,
{\it Phys. Lett.} B {\bf 260}, 131 (1991);
U.~Amaldi, W.~de Boer and H.~Furstenau,
{\it Phys. Lett.} B {\bf 260}, 447 (1991);
P.~Langacker and M.~x.~Luo,
{\it Phys. Rev.} D {\bf 44}, 817 (1991);
C.~Giunti, C.~W.~Kim and U.~W.~Lee,
{\it Mod. Phys. Lett.} A {\bf 6}, 1745 (1991).

\bibitem{GQW}
H.~Georgi, H.~R.~Quinn and S.~Weinberg,
{\it Phys. Rev. Lett.} {\bf 33}, 451 (1974).

\bibitem{susyHiggs}
Y.~Okada, M.~Yamaguchi and T.~Yanagida,
{\it Prog. Theor. Phys.} {\bf 85}, 1 (1991);
J.~R.~Ellis, G.~Ridolfi and F.~Zwirner,
{\it Phys. Lett.} B {\bf 257}, 83 (1991);
H.~E.~Haber and R.~Hempfling,
{\it Phys. Rev. Lett.} {\bf 66}, 1815 (1991).

\bibitem{EHNOS}
J. Ellis, J. S. Hagelin, D. V. Nanopoulos, K. A. Olive
and M. Srednicki, {\it Nucl. Phys.} B {\bf 238}, 453 (1984).

\bibitem{Goldberg}
H.~Goldberg, {\it Phys. Rev. Lett.} {\bf 50}, 1419 (1983).

\bibitem{BNL}
G.~W.~Bennett {\it et al.}  [Muon g-2 Collaboration],
{\it Phys. Rev. Lett.}  {\bf 89}, 101804 (2002) 
[Erratum-ibid. {\bf 89}, 1219903 (2002)]
[arXiv:hep-ex/0208001].

\bibitem{Davier}
M.~Davier, S.~Eidelman, A.~Hocker and Z.~Zhang,
arXiv:hep-ph/0208177; see also
K.~Hagiwara, A.~D.~Martin, D.~Nomura and T.~Teubner,
arXiv:hep-ph/0209187;
F.~Jegerlehner, unpublished, as reported in
M.~Krawczyk,
arXiv:hep-ph/0208076.

\bibitem{LEPsusy}
Joint LEP~2 Supersymmetry Working Group,
{\it Combined LEP Chargino Results, up to 208 GeV}, \\
{\tt http://lepsusy.web.cern.ch/lepsusy/www/inos{\_}moriond01/}
{\tt charginos{\_}pub.html}.

\bibitem{LEPSUSYWG_0101}
Joint LEP~2 Supersymmetry Working Group,
{\it Combined LEP
Selectron/Smuon/Stau Results, 183-208 GeV}, \\
{\tt 
http://lepsusy.web.cern.ch/lepsusy/www/sleptons{\_}summer02/}
{\tt slep{\_}2002.html}.

\bibitem{EOSS}
J.~Ellis, K.~A.~Olive, Y.~Santoso and V.~C.~Spanos,
arXiv:hep-ph/0303043.

\bibitem{bsg}
M.~S.~Alam {\it et al.}, [CLEO Collaboration], {\it Phys. Rev. Lett.} 
{\bf 74}, 2885 (1995), as updated in
S.~Ahmed et al., {CLEO CONF 99-10};
BELLE Collaboration, BELLE-CONF-0003, contribution to the 30th
International conference on High-Energy Physics, Osaka, 2000.
See also
K.~Abe {\it et al.},  [Belle Collaboration],
arXiv:hep-ex/0107065;
L.~Lista  [BaBar Collaboration],
arXiv:hep-ex/0110010;
C. Degrassi, P. Gambino and G.~F. Giudice,
JHEP {\bf 0012}, 009 (2000) [arXiv:hep-ph/0009337];
M.~Carena, D.~Garcia, U.~Nierste and C.~E.~Wagner,
{\it Phys. Lett.} B {\bf 499}, 141 (2001)
[arXiv:hep-ph/0010003].

\bibitem{EGNO}
J.~R.~Ellis, G.~Ganis, D.~V.~Nanopoulos and K.~A.~Olive,
{\it Phys. Lett.} B {\bf 502}, 171 (2001)
[arXiv:hep-ph/0009355].

\bibitem{BNL1}
H.~N.~Brown {\it et al.}  [Muon g-2 Collaboration],
{\it Phys. Rev. Lett.} {\bf 86}, 2227 (2001)
[arXiv:hep-ex/0102017].

\bibitem{lightbylight}
M.~Knecht and A.~Nyffeler,
arXiv:hep-ph/0111058;
M.~Knecht, A.~Nyffeler, M.~Perrottet and E.~De Rafael,
arXiv:hep-ph/0111059;
M.~Hayakawa and T.~Kinoshita,
arXiv:hep-ph/0112102;  
I.~Blokland, A.~Czarnecki and K.~Melnikov,
arXiv:hep-ph/0112117;
J.~Bijnens, E.~Pallante and J.~Prades,
arXiv:hep-ph/0112255.

\bibitem{susygmu}
L.~L.~Everett, G.~L.~Kane, S.~Rigolin and L.~Wang,
{\it Phys. Rev. Lett.} {\bf 86}, 3484 (2001)
[arXiv:hep-ph/0102145];  
J.~L.~Feng and K.~T.~Matchev,
{\it Phys. Rev. Lett.} {\bf 86}, 3480 (2001)
[arXiv:hep-ph/0102146];
E.~A.~Baltz and P.~Gondolo,
{\it Phys. Rev. Lett.}  {\bf 86}, 5004 (2001)
[arXiv:hep-ph/0102147];
U.~Chattopadhyay and P.~Nath, 
{\it Phys. Rev. Lett.}  {\bf 86}, 5854 (2001)
[arXiv:hep-ph/0102157];
S.~Komine, T.~Moroi and M.~Yamaguchi,
{\it Phys. Lett.} B {\bf 506}, 93 (2001)
[arXiv:hep-ph/0102204];
J.~Ellis, D.~V.~Nanopoulos and K.~A.~Olive,
{\it Phys. Lett.} B {\bf 508}, 65 (2001)
[arXiv:hep-ph/0102331];
R.~Arnowitt, B.~Dutta, B.~Hu and Y.~Santoso,
{\it Phys. Lett.} B {\bf 505}, 177 (2001)
[arXiv:hep-ph/0102344]
S.~P.~Martin and J.~D.~Wells,
{\it Phys. Rev.} D {\bf 64}, 035003 (2001)
[arXiv:hep-ph/0103067];
H.~Baer, C.~Balazs, J.~Ferrandis and X.~Tata,
{\it Phys. Rev.} D {\bf 64}, 035004 (2001)
[arXiv:hep-ph/0103280].

\bibitem{coann}
S.~Mizuta and M.~Yamaguchi,
{\it Phys. Lett.} B {\bf 298}, 120 (1993)
[arXiv:hep-ph/9208251];
J.~Edsjo and P.~Gondolo,
{\it Phys. Rev.} D {\bf 56}, 1879 (1997)
[arXiv:hep-ph/9704361].

\bibitem{ourcoann}
J.~Ellis, T.~Falk and K.~A.~Olive, {\it Phys. Lett.} B {\bf
444}, 367 (1998) [arXiv:hep-ph/9810360];
J.~Ellis, T.~Falk, K.~A.~Olive and M.~Srednicki,
{\it Astropart. Phys.} {\bf 13}, 181 (2000) [arXiv:hep-ph/9905481];
M.~E.~G\'omez, G.~Lazarides and C.~Pallis,
{\it Phys. Rev.} D {\bf 61}, 123512 (2000)
[arXiv:hep-ph/9907261]
and
{\it Phys. Lett.} B {\bf 487}, 313 (2000) [arXiv:hep-ph/0004028];
R.~Arnowitt, B.~Dutta and Y.~Santoso,
{\it Nucl. Phys.} B {\bf 606}, 59 (2001) [arXiv:hep-ph/0102181].

\bibitem{funnel}
M.~Drees and M.~M.~Nojiri,
{\it Phys. Rev.} D {\bf 47}, 376 (1993) [arXiv:hep-ph/9207234];
H.~Baer and M.~Brhlik,
{\it Phys. Rev.} D {\bf 53}, 597 (1996) [arXiv:hep-ph/9508321]
and {\it Phys. Rev.} D {\bf 57}, 567 (1998) [arXiv:hep-ph/9706509];
H.~Baer, M.~Brhlik, M.~A.~Diaz, J.~Ferrandis, P.~Mercadante, P.~Quintana
and X.~Tata,
{\it Phys. Rev.} D {\bf 63}, 015007 (2001) [arXiv:hep-ph/0005027];
A.~B.~Lahanas, D.~V.~Nanopoulos and V.~C.~Spanos, 
{\it Mod. Phys. Lett.} A {\bf 16}, 1229 (2001) [arXiv:hep-ph/0009065].

\bibitem{EFGOSi}
J.~R.~Ellis, T.~Falk, G.~Ganis, K.~A.~Olive and M.~Srednicki,
{\it Phys. Lett.} B {\bf 510}, 236 (2001)
[arXiv:hep-ph/0102098].

\bibitem{focus}
J.~L.~Feng, K.~T.~Matchev and T.~Moroi,
{\it Phys. Rev. Lett.} {\bf 84}, 2322 (2000)
[arXiv:hep-ph/9908309];
J.~L.~Feng, K.~T.~Matchev and T.~Moroi,
{\it Phys. Rev.} D {\bf 61}, 075005 (2000)
[arXiv:hep-ph/9909334];
J.~L.~Feng, K.~T.~Matchev and F.~Wilczek,
{\it Phys. Lett.} B {\bf 482}, 388 (2000)
[arXiv:hep-ph/0004043].

\bibitem{Bench}
M.~Battaglia {\it et al.}, {\it Eur. Phys. J.} C {\bf 22}, 535 (2001)
[arXiv:hep-ph/0106204].

\bibitem{EO}
J.~R.~Ellis and K.~A.~Olive,
{\it Phys. Lett.} B {\bf 514}, 114 (2001)
[arXiv:hep-ph/0105004].

\bibitem{EENZ}
J.~Ellis, K.~Enqvist, D.~V.~Nanopoulos and F.~Zwirner,
{\it Mod. Phys. Lett.} A {\bf 1}, 57 (1986);
R.~Barbieri and G.~F.~Giudice,
{\it Nucl. Phys.} B {\bf 306}, 63 (1988).

\bibitem{BenchKane}
G.~L.~Kane, J.~Lykken, S.~Mrenna, B.~D.~Nelson, L.~T.~Wang and T.~T.~Wang,
arXiv:hep-ph/0209061.

\bibitem{Tovey}
D.~R.~Tovey, {\it Phys. Lett.} B {\bf 498}, 1 (2001)
[arXiv:hep-ph/0006276].

\bibitem{Paige}
F.~E.~Paige, hep-ph/0211017.

\bibitem{CMS}
ATLAS Collaboration, {\it ATLAS detector and physics
performance
Technical Design Report}, CERN/LHCC 99-14/15 (1999);
S.~Abdullin {\it et al.}  [CMS Collaboration],
arXiv:hep-ph/9806366;
S.~Abdullin and F.~Charles,
{\it Nucl. Phys.} B {\bf 547}, 60 (1999) [arXiv:hep-ph/9811402];
CMS Collaboration, Technical Proposal, CERN/LHCC 94-38 (1994).

\bibitem{HP}
I.~Hinchliffe, F.~E.~Paige, M.~D.~Shapiro, J.~Soderqvist and W.~Yao,
{\it Phys. Rev.} D {\bf 55}, 5520 (1997).

\bibitem{EGO}
J.~R.~Ellis, G.~Ganis and K.~A.~Olive,
{\it Phys. Lett.} B {\bf 474}, 314 (2000)
[arXiv:hep-ph/9912324].

\bibitem{SS}
J.~Silk and M.~Srednicki, {\it Phys. Rev. Lett.} {\bf 53}, 624 (1984).

\bibitem{EFFMO}
J.~Ellis, J.~L.~Feng, A.~Ferstl, K.~T.~Matchev and K.~A.~Olive,
arXiv:astro-ph/0110225.

\bibitem{SOS}
J.~Silk, K.~A.~Olive and M.~Srednicki, {\it Phys. Rev. Lett.}
{\bf 55}, 257 (1985).

\bibitem{GW}
M.~W.~Goodman and E.~Witten,
{\it Phys. Rev.} D {\bf 31}, 3059 (1985).

\bibitem{DAMA}
R.~Bernabei {\it et al.} [DAMA Collaboration],
{\it Phys. Lett.} B {\bf 436}, 379 (1998).

\bibitem{unDAMA}
D.~Abrams {\it et al.}  [CDMS Collaboration],
arXiv:astro-ph/0203500;
A.~Benoit {\it et al.}  [EDELWEISS Collaboration],
{\it Phys. Lett.} B {\bf 513}, 15 (2001)
[arXiv:astro-ph/0106094].   

\bibitem{Schnee:1998gf}
R.~W.~Schnee {\it et al.} [CDMS Collaboration],
{\it Phys. Rept.} {\bf 307}, 283 (1998).

\bibitem{Bravin:1999fc}
M.~Bravin {\it et al.} [CRESST Collaboration],
{\it Astropart. Phys.} {\bf 12}, 107 (1999)
[arXiv:hep-ex/9904005].

\bibitem{GENIUS}
H.~V.~Klapdor-Kleingrothaus,
arXiv:hep-ph/0104028.  

\bibitem{neut}
G.~Jungman, M.~Kamionkowski and K.~Griest,
{\it Phys. Rept.} {\bf 267}, 195 (1996)
[arXiv:hep-ph/9506380];
{\tt http://t8web.lanl.gov/people/jungman/neut-package.html}.

\bibitem{inflation}
D.~H.~Lyth and A.~Riotto,
{\it Phys. Rept.} {\bf 314}, 1 (1999)
[arXiv:hep-ph/9807278].

\bibitem{Guth}
A.~H.~Guth,
{\it Phys. Rev.} D {\bf 23}, 347 (1981).

\bibitem{SN}
A.~G.~Riess {\it et al.}  [Supernova Search Team Collaboration],
{\it Astron. J.} {\bf 116}, 1009 (1998)
[arXiv:astro-ph/9805201];
S.~Perlmutter {\it et al.}  [Supernova Cosmology Project Collaboration],
{\it Astrophys. J.} {\bf 517}, 565 (1999)
[arXiv:astro-ph/9812133];
Perlmutter, S. \& Schmidt, B.~P. 2003 arXiv:astro-ph/0303428;
J.~L.~Tonry {\it et al.}, arXiv:astro-ph/0305008.

\bibitem{Bahcall}
N.~A.~Bahcall, J.~P.~Ostriker, S.~Perlmutter and P.~J.~Steinhardt,
{\it Science} {\bf 284}, 1481 (1999)
[arXiv:astro-ph/9906463].

\bibitem{Linde}
A.~D.~Linde,
{\it Phys. Lett.} B {\bf 108}, 389 (1982).

\bibitem{chaos}
A.~D.~Linde,
{\it Phys. Lett.} B {\bf 129}, 177 (1983).

\bibitem{extended}
D.~La and P.~J.~Steinhardt,
{\it Phys. Rev. Lett.} {\bf 62}, 376 (1989)
[Erratum-ibid.\  {\bf 62}, 1066 (1989)].

\bibitem{susyinf}
J.~R.~Ellis, D.~V.~Nanopoulos, K.~A.~Olive and K.~Tamvakis,
{\it Phys. Lett.} B {\bf 118}, 335 (1982) and
{\it Nucl. Phys.} B {\bf 221}, 524 (1983).

\bibitem{perts}
J.~M.~Bardeen, P.~J.~Steinhardt and M.~S.~Turner,
{\it Phys. Rev.} D {\bf 28}, 679 (1983).

\bibitem{Kinney}
W.~H.~Kinney,
arXiv:astro-ph/0301448.

\bibitem{ERY}
J.~R.~Ellis, M.~Raidal and T.~Yanagida,
arXiv:hep-ph/0303242.

\bibitem{realbarger}
V.~Barger, H.~S.~Lee and D.~Marfatia,
arXiv:hep-ph/0302150.

\bibitem{MY}
H.~Murayama, H.~Suzuki, T.~Yanagida and J.~Yokoyama,
{\it Phys. Rev. Lett.} {\bf 70}, 1912 (1993);
H.~Murayama, H.~Suzuki, T.~Yanagida and J.~Yokoyama,
{\it Phys. Rev.} D {\bf 50}, 2356 (1994)
[arXiv:hep-ph/9311326].

\bibitem{Sakharov}
A.~D.~Sakharov,
{\it Pisma Zh. Eksp. Teor. Fiz.} {\bf 5}, 32 (1967)

\bibitem{UHECR}
M.~Takeda {\it et al.}, arXiv:astro-ph/0209422;
T.~Abu-Zayyad {\it et al.}  [High Resolution Fly's Eye Collaboration],
arXiv:astro-ph/0208243 and
arXiv:astro-ph/0208301.

\bibitem{GG}
H. Georgi and S.L. Glashow, {\it Phys. Rev. Lett.} {\bf 32}, 438 (1974).

\bibitem{DRW}
S. Dimopoulos and H. Georgi, {\it Nucl. Phys.} B {\bf 193}, 50 (1981);
S.~Dimopoulos, S.~Raby and F.~Wilczek, {\it Phys. Rev.} D {\bf 24}, 1681 
(1981); L.~Ib\`a\~nez and G.~G.~Ross, {\it Phys. Lett.} B {\bf 105}, 439 
(1981).

\bibitem{SU5mix}
J. Ellis, M. K. Gaillard and D. V. Nanopoulos, {\it Phys. Lett.} B {\bf
91}, 67 (1980).

\bibitem{BEGN}
A. J. Buras, J. Ellis, M. K. Gaillard and D. V. Nanopoulos,
{\it Nucl. Phys.} B {\bf 135}, 66 (1978).

\bibitem{SKpdk}
M. Shiozawa {\it et al.} [Super-Kamiokande collaboration], {\it Phys. 
Rev. Lett.} {\bf 81}, 3319 (1998).

\bibitem{susySU5pdk}
J. Ellis, D. V. Nanopoulos and S. Rudaz, {\it Nucl. Phys.} B {\bf 202}, 43
(1982);
S.~Dimopoulos, S.~Raby and F.~Wilczek, {\it Phys. Lett.} B {\bf 112}, 133 
(1982).

\bibitem{WSY}
S.~Weinberg, {\it Phys. Rev.} D {\bf 26}, 287 (1982);
N.~Sakai and T.~Yanagida, {\it Nucl. Phys.} B {\bf 197}, 533 (1982).

\bibitem{Yoshimura}
M.~Yoshimura,
{\it Phys. Rev. Lett.} {\bf 41}, 281 (1978)
[Erratum-ibid.  {\bf 42}, 746 (1979)].

\bibitem{EGNbnotu}
J.~R.~Ellis, M.~K.~Gaillard and D.~V.~Nanopoulos,
{\it Phys. Lett. B} {\bf 80}, 360 (1979)
[Erratum-ibid. B {\bf 82}, 464 (1979)].

\bibitem{GKZ}
K. Greisen, {\it Phys.\ Rev.\ Lett.} {\bf 16}, 748 (1966);
G.~T.~Zatsepin and V.~A.~Kuzmin,
{\it Pisma Zh. Eksp. Teor. Fiz.} {\bf 4}, 114 (1966).

\bibitem{Igor}
P.~G.~Tinyakov and I.~I.~Tkachev,
{\it Pisma Zh. Eksp. Teor. Fiz.} {\bf 74}, 3 (2001)
[arXiv:astro-ph/0102101].

\bibitem{Igor2}
P.~G.~Tinyakov and I.~I.~Tkachev,
{\it Pisma Zh. Eksp. Teor. Fiz.} {\bf 74}, 499 (2001)
[arXiv:astro-ph/0102476] and
arXiv:astro-ph/0301336. See, however,
W.~Evans, F.~Ferrer and S.~Sarkar,
arXiv:astro-ph/0212533.

\bibitem{aniso}
N.~W.~Evans, F.~Ferrer and S.~Sarkar,
{\it Astropart. Phys.} {\bf 17}, 319 (2002)
[arXiv:astro-ph/0103085].

\bibitem{Navarro}
M.~G.~Abadi, J.~F.~Navarro, M.~Steinmetz and V.~R.~Eke,
arXiv:astro-ph/0211331 and
arXiv:astro-ph/0212282.

\bibitem{cryptons}
J.~R.~Ellis, J.~L.~Lopez and D.~V.~Nanopoulos,
{\it Phys. Lett.} B {\bf 247}, 257 (1990);
V.~Berezinsky, M.~Kachelriess and A.~Vilenkin,
{\it Phys. Rev. Lett.} {\bf 79}, 4302 (1997)
[arXiv:astro-ph/9708217].

\bibitem{BEN}
K.~Benakli, J.~R.~Ellis and D.~V.~Nanopoulos,
{\it Phys. Rev.} D {\bf 59}, 047301 (1999)
[arXiv:hep-ph/9803333].

\bibitem{EGLNS}
J.~R.~Ellis, G.~B.~Gelmini, J.~L.~Lopez, D.~V.~Nanopoulos and S.~Sarkar,
{\it Nucl. Phys.} B {\bf 373}, 399 (1992).

\bibitem{Sarkar}
See, for example,
M.~Birkel and S.~Sarkar,
{\it Astropart. Phys.} {\bf 9}, 297 (1998)
[arXiv:hep-ph/9804285].

\bibitem{Chung}
See, for example,
D.~J.~Chung, P.~Crotty, E.~W.~Kolb and A.~Riotto,
{\it Phys. Rev.} D {\bf 64}, 043503 (2001)
[arXiv:hep-ph/0104100].

\bibitem{Auger}
A.~Letessier-Selvon,
arXiv:astro-ph/0208526; 
J. Cronin {\it et al.}, \\
{\tt http://www.auger.org/}.

\bibitem{EUSO}
L.~Scarsi,      
{\it EUSO: Using high energy cosmic rays and neutrinos as messengers from
the unknown universe}, in
{\it Metepec 2000, Observing ultrahigh energy cosmic rays from space
and earth}, p113.

\end{thebibliography}
\end{document}